\documentclass[prx,aps,twocolumn,floatfix,longbibliography,nofootinbib]{revtex4-1}
\usepackage[utf8]{inputenc}
\usepackage{natbib}
\usepackage{graphicx}
\usepackage{color}
\usepackage[tbtags]{amsmath}
\usepackage{hyperref}
\hypersetup{colorlinks,allcolors=black}
\usepackage{amssymb}
\usepackage{gensymb}
\usepackage{float}
\usepackage{amsmath}
\usepackage{tabularx,graphicx}
\usepackage{epstopdf}
\usepackage{latexsym}
\usepackage{color, colortbl}
\usepackage{psfrag}
\usepackage{bbm}
\usepackage{bm}
\usepackage{titlesec}
\usepackage{dsfont}
\usepackage{feynmp}
\usepackage{slashed}
\usepackage{multirow}
\usepackage[normalem]{ulem}
\newcommand{\be}{\begin{equation}}
\newcommand{\ee}{\end{equation}}
\newcommand{\beq}{\begin{eqnarray}}
\newcommand{\eeq}{\end{eqnarray}}
\newcommand{\ba}{\[\begin{aligned}}
\newcommand{\ea}{\end{aligned}\]}
\newcommand{\HH}{{\cal H}}

\newcommand{\s}{\sigma}
\newcommand{\la}{\langle}
\newcommand{\ra}{\rangle}

\newcommand{\V}{{\cal V}}

\renewcommand{\Im}{{\rm Im\,}}
\renewcommand{\vec}[1]{{\bf #1}}

\renewcommand{\phi}{\varphi}
\renewcommand{\epsilon}{\varepsilon}
\renewcommand{\dag}{\dagger}
\def\nn{\nonumber}

\renewcommand{\vec}[1]{\boldsymbol{#1}}

\def \W{{\Omega}}
\def \e{{\epsilon}}
\def \ve{{\varepsilon}}

\def \a{{\alpha}}

\def \g{{\gamma}}
\def \D{{\Delta}}
\def \d{{\delta}}
\def \w{{\omega}}
\def \s{{\sigma}}

\def \e{{\epsilon}}

\def \ra{{\rangle}}
\def \la{{\langle}}
\def \tn{\textnormal}

\def \ba{\begin{align*}}
\def \ea{\end{align*}}

\newcounter{indice}

\def \mrm{\mathrm}

\def \bs{\boldsymbol}
\def \mc{\mathcal}

\begin{document}
\title{Pairing in magic-angle twisted bilayer graphene: role of phonon and plasmon umklapp}
\author{Cyprian Lewandowski}
\affiliation{Department of Physics, California Institute of Technology, Pasadena, CA 91125, USA}
\affiliation{Department of Physics, Massachusetts Institute of Technology, Cambridge MA 02139, USA}

\author{Debanjan Chowdhury}
\affiliation{Department of Physics, Cornell University, Ithaca NY 14853, USA}

\author{Jonathan Ruhman}
\affiliation{Department of Physics, Bar-Ilan University, Ramat Gan 52900, Israel}
\affiliation{Center for Quantum Entanglement Science and Technology, Bar-Ilan University, Ramat Gan 52900, Israel}

\begin{abstract}
Identifying the microscopic mechanism for superconductivity in magic-angle twisted bilayer graphene (MATBG) is an outstanding open problem. While MATBG exhibits a rich phase-diagram, driven partly by the strong interactions relative to the electronic bandwidth, its single-particle properties are unique and likely play an important role in some of the phenomenological complexity. Some of the salient features include an electronic bandwidth smaller than the characteristic phonon bandwidth and a non-trivial structure of the underlying Bloch wavefunctions.
We perform a theoretical study of the cooperative effects due to phonons and plasmons on pairing in order to disentangle the distinct role played by these modes on superconductivity. We consider a variant of MATBG with an enlarged number of fermion flavors, $N \gg 1$, where the study of pairing instabilities reduces to the conventional (weak-coupling) Eliashberg framework. In particular, we show that certain {\it umklapp} processes involving mini-optical phonon modes, which arise physically as a result of the folding of the original acoustic branch of graphene due to the moir{\'e} superlattice structure, contribute significantly towards enhancing pairing.
We also investigate the role played by the dynamics of the screened Coulomb interaction on pairing, which leads to an enhancement in a narrow window of fillings, and study the effect of external screening due to a metallic gate on superconductivity. At strong coupling the dynamical pairing interaction leaves a spectral mark in the single particle tunneling density of states. We thus predict such features will appear at specific frequencies of the umklapp phonons corresponding to the sound velocity of graphene times an integer multiple of the Brillouin zone size.  
\end{abstract}
\maketitle


\section{Introduction}

Since the discovery of superconductivity (SC) \cite{cao2} near an interaction-induced  insulator \cite{cao1} in magic-angle twisted bilayer graphene (MATBG), the field has evolved dramatically \cite{Yankowitz1059,efetov2019,Balents2020}.
One of the defining characteristic features of MATBG is the emergence of isolated, (nearly) flat bands separated by a large energy gap to higher dispersive bands~\cite{Castro,suarez,macdonald11}. A variety of symmetry broken states and phase transitions have been observed experimentally~\cite{AP19,SNP19,EA19,efetov2019,SI20,AY20,PJH20} as a function of electron filling ($\nu\in [-4,4]$) measured in units of electrons per moir\'e unit cell~\cite{Cao2016}. In spite of some similarities between the phenomenology in MATBG and copper-oxide based materials \cite{Keimer15}, MATBG bears its own set of unique properties, which likely play a central role in the underlying microscopic origins of superconductivity and other experimentally observed features. For example, MATBG has spatially extended Wannier wavefunctions \cite{TS1}, a non-trivial band topology \cite{TS2,TS3,AB1,BJY1}, a complex phonon band structure \cite{MK19,HO19} that may extend beyond the electronic bandwidth, and plasmons that decouple from the particle-hole continuum associated with the narrow bands \cite{Lewandowski20869}.

These unique features call for a detailed analysis of their influence on the low-energy electronic properties in MATBG. For example, while the enhancement of Coulomb interactions relative to the narrow bandwidth undoubtedly plays an important role in stabilizing the insulators at various commensurate fillings \cite{cao1,Yankowitz1059,efetov2019}, the extent to which they are crucial for superconductivity remains unclear. A number of recent experiments have attempted to study the role of Coulomb interactions, either by varying the distance between the MATBG layer and a nearby metallic screening gate across different devices \cite{stepanov2019interplay,saito2019decoupling} 
{\footnote{The devices are nominally at different twist angles and have varying levels of disorder; see \cite{senthil2020jccm} for a further discussion.}}, by independently varying the density of carriers in a layer of Bernal-stacked bilayer graphene \cite{liu2020tuning}, situated a fixed distance away from MATBG in the same device, or by stabilizing superconductivity at angles far away from the magic angle where electronic correlations are suppressed\cite{10.1038/s41586-020-2473-8}. Not surprisingly, the enhanced screening suppresses the various insulating phases, that develop originally at a sequence of commensurate fillings. On the other hand, SC is affected only weakly, if at all, by the gate.  A number of recent theoretical studies of superconductivity in MATBG have relied on a variety of BCS mean-field and other weak-coupling based approaches \cite{Fu18,TH18a,Wu18,CK18,HJC18,You2019,BAB19,RN19a,PhysRevB.102.064501}{, whilst other works extend the treatment to an Eliashberg framework \cite{TH18b,SA19,PO20}}.  In particular, Ref.\cite{PO20} applied Eliashberg theory to study pairing mediated by a single Einstein Einstein phonon, where the electron-phonon coupling strength was chosen phenomenologically. Ref.\cite{SA19} considered plasmon-mediated pairing in a hexagonal lattice model for MATBG. See also two recent studies \cite{AV20,SS20} focusing on the interplay of insulating and superconducting states, where the underlying band topology plays a crucial role.

Inspired by the growing number of such interesting experiments, we focus on a possible microscopic mechanism for the superconducting instabilities in MATBG due to an interplay of attraction generated by phonons and purely electronic collective modes, such as plasmons.
Importantly, we take into account the retarded interaction mediated by the acoustic phonons and incorporate the dynamics of the screened Coulomb interaction within Eliashberg theory \cite{DJSeliash}. In the experimental regime of interest, MATBG is likely defined by an ``intermediate-coupling" problem with no natural small parameter. In order to make theoretical progress, we introduce a large$-N$ expansion, such that our results correspond to the weak-coupling limit (obtained by taking $N\gg 1$). We argue that in spite of this approximation, the results shed light on the interplay of different sources of attraction on SC in MATBG. Our theoretical approach is not suited to discuss the insulating states observed in MATBG at the integer fillings \cite{efetov2019,SI20,AY20}. However, it is entirely possible that the interactions responsible for SC are distinct from those responsible for the insulating phases \cite{SDSRN19}. This work addresses only the possible role played by the interplay of phonons and screened Coulomb interactions on SC in MATBG in a systematic fashion. It is entirely conceivable that the strong-interaction induced insulating states, that develop at commensurate fillings, simply punctuate the SC phase-diagram that we discuss and present in this study.    

\begin{figure}[!tb]
    \centering
    \includegraphics[width=\linewidth]{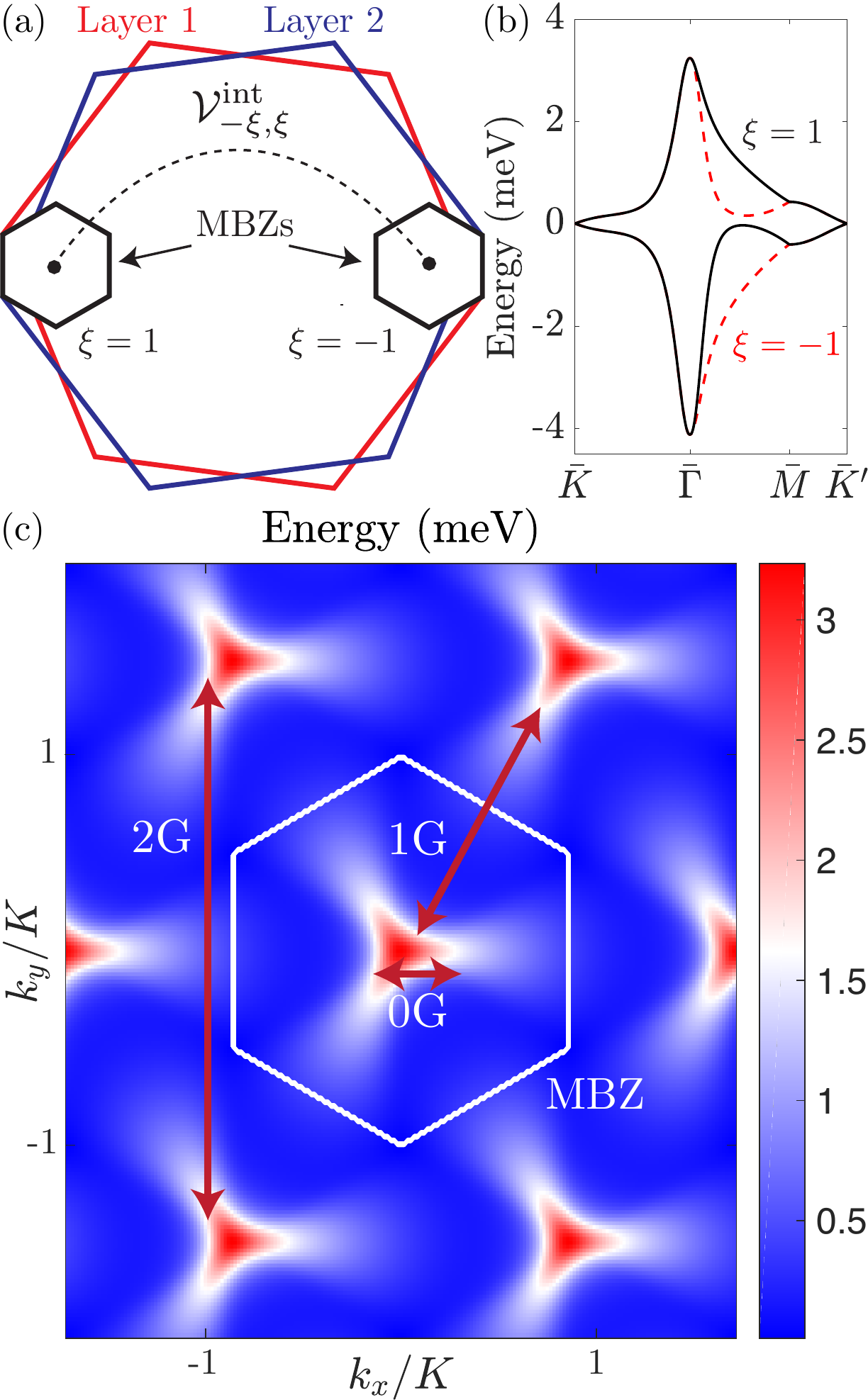}
    \caption{(a) A small twist-angle between two graphene layers leads to formation of mini-Brillouin zones located in the vicinity of the $\bs K$ and $\bs K'$ points. We focus on intervalley ($\xi \neq \xi'$) pairing (dashed line) to form Cooper pairs at zero center of mass momentum. (b) The moir{\'e} interlayer coupling gives rise to formation of two (spin degenerate) nearly-flat bands for each valley $\xi$. $\bar{K},\bar{\Gamma},\bar{M},\bar{K}'$ correspond to high symmetry points of the MBZ. Positive ($\nu>0$) and negative ($\nu<0$) fillings correspond to the electron and hole band, respectively. (c) The energy landscape of an electron band for $\xi=1$. The red arrows depict the umklapp processes ($0$G, $1$G, $2$G) between states originating from different MBZs (one shown as white hexagon) in the extended zone scheme. Here $K=|\bar{K}-\bar{\Gamma}|$.}
    \label{fig:fig_1}
\end{figure}

One of the unique features of MATBG is the large size of the unit-cell, compared to the inter atomic distance. Consequently, higher order {\it umklapp} processes involve momenta which are still small compared to the original Brillouin zone (BZ) and thus, play an important role in phonon-mediated SC {\footnote{A detailed study of these umklapp processes and on their role in superconductivity has not been clarified in earlier studies \cite{Wu18,BAB19,PhysRevB.99.165112} {- we relate to these works when discussing our results.}}}  Below, we show that the SC transition temperature, $T_c$, due to acoustic phonons of graphene is enhanced significantly upon including the effects of these umklapp processes for a wide range of dopings. On the other hand, pairing due to plasmons is dominant only in a narrow range of fillings $\nu \approx 2-3$ and is much less sensitive to the inclusion of umklapp processes.

To verify our predictions, we propose a tunneling experiment.  As is well known when the coupling is strong enough ``dip-hump" features in the tunneling spectra appear at energies that can be associated with the frequency of the bosonic modes that contribute most to pairing~\cite{Schrieffer1963}. 
In the present case the umklapp phonons have a very distinct frequency scale associated with the speed of sound in graphene and an integer multiple of the MBZ size. Thus, this allows us to make a rather sharp prediction for the energy at which such features will appear. Note that this prediction does not depend on the model for the electronic band structure.  

Finally, we also study the effect of a screening layer, coupled to MATBG by Coulomb interactions, on the density dependence of $T_c$. The distinct effect of screening on the plasmonic modes in graphene is discussed and proposed as a method to distinguish their contribution to pairing.

The remainder of this paper is organized as follows: in Sec.~\ref{sec:model}, we review the continuum model for MATBG and introduce the Coulomb and electron-phonon interactions, projected to the nearly flat bands. We also highlight some of the technical aspects associated with umklapp scattering and the large$-N$ formulation in this section. In Sec.~\ref{sec:eliashberg}, we focus on setting up the (linearized) Eliashberg equations for an inter-valley, spin-singlet $s-$wave pairing gap and highlight some salient features associated with the pairing kernel. Sec.~\ref{sec:results} is devoted to our results, which includes a comprehensive study of the filling dependence of $T_c$ due to the different sources of attraction, the role of external screening and a discussion of our proposed experimental setup for investigating the fingerprint of umklapp phonon processes. Sec.~\ref{sec:dis} concludes with a discussion and future outlook for interesting directions. We describe some of the technical aspects of our work in appendices \ref{app:bandstructure} - \ref{app:dos}.

\section{Model} 
\label{sec:model}
The non-interacting part of our model, $\mathcal{H}_0$, leads to the action,
\be
\mathcal{S}_{0}=\sum_{i=1}^N\sum_{\vec{k},\w,\{\g\}} \left(-i\w+E_{\vec{k}\{\g\}}\right) c^\dag_{\w\vec{k}\{\g\};i}   c_{\w\vec{k}\{\g\};i},
\ee
where $c_{\w\vec{k}\{\g\};i}$ is the electron annihilation operator in the eight-dimensional space of $\{\g\}\equiv\{\xi,\s,n\}$, with $\xi$ being the valley index of the original graphene layers, $\sigma$ labeling the electron spin and $n$ tracking the electronic band index of MATBG. 
We have introduced a fictitious index $i=1,..,N$ which labels $N$ identical copies of the non-interacting MATBG Hamiltonian, with an eye for carrying out a $1/N$ expansion {\footnote{$N = 1$ corresponds to the physical limit of MATBG.}}. 
We denote $\vec{k}$ as the crystal momentum in the original graphene layer, such that the two mini-Brillouin zones (MBZ) are located near $\vec{K}$ and $\vec{K'}$ points of the original graphene layers, and the $\vec{k}$ summation ranges over the MBZ, see schematic in Fig.\ref{fig:fig_1}a.

The energy bands, $E_{\vec{k} \{\g\} }$, are computed from the effective continuum Hamiltonian introduced in Ref.~\cite{MK18} for MATBG at a twist angle  $\theta=1.05^\circ$  {\footnote{We note that the eigenstates and eigenvalues of the two valleys $\xi=\pm$ are time-reversed partners and accompanied by a simultaneous complex conjugation and $\vec{k}\to-\vec{k}$ transformation. }}. The band structure of the two narrow bands, which are relevant for superconductivity are shown in Fig.\ref{fig:fig_1}b. In the analysis that follows, we introduce a chemical potential ($\mu$) in the usual way by shifting the single-particle energies, $E_{\vec{k} \{\g\} } \to E_{\vec{k} \{\g\} } - \mu$, and control the electron filling. Additional details associated with the bandstructure appears in Appendix \ref{app:bandstructure}.

Let us now turn our attention to the interaction terms in our model. As a result of the large momentum separation between $\vec{K}$ and $\vec{K'}$ in the original BZ (see Fig.\ref{fig:fig_1}a), the states near $\xi=\pm$ valleys in the two MBZs are effectively decoupled. The interaction is then given by, 
\begin{align}\label{eq:ham_int}
&\mc S_{\tn{int}} = \frac{1}{2}\sum_{\substack{\vec{q},\w\\ \xi,\xi',i}}\V^{\mrm{int}}_{\xi,\xi'}(\vec{q},i\w)\nn \rho_{\xi; i}(\vec{q},i\w)~\rho_{\xi';i}(-\vec{q},-i\w),\\
&\rho_{\xi; i}(\vec{q},i\w) = \sum_{\bs k, \nu} \Lambda_{\g\g'}(\bs k+\bs q,\bs k)~c^\dagger_{\nu+\w\vec{k}+\vec{q}\{\g\};i} c_{\nu\vec{k}\{\g'\};i}\,, 
\end{align}
where the density operators, $\rho_{\xi; i}(\vec{q},i\w)$, include the form-factors,
$\Lambda_{\g\g'}(\bs p,\bs k)=\delta_{\xi\xi'}\delta_{\s\s'}\left\langle \bs p,\{\g\}\left|e^{i (\vec{p-k})\cdot \vec{r}} \right|\bs k,\{\g'\}\right \rangle$, which necessarily involve states located near the same valley $\xi$ as the Hamiltonian is block-diagonal in the valley space. Here $|\bs k,\{\g'\} \rangle$ denotes a Bloch wavefunction of the non-interacting Hamiltonian (see Appendix \ref{app:bandstructure}). Note that we have assumed explicitly that the interaction term is diagonal in the fictitious $i=1,\dots,N$ index.  
The interaction vertex above is a sum of two contributions --- the dynamically screened Coulomb interaction ($\V^{\tn{C}}_{\xi,\xi'}$) and the phonon mediated interaction ($\V^{\tn{ph}}_{\xi,\xi'}$):
\be \label{eq:Int}
\V^{\mrm{int}}_{\xi,\xi'}(\vec{q},i\w) = \V^{\tn{C}}_{\xi,\xi'}(\vec{q},i\w)+\V^{\tn{ph}}_{\xi,\xi'}(\vec{q},i\w)\,.
\ee
To stress the intervalley nature of the pairing in what follows, we make the labels $\xi$ and $\xi'$ explicit in both the Coulomb and phonon interactions as well as make the Coulomb and phonon interaction manifestly intervalley only through the $\delta_{\xi,-\xi'}$ (see below). The microscopic form of the interaction however is the same for both inter-/intravalley interactions with the dependence on valley indices $\xi,\xi'$ contained only in the density operators $\rho_{\xi}(\vec{q},i\omega)$ introduced in Eq.~\eqref{eq:ham_int}.

The first term in Eq.~\eqref{eq:Int} is the dynamically screened Coulomb interaction, which is given by 
\be  \label{eq:V_C}
\V^{\tn{C}}_{\xi,\xi'}(\vec{q},i\w) = {1\over N}{2\pi e^2\over \ve_{\mrm{RPA}}(\bs q ,i\w) q}\delta_{\xi,-\xi'}\,,
\ee
where the RPA dielectric function is given by $\ve_{\mrm{RPA}}(\bs q,i\w) = \kappa - 2\pi e^2 \Pi_{ee}(\bs q ,i\w)/q$. Here $\Pi_{ee}(\bs q, i\w)$ is the electronic polarization of MATBG \cite{Lewandowski20869} and $\kappa$ is the background dielectric constant, which can in principle be frequency and momentum dependent as well.
 
The second term, the phonon-mediated interaction, arises from the electron-phonon coupling Hamiltonian,
\be\label{eq:ham_phonon}
\HH_{\tn{el-ph}} = -i\sqrt{\frac{gc_s}{2N}}\sum_{\xi, \vec{q}} \sqrt{ q}~ \rho_{\xi; i}(\vec{q}) (a_{\vec{q}}+a^\dagger_{-\vec{q}}),
\ee
where $a_{\vec{q}},~a^\dagger_{\vec{q}}$ represent the phonon annihilation and creation operators, respectively{\footnote{Here we use the notation that $\rho_{\xi; i}(\vec{q})\equiv\rho_{\xi; i}(\vec{q},i\omega=0)$ as per the definition in Eq.~\eqref{eq:ham_int}}}. The phonon coupling constant, $g = D^2/\rho_m c_s^2$, is related to the deformation potential, which we set as $D~(= 25~ \tn{eV})$, and we also use $c_s~ (= 12 000~\tn{m/s})$ for the speed of sound in graphene and $\rho_m ~(= 7.6\times 10^{-8} ~\mathrm{g}/\mathrm{cm}^2)$ for the atomic mass density \cite{PhysRevLett.105.256805,Chen2008}. After integrating out the phonons, we obtain the density-density interaction term of the form Eq.~\eqref{eq:ham_int} where, 
\be \label{eq:V_ph}
\V_{\xi,\xi'}^{\tn{ph}}(\vec{q},i\w) = -{g\over N}{\omega^2_{\tn{ph}}(q) \over \w^2 + \omega^2_{\tn{ph}}(q)} \delta_{\xi,-\xi'},
\ee
and $\omega_{\tn{ph}}(q)=c_s q$ is the acoustic phonon dispersion for graphene. Note that we have chosen the individual $N-$dependent normalizations such that both $\V^{\tn{c}}_{\xi,\xi'}$ and $\V_{\xi,\xi'}^{\tn{ph}}$ have the same $1/N$ prefactor and can therefore, be compared in a meaningful fashion in the large$-N$ limit. 

Let us now clarify an important aspect of the underlying scattering processes, that play a crucial role in our subsequent analysis. The summation over the crystal momenta, $\vec{k}$, in Eq.~\eqref{eq:ham_int} is limited to the first MBZ. However, the momentum transfer, $\vec{q} \equiv \bs G + \tilde {\bs q}$,
can scatter to Bloch states belonging to different MBZs, where $\tilde {\bs q}$ is now restricted to lie in the first MBZ. The lattice vector, $\vec{G}=m_1 \vec{G}_{1}^{M}+ m_2 \vec{G}_{2}^{M}$ ($m_1,m_2 \in \mathbb{Z}$), accounts for  scattering by the multiplicity of reciprocal moir{\'e} lattice vectors,  $\vec{G}_{1}^M$, $\vec{G}_{2}^M$. In order to evaluate the importance of the umklapp processes described above, from now on we restrict our analysis to the $m^{\tn{th}}$ MBZ, such that $m_1,m_2 \in [-m,m]$. We refer to such an analysis as including  $m$G processes, see Fig.\ref{fig:fig_1}c. Varying $m$ allows us to assess the relative importance of the successive umklapp processes. 
As we shall highlight later on, while including a few umklapp processes beyond the simplest 0G process leads to qualitatively new features, there is also rapid convergence with increasing $m$, such that we do not need to include processes with arbitrarily large values of $m$. 

It is worth noting that these umklapp processes can be equivalently viewed as processes where {\it mini-optical} phonons are exchanged. These optical-like modes are a natural consequence of folding of the original acoustic phonon branch of graphene due to the moir\'e potential \cite{MK19}, and have no relation to the optical phonons of the original (decoupled) graphene layers. {In what follows we will also relate our findings to previous works that focused on pairing mechanisms mediated by optical phonons\cite{HJC18,Wu18,BAB19,PhysRevB.99.165112,PO20} irrespective of their microscopic origins or frequency, i.e. whether they are mini-optical phonons arising from folding of the acoustic dispersion, actual optical phonons of graphene or just phenomenological Einstein modes. We also clarify that the phonons we consider correspond to symmetric motion of atoms in both graphene layers, but as Ref. \cite{MK19} points out, there are also in-plane phonon modes which we ignore in our study corresponding to asymmetric motion of atoms in both layers that can develop small ``gaps'' as a result of the moir\'e potential.}

\section{Eliashberg Equations}
\label{sec:eliashberg}

We are interested in the linearized gap equation in the pairing channel, ignoring the self-energy corrections to the electron dispersion and quasiparticle weight (for details on how to include them we refer the reader to Ref.~\cite{takada1992insignificance}). The gap function in the spin-singlet, $s$-wave pairing channel with zero center of mass momentum is defined as,
\beq
 \Delta(i\omega,{k}) \equiv \la\ve_{\s\s'} c_{\omega,\vec{k}\{\g\}} c_{-\omega,-\vec{k}\{\g'\}}\ra,
\eeq
where $\ve_{\alpha\beta}$ is the fully antisymmetric tensor in the indices $\alpha,~\beta$ and we have further assumed that the gap function has no explicit dependence on the angle of $\vec{k}$. The latter assumption is consistent with our assumption of intervalley Cooper pairing from valleys that are time-reversed partners, as explained before. Note that the above requires an intervalley scattering, Fig.\ref{fig:fig_1}a, which imposes the condition $\xi \neq \xi'$ in the interaction Eq.~\eqref{eq:ham_int}. The gap equation thereby reduces to an eigenvalue problem
\begin{align}
 \Delta(i\omega,{k}) = -T \sum_{\nu} \sum_{{p}} K(i\w,{k};i\nu,{p})\Delta(i\nu,{p}) \label{eq:sc_gap_equation},
\end{align}
where the kernel $K(...)$ is given by 
\begin{align}\label{eq:Kernel}
K(i\w,&{k};i\nu,{p}) \equiv \\&{1\over (2\pi)^2} \int d\W_{\bs p} \V^{\tn{int}}_{-\xi,\xi}(\bs k-\bs p,i\omega-i\nu) \frac{\Lambda( \bs p ,\bs k)\Lambda(-\bs p,\bs k) }{\nu^2+{E}^2_{\xi,\vec{p}}}\nn
\end{align}
and where $\int d \W_{\bs p}$ denotes integration over the angle between vector $\vec{k}$ and $\vec{p}$ for a fixed direction of $\vec{k}$.  \footnote{Note that here we have neglected the extended $s$-wave of the $D_3$ symmetry group.} In the above expression, as we are focusing on an intervalley pairing, we have suppressed the indices $\gamma,\gamma'$ on each of the two form-factors for clarity, but the two $\Lambda(...)$ terms carry opposite time-reversed labels. Accordingly the valley indices are omitted henceforth.

The Eliashberg equation for the pairing gap in Eq.~\eqref{eq:sc_gap_equation} ignores a number of possibly important contributions. We ignore an interaction induced momentum-dependent (``Hartree'') renormalization of the MATBG dispersion, which has been argued to lead to qualitative changes in the bandstructure \cite{Guinea13174,2020arXiv200414784G}. We have (artificially) introduced a control parameter that selects a subset of the relevant processes {and weakens the strength of the pairing mechanism. This places our analysis in the weak-coupling regime allowing us to focus on solving one equation only for the superconducting order parameter.} Namely, we have neglected the mass and dispersion renormalizations, $Z$ and $\chi$ (as these are subleading in $1/N$). For more details on their inclusion we refer the readers to Ref.~\cite{takada1992insignificance}. We also note that the mass renormalization $Z$ tends to reduce $T_c$ by a factor~\cite{PhysRevB.101.024503}, while the $\chi$ becomes most important when $T_c$ is of the order of the Fermi energy $\e_F$~\cite{PhysRevB.99.094524,PhysRevB.101.024503} (i.e. at strong coupling).  

With the possible qualitative changes in mind, in the analysis that follows we clearly identify and distinguish the features that follow from generic properties associated with moir{\'e} narrow-band systems and the ones that are tied to the specific aspects of the model and approximations used. This allows us to shed interesting light on the importance of different interactions on the origin of pairing and various qualitative aspects thereof in the limit where the Eliashberg framework is applicable. On the other hand,
it is important to recall that the large-$N$ model we consider here is somewhat different from the experimental situation. Therefore, the actual value of $T_c$ takes meaning only by comparison and not by its absolute value.

\begin{figure}[tb]
    \centering
    \includegraphics[width=\linewidth]{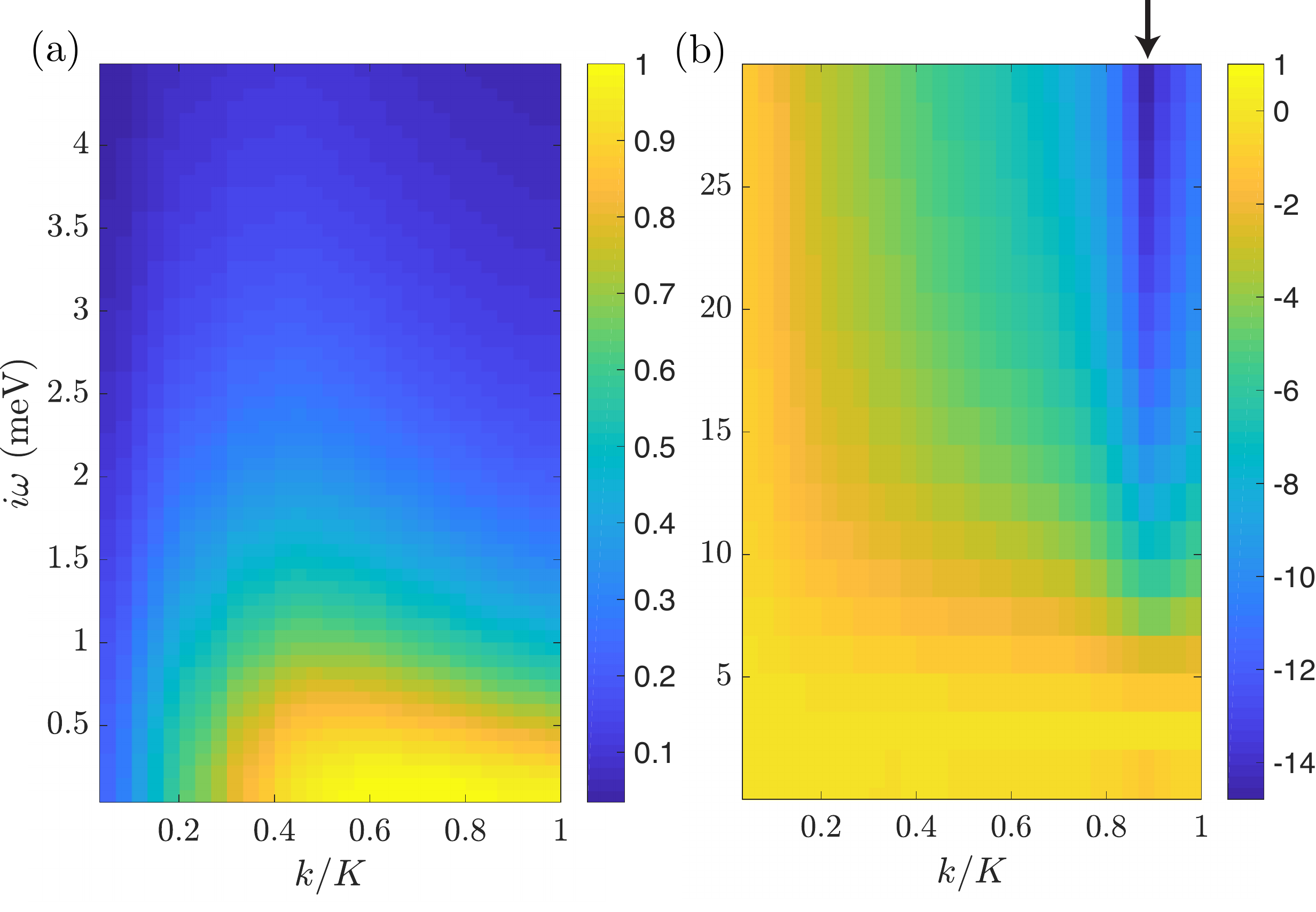}
    \caption{Solution to the gap-equation in Eq.~\eqref{eq:sc_gap_equation} due to (a) phonon and (b) plasmon for $\nu\approx 1$. In both cases we consider only $0$G processes giving $T_C=0.0118$ meV and $T_C=0.0072$ meV respectively. The phonon-induced solution is always positive, indicative of a purely attractive pairing, whilst the plasmon-induced solution has a sign-change associated with the gap --- a characteristic behavior for an overall repulsive interaction (see main text). The solutions are normalized such that ${\rm max}\{\Delta(i\pi T,k)\}=1$. We intentionally show two different matsubara grids: linearly spaced in (a), and, variably spaced (exact matsubara at low frequencies and then approximate) in (b); see Appendix \ref{app:numsol}. A wide range of matsubara frequencies is required to capture the {\it plasmonic-gorge} (black arrow) in (b). Note that momenta, $k$, are measured with respect to the center of the MBZ, see Fig.~\ref{fig:fig_1}b.}
    \label{fig:sols}
\end{figure}

To solve for $T_c$ we seek the temperature, at which the kernel has an eigenvector with corresponding eigenvalue $-1/T_c$. To perform the analysis numerically we must first perform the angular average Eq.~\eqref{eq:Kernel}. We then need to perform two additional approximations:
The first is to introduce frequency and momentum cutoffs and choose an appropriate mesh discretization.  
Furthermore, we select only a subset of momentum and frequency points under the sum. For more details on the numerical procedure we refer the reader to Appendix \ref{app:numsol}.

Before proceeding with our analysis of the results, let us pause to discuss some of the properties of the eigenvector in Eq.~\eqref{eq:sc_gap_equation}. The optimal solution has a large negative weight where the interaction is most repulsive. Therefore, the optimal solution for a phonon vs. plasmon mechanism is qualitatively different; see Fig.~\ref{fig:sols}a,b for the solutions to Eq.~\eqref{eq:sc_gap_equation} due to a phonon and plasmon solution, respectively. While the phonon solution is almost featureless at frequencies below the characteristic pairing energy (see later), it has a suppression of the solution at small $k$. As discussed later, this is connected to a suppression of the superconducting dome at large values of the filling. On the other hand, the plasmon solution presents a sharp resonance-like feature at a characteristic pairing momentum (denoted the {\it plasmon gorge}) \cite{doi:10.1143/JPSJ.45.786} as well as an eigenstate solution that changes sign --- a necessary requirement to satisfy the gap equation \eqref{eq:sc_gap_equation} when the Coulomb interaction is repulsive at all frequencies. The observed frequency dependencies, of the phonon solution in particular, leaves behind a resonant feature in the real-time Green's function of single-fermion excitations, which can be detected in the single particle density of states.

\section{Results}
\label{sec:results}

Let us now present our results for the superconducting transition temperature, obtained by solving the eigenvalue problem in Eq.~\eqref{eq:sc_gap_equation} numerically.
In order to disentangle the effects of a purely phonon mediated attraction, including the effects of umklapp scattering (highlighted above), vs. the combination of phonon and plasmon mediated superconductivity, we study their effects individually in the next two sections. We also investigate various spectral features associated with the electron-phonon coupling and the effect of a metallic screening gate in subsequent sections. In the following analysis, we set the parameter $N=20$ to ensure that we are firmly in the weak-coupling regime {\footnote{We stress that if the resulting $T_C$ for a given filling becomes of order of the chemical potential, $\mu$, then it is necessary to determine the chemical potential self-consistently \cite{PhysRevB.101.024503}. In our results, however, this limit is never reached, but the theory at very low fillings $|\nu| \lesssim 0.2$ for higher-order ($m \geq 2$) umklapp processes starts to approach, $T_C\approx 0.5 \mu$.}}.

\subsection{Phonon mediated superconductivity}
\label{subsec:phonon}

\begin{figure}[tb]
    \centering
    \includegraphics[width=\linewidth]{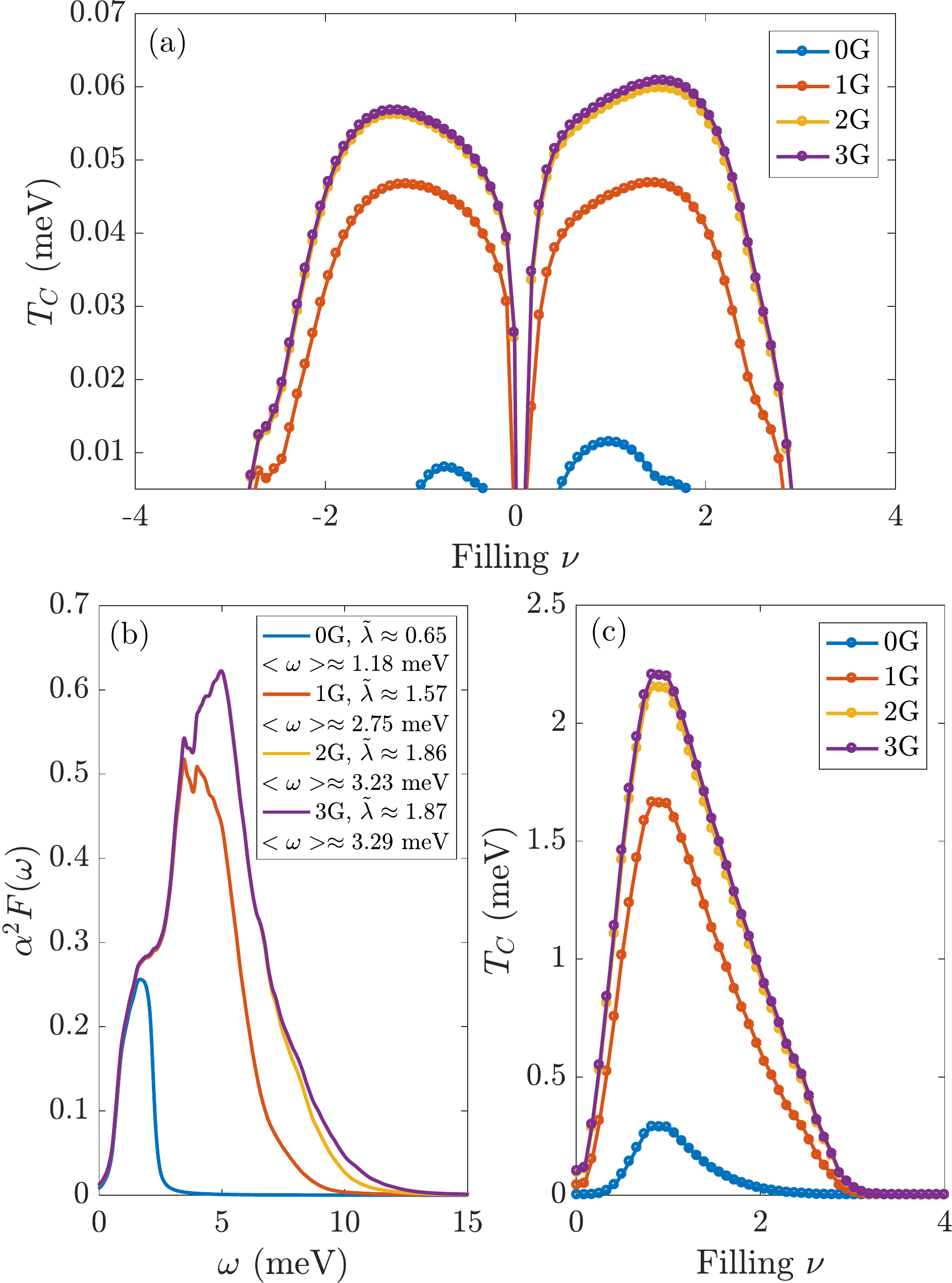}
    \caption{(a) Superconducting dome due to a phonon-only mechanism ($N=20$). Colored lines correspond to a different number of umklapp processes ($m$G with $m=0,1,2,3$) involved in the pairing interaction. The $T_C$ values for $\nu >0$ are higher than those for $\nu<0$ as the conduction band, Fig.~\ref{fig:fig_1}b, of the model in Ref.~\cite{MK18} is flatter than the valence band. The $0$G dome peaks near the van-Hove singularities of the bands. Saturation of the $T_C$ increase with $3$G umklapp processes can be traced back to the nature of the Bloch wavefunctions (see main text and details of the bandstructure discussed in Appendix \ref{app:bandstructure}). (b) Electron-phonon spectral function for $\nu\approx 1$ showing the importance of umklapp processes. (c) A plot of $T_C$ obtained from a BCS-type formula, Eq.~\eqref{eq:BCS_formula}, with parameters obtained from the MATBG spectral function, Eq.~\eqref{eq:spectral_function}. As expected, the BCS dome peaks in the vicinity of the van-Hove singularity, which by virtue of the specific details of the model in Ref.~ \cite{MK18}, occurs at $\nu\approx 0.63$. Note the similar trend of $T_C$ enhancement between panels (a) and (c) with umklapp processes.}
    \label{fig:phonon_figure}
\end{figure}

We start our discussion by focusing exclusively on the effects of the phonon mediated interaction, Eq.~\eqref{eq:V_ph}. In order to investigate the importance of phonons in a controlled fashion, we first set the Coulomb interaction ($\V^{\tn{C}}_{\xi,\xi'}$) in Eq.~\eqref{eq:V_C} to zero and compute $T_c$ due to the phonon mediated interaction ($\V_{\xi,\xi'}^{\tn{ph}}$) in Eq.~\eqref{eq:ham_phonon}. Moreover, we include a sequence of umklapp processes, $m\mrm G$, with $m=0,1,2$ and $3$; the results for the transition temperature ($T_c$) as a function of $\nu$ are shown in  Fig.~\ref{fig:phonon_figure}a.

As is evident from our results, umklapp scattering processes up to $m=2$ have a dramatic effect on $T_c$, which saturates for $m\geq3$. These umklapp phonons are essentially the lowest optical modes resulting from the folding of the original acoustic phonon branch into the MBZ. Thus, we find that the interaction with these lowest optical modes are crucial for understanding the electronic properties of MATBG. In order to clearly demonstrate the effect of umklapp processes on the pairing interaction, it is useful to study the phonon spectral function as a function of energy, defined as 
\beq
\a^2 F(\w) &=& {N_0(0) g \over 2N}\times\\\nn 
&&\left\langle \sum_{\vec{q}}  \omega_{\tn{ph}}(q) \lvert\Lambda( \bs k + \vec{{q}} ,\bs k)\rvert^2 ~\d(\w - \omega_{\tn{ph}}(q))\right\rangle_{\tn{FS}},\label{eq:spectral_function}
\eeq
where $\vec{q}=\vec{\tilde{q}}+\vec{G}$, as defined previously (see paragraph following Eq.~\eqref{eq:V_ph}). In the above, $N_0(0)$ is the density of states at the Fermi surface (FS) and $\langle \dots \rangle_{\tn{FS}}$ denotes averaging $\vec{k}$ over the FS. We stress the presence of the form-factors, $\Lambda(\bs p, \bs k)$, which encode the unusual dependence of the superconducting kernel on the large momentum umklapp processes. As can be seen in Fig.~\ref{fig:phonon_figure}b, the inclusion of umklapp processes have a clear effect on the spectral function, most prominent of which include a significant enhancement in the electron-phonon coupling strength, combined with a drastic rearrangement of the spectral weight to higher energies. These same umklapp processes are responsible for increasing $T_c$.

The enhancement of $T_C$ with umklapp processes can be anticipated by looking at the form of Eq.~\eqref{eq:V_ph}. {\footnote{We thank an anonymous referee for sharing this interesting observation.}} The effective electron-electron interaction at the lowest frequency, $\omega=0$, is a constant independent of $q$. Momentum dependence enters in Eq.\eqref{eq:V_ph} only for finite frequency. For the first such frequency ($n=1$, $\omega= 2\pi T$), the interaction has a reversed lorentzian shape in $q$-space where it is zero for $q\to0$ and saturates to the ($\omega=0$) value as momentum becomes large. With the above observations, one can predict that as $q$ increases, the coupling strength and therefore the pairing tendency will be enhanced until some value where the effect of taking higher umklapp contributions into account will lead to a saturation. We highlight, however, that this behavior alone is not behind the sharp saturation seen in Fig.\ref{fig:phonon_figure}a, but rather occurs in combination with the presence of form-factors that rapidly vanish past some critical $mG$ (here $3$G).

The characteristic energy scale associated with the peak of the above spectral function, provided that the form-factors do not lead to a suppression, is set by the graphene sound velocity ($c_s$) and the length-scales stemming from the moir{\'e} period. Using the magnitude of the moir{\'e} reciprocal vector $G=4\pi/(\sqrt{3}L_M)$ as the characteristic size of the MBZ ($L_M = \frac{a}{2}\sin(\frac{\theta}{2})$), we find the typical frequencies as
\be
\frac{1}{2} c_s (m+1) G = c_s (m+1) \frac{2 \pi}{\sqrt{3} L_M} \approx 2.1\,, 4.3\,, 6.4 ~{\rm meV}
\ee
for the first three $m$G umklapp processes: $0$G, $1$G, $2$G. These estimates are in reasonably good agreement with the location of the peaks in Fig.~\ref{fig:phonon_figure}b. In MATBG, as mentioned in the introduction, the precise values of these characteristic frequency scales can be affected by the details of the phonon dispersion (including, for example, the presence of a gap \cite{MK19,HO19}).  {It is worthwhile to comment that similar multi-peak features in el-ph spectral functions were seen also in previous ab-initio calculations\cite{HJC18}, however the precise connection to umklapp processes was not emphasized.}

The microscopic mechanism responsible for the significant contribution of umklapp processes to the spectral function, which leads to the enhancement of superconducting $T_C$, is intimately tied to the origins of the flat moir{\'e} bands. To obtain the electronic dispersion, $E_{\bs k}$, of carriers in these narrow-bands (see Fig.~\ref{fig:fig_1}b) due to the slowly varying moir{\'e} interlayer potential, it is not sufficient to consider only plane wave Bloch states with crystal momentum $\vec{k}$, but also those of nearby states that are coupled by multiple of the moir{\'e} reciprocal vectors, $\vec{G}$ (see Appendix \ref{app:bandstructure}). As a result, the spectral weight of the resulting Bloch wavefunction is extended across several plane wave states. This results in a slowly vanishing Bloch wavefunction overlap, on the scale of the moir{\'e} reciprocal momentum scale $G$, that enters into the form-factors, $\Lambda_{\gamma\gamma'}(\vec{p},\vec{k})$. We stress that this property of narrow-band wavefunctions is independent of the finer details of the bandstructure and is intimately tied to many of the unusual properties of MATBG (e.g. it is partly responsible for the extended Wannier functions \cite{MK18,OV18}).

Let us now briefly comment on the overall similarity of the dome shape to that of other phonon-pairing based approaches. This is best seen by a comparison with, for example Ref. \cite{PO20}, where an Eliashberg treatment together with a pairing mechanism relying on an optical-like phonon is considered. This similarity simply stems from the fact that in a phonon-only mechanism that produces an attractive interaction over all frequency range, the overall shape of the dome is largely dictated by the electron bandstructure rather than the details of the pairing. As such, any analysis based on a bandstructure model, either tight-binding/continuum/ab-initio in nature, will produce similar-looking domes if the underlying bandstructures are qualitatively similar.

Let us now place the essence of the phonon umklapp-driven enhancement of the critical temperature, $T_C$ in the context of BCS theory. We note upfront, however, that although it captures some of the trends seen in Fig. \ref{fig:fig_3}a, it is by no means a replacement for the full Eliashberg approach (see later). Within standard BCS theory with a coupling constant, $g$, the expected transition temperature is given  by \cite{PhysRev.108.1175},
\be
T_C \approx 1.14 \langle\omega\rangle ~e^{-1/\tilde{\lambda}}\label{eq:BCS_formula}
\ee
where $\langle\omega\rangle$ corresponds to a pairing energy range (usually the Debye frequency) and $\tilde{\lambda}=g N_0(0)$. We can now use the spectral function defined in Eq.~\ref{eq:spectral_function} to recast the effective coupling constant as \cite{PhysRev.167.331}
\be
\tilde{\lambda} = 2 \int d \omega ~\frac{\a^2 F(\w)}{\omega},
\ee
and similarly denote the characteristic pairing frequency  $\la\omega\ra$ in terms of the moment of the same distribution as
\be
\la\omega\ra = \frac{2}{\tilde{\lambda}} \int d\omega ~\a^2 F(\w)\,.
\ee
For the MATBG phonon spectral function plotted in Fig.~\ref{fig:phonon_figure}b, corresponding to an electron filling near $\nu\approx 1$, we list each of the two parameters, $\tilde{\lambda}$ and $\la\omega\ra$, in the legend for all of the umklapp processes ($m$G, with $m=0-3$). A simple application of the BCS formula in  Eq.~\eqref{eq:BCS_formula} in terms of the two filling dependent parameters, $\tilde{\lambda}$ and $\la\omega\ra$, captures many of the interesting trends that we saw previously, as shown in Fig.~\ref{fig:phonon_figure}c, including the relevance of umklapp processes. The simple BCS formula, parametrized by $\tilde{\lambda}=g N_0(0)$, predicts the $T_C$ dome to peak at the location of the van Hove singularity. To a large extent, this is also reproduced in the domes of Fig.\ref{fig:phonon_figure}a; we note, however, that the peak of the dome shifts towards higher fillings upon including successive umklapp processes. {\footnote{The exact location of the van-Hove singularity at $\nu\approx 0.63$ is a consequence of the underlying details of the bandstructure in Ref.~\cite{MK18}.}}  

The simple BCS formula, where the characteristic pairing energy is naively given by $\la\omega\ra$, was used here only for a rough estimate and comparison with the results of the Eliashberg theory. It should be mentioned that in this case, where $\langle \w \rangle \gtrsim \e_F$, the frequency cutoff is sometimes substituted by $\e_F$~\cite{gor1961contribution,ikeda1992migdal,Kozii2019}. However, given that this substitution is also non-rigorous~\cite{PhysRevB.101.024503}, we do not perform it here.   

Finally, we note that the umklapp-driven enhancement of phonon-mediated superconductivity is not model specific and is a generic property of any moir{\'e} narrow-band system with a Bloch wavefunction overlap that is vanishing slowly, on a moir{\'e} momentum scale, as a function of the momentum exchange, $\vec{q}$.  In Sec.~\ref{subsec:spec}, we shall return to a discussion of the electron-phonon spectral function and consider the experimental fingerprints of these modes in the tunneling density of states.

\subsection{Plasmon mediated superconductivity}
\label{subsec:plasmon_sc}

We are now in a position to include the effects of Coulomb interactions on pairing. To that end, we begin by evaluating the dielectric function $\ve_{\tn{RPA}}(i\w,\bs q)$ numerically~\cite{Lewandowski20869} (see also Appendix \ref{app:RPA}) and reinstating $\V^{\tn{C}}_{\xi,\xi'}$ to the pairing kernel in Eq.~\eqref{eq:Kernel}.

\begin{figure}[tb]
    \centering
    \includegraphics[width=\linewidth]{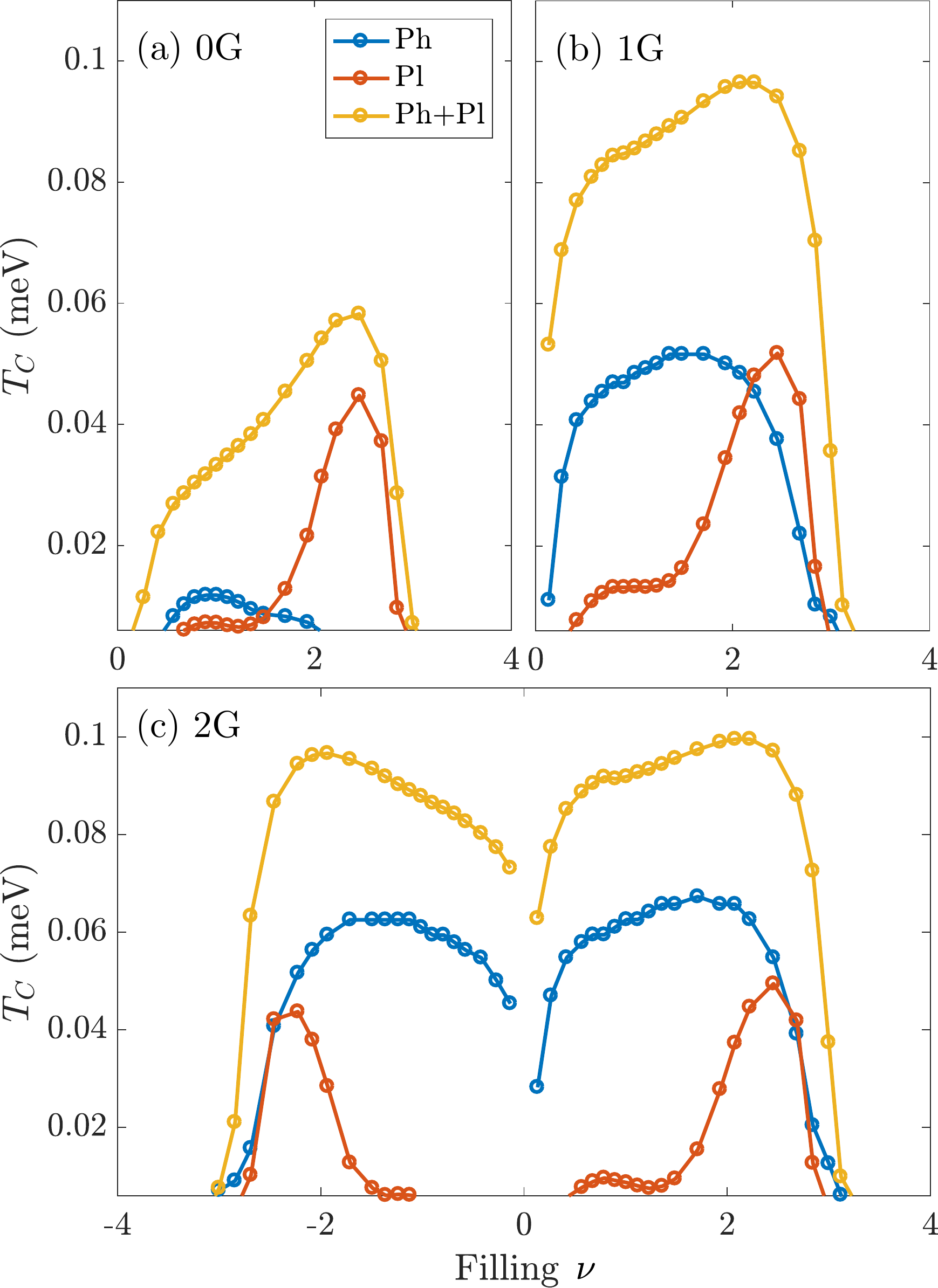}
    \caption{Comparison of phonon and plasmon mediated pairing mechanisms as a function of an increasing number of umklapp processes. (a) $0$G (no umklapp), (b) $1$G and (c) $2$G. Blue curves correspond to a phonon only mechanism (Sec.~\ref{subsec:phonon}), red curves correspond to a plasmon only  mechanism (Sec.~\ref{subsec:plasmon_sc}), and, yellow curves include effects of both phonon and plasmon on pairing (see Sec.~\ref{subsec:plasmon_sc}). The phonon-driven attraction is strongly enhanced with the inclusion of umklapp scatterings; the plasmon mechanism is largely insensitive to umklapp. We choose $N=20$.}
    \label{fig:fig_3}
\end{figure}

Let us begin by studying the problem in the absence of the phonon mediated interaction, i.e. set $\V_{\xi,\xi'}^{\tn{ph}}=0$, and include only the effects of the Coulomb interaction. In this case, the origin of pairing lies in the frequency dependence of the dielectric function close to the plasma resonance \cite{doi:10.1143/JPSJ.45.786,ruhman2017pairing}.  Moreover, just as in the case of phonons, we include a sequence of $m$G umklapp processes for the plasmons (we choose $N=20$). The result for $T_c$, due purely to the plasmonic mechanism, is shown in red in Fig.~\ref{fig:fig_3}(a)-(c) as a function of filling. This analysis leads us to conclude that (i) the plasmonic mechanism of pairing leads to an enhancement of $T_c$ for a narrow range of fillings near $\nu \approx 2-3$, and, (ii) successive $m>0$ umklapp processes have no appreciable effect on $T_c$.

We now explain the microscopic origin for both of these observations. In contrast to a purely attractive phonon-mediated interaction, the dynamically screened Coulomb interaction is always repulsive. However, as argued in Sec.~\ref{sec:eliashberg}, the frequency dependent dynamically screened interaction becomes weak enough at certain momenta, such that the sign-changing gap function can minimize the overall effect of repulsive Coulomb interaction \cite{coleman_2015}, giving rise to an effective attractive part. 

\begin{figure}[!tb]
    \centering
    \includegraphics[width=\linewidth]{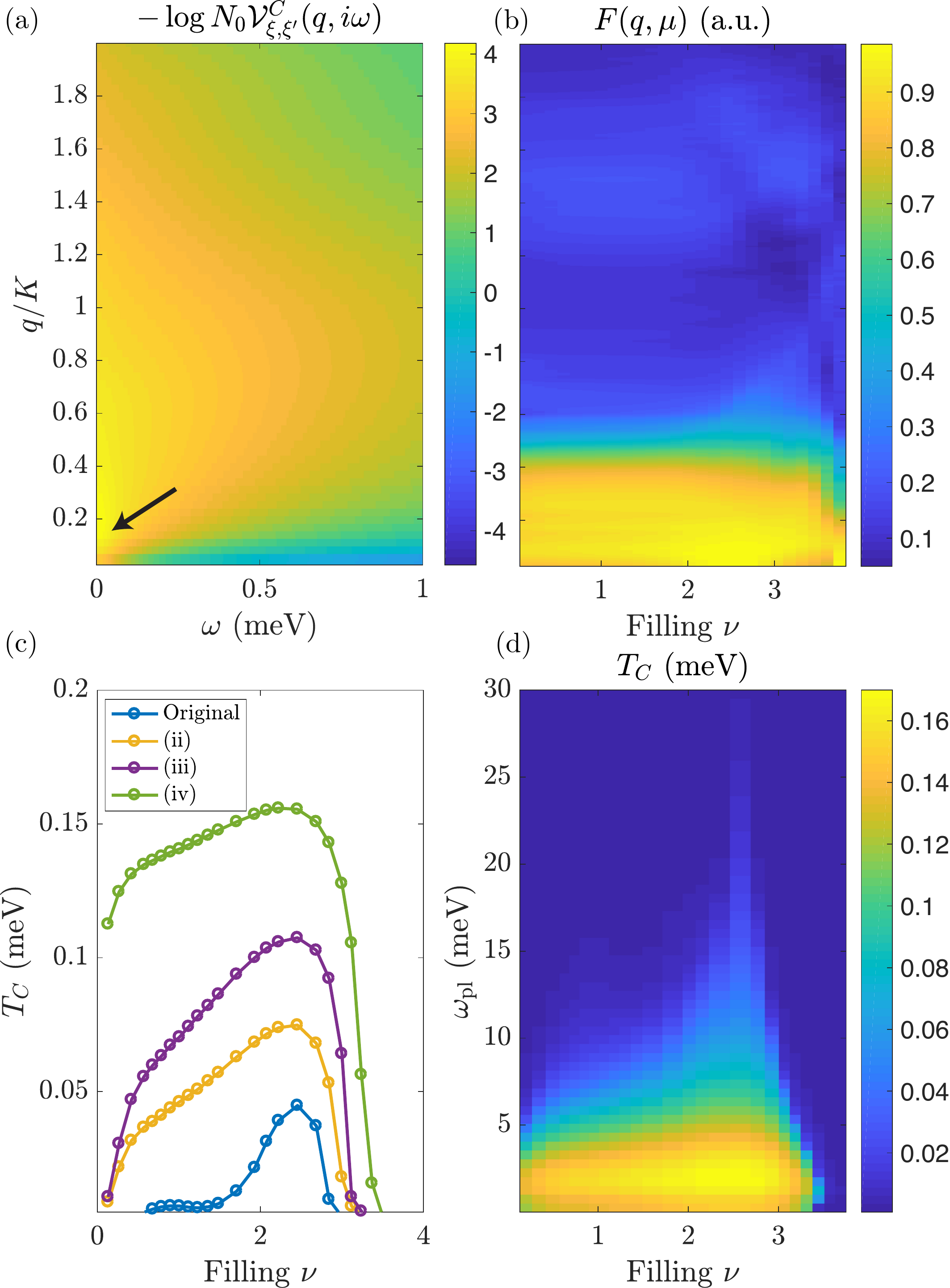}
    \caption{(a) Plot of $-\log\left[N_0(0) \V^{\tn{C}}_{\xi,\xi'}(\vec{q},i\w)\right]$ for $\nu\approx 1$. Note that a local minimum (black arrow) of the interaction occurs at a finite momentum (see discussion in the main text). (b) Fermi-surface averaged dependence of form-factors, Eq.~\eqref{eq:coherence_factor_overlap}, with the same vertical axis as in panel (a). Note how small momentum ($q < K$) processes dominate over the large momentum terms ($q > K$). Near a filling, $\nu\approx 2.45$, the form-factors are slightly larger for a wider range of momenta than at lower fillings. This behavior lies behind the narrow plasmon peak of the $T_C$ dome, Fig.~\ref{fig:fig_3}, which is exponentially sensitive to the coupling strength, c.f. Eq.~\eqref{eq:BCS_formula}. (c) Plasmon-mediated SC under different approximation schemes. For the plasmon-pole approximations (iii,iv) we use $\omega_{\tn{pl}}\approx 6$ meV. There is hardly any difference between (i) and the original result. (d) Superconducting temperature in the plasmon pole approximation as a function of plasma frequency $\omega_{\tn{pl}}$. Pairing occurs from frequencies close to $\omega_{\tn{pl}}$, and leads to an increase in $T_C$ as $\omega_{\tn{pl}}$ approaches the chemical potential.}
    \label{fig:plasmon}
\end{figure}

In order to understand the structure of the frequency-momentum regions where the screened interaction becomes weak, it is helpful to plot $\V^{\tn{C}}_{\xi,\xi'}(\vec{q},i\w)$ in Eq.~\eqref{eq:V_C}, as shown in Fig.~\ref{fig:plasmon}a at a fixed filling ($\nu\approx1$). The behavior differs from the one in a conventional 2D Fermi gas \cite{PhysRevLett.18.546}. Most crucially, in the latter, given the form of the polarization function, one would expect the interaction to be most repulsive as $q\to 0$ and then weaken with increasing $q$. On the other hand, in MATBG, there is a local minimum of the interaction at a finite momentum $q$.  We can understand this behavior by focusing on the limit of $\omega\to 0$, as discussed in Ref.~\cite{PhysRevB.100.161102, PhysRevB.100.235424}. At the magic angle and at low fillings $\nu\approx 0$, the static polarization function behaves as $\Pi_{ee}(\vec{q},\omega\to 0) \propto q/v_F$. This form is reminiscent of the polarization function in a Dirac-like system \cite{PhysRevB.75.205418,Wunsch_2006}. At momenta smaller than the moir{\'e} reciprocal momentum ($G$), $v_F$ corresponds to the renormalized MATBG Fermi-velocity near $\nu\approx0$ with $v_F\sim 10^4$ m/s. The polarization function is dominated by the inter flat-band transitions. On the other hand, at momenta comparable to and larger than the moir{\'e} scale (and at a similar filling), $v_F\sim 10^6$ m/s, the bare graphene velocity. The polarization function is now dominated by the inter-band transitions between the flat and dispersive bands as the effect of the moir{\'e} interlayer potential becomes less relevant. As a result, the plasmons in MATBG \cite{Lewandowski20869} have interesting properties. As long as plasmons rise above the particle-hole continuum, its dispersion is controlled by the energy scale associated with inter-band transitions between flat-bands, and between the flat and the dispersive bands. As such, it therefore becomes weakly sensitive to the filling value. 

The aforementioned local minimum of the pairing interaction occurs at momenta smaller that the moir{\'e} reciprocal lattice scale, $G$. At momenta larger than $G$, the interaction in Eq.~\eqref{eq:V_C} reduces to the simple unscreened form, $2\pi e^2/q$, suppressing any dynamical contribution to the superconducting gap. As a result, any contribution due to higher umklapp processes involving plasmons does not enhance $T_C$ drastically; see Fig.~\ref{fig:fig_3}a-b. In fact even at large enough momenta, on the scale of $2$G, as the form-factors are still non-vanishing it can suppress it slightly; see Fig.~\ref{fig:fig_3}c.

While the specific form of the screened interaction, Fig.~\ref{fig:plasmon}a, affects the shape of the plasmon-induced superconducting dome, it is useful to disentangle it from the role played by other elements that appear in the pairing kernel, Eq.\eqref{eq:Kernel}: the form-factors $\Lambda_{\gamma\gamma'}(...)$ and the energy denominator,  $1/(\nu^2+E^2_{\xi,\vec{p}})$.  To that end, we focus on the $0$G processes and selectively modify the different elements that enter into the kernel, Eq.~\eqref{eq:Kernel}. In particular, we consider the following modifications: (i) we compute the dielectric function at one specific filling, $\nu$, and use it for all fillings (thereby ignoring the $\nu-$dependence of the dielectric function), (ii) we ignore the momentum dependence of the form-factors and replace them with unity, except for restricting the interaction to intervalley pairing, (iii) we use the ``plasmon-pole'' approximation instead of using the full RPA dielectric function, and, (iv) we invoke the same approximation as in (iii) above, but make the substitution for the form-factors as in (ii). A figure demonstrating all of these cases is shown in Fig.~\ref{fig:plasmon}c. Notably, the results under the approximation in (i) above do not affect the results at all (not shown), thereby indicating that the density dependence of the dielectric function does not play a significant role. 

Let us begin with a discussion of points (iii) and (iv) above, that rely on the plasmon-pole approximation. As described previously, the Coulomb interaction is always repulsive; the dynamical screening can, however, lead to a possible superconducting solution that overcomes the effect of the repulsion. To investigate this matter further, we explore here an idealized limit --- the so-called ``plasmon-pole'' approximation--- where the Coulomb interaction in Eq.~\eqref{eq:V_C} is replaced by
\be
\V^{\tn{C}}_{\xi,\xi'}(\vec{q},i\w) \approx {1\over N}{2\pi e^2\over \kappa q} \left(1-\frac{\omega^2_{\tn{pl}}}{\omega^2_{\tn{pl}}+\omega^2}\right) \delta_{\xi,-\xi'}\,.
\ee
Here, $\omega_{\tn{pl}}$ is the plasmon frequency. We note that the ``plasmon-pole'' approximation is strictly valid only in the $\omega\gg v_Fq$ limit, which will be of interest to us below. For $\omega \ll v_F q$, the interaction can be approximated by  its purely repulsive and static Thomas-Fermi screened form. For a wide range of momenta, the plasmons of interest to us originate primarily from inter-band transitions between the nearly-flat and dispersive bands. As argued above, the resulting plasma frequency is independent of the filling and is instead set by the bandwidth, $W$, of the nearly-flat band and the gap, $\Delta_{\tn{band}}$, between flat and dispersive bands as $\omega_{\tn{pl}}\approx\sqrt{W \Delta_{\tn{band}}}$ (see Ref.~\cite{Lewandowski20869} and App. \ref{app:RPA} for details). We therefore choose a constant $\omega_{\tn{pl}}$ in our calculation. As is evident from Fig.~\ref{fig:plasmon}c, the plasmon-pole result exceeds the RPA result. With increasing filling, $T_C$ rises almost linearly as the Fermi energy approaches the plasma frequency, thereby enhancing the effect of the plasmon in driving pairing. However, for $\nu\gtrsim 3$, $T_C$ starts to drop rapidly --- a behavior we attribute to the bands being more dispersive at these fillings, c.f. Fig.~\ref{fig:fig_1}b. 

We can verify our understanding of the interplay of these two results by varying $\omega_{\tn{pl}}$ as an external phenomenological parameter, as done in Fig.~\ref{fig:plasmon}d. We notice immediately that the closer $\omega_{\tn{pl}}$ is to the relevant chemical potential, the higher is the $T_C$. This is however not sufficient at large fillings, $\nu > 3$. We note that the sharp fall-off at $\nu>3$ is a property of the continuum model used. This fact is precisely the reason for the form of the gap solutions obtained in Fig.~\ref{fig:sols}, where for $k\lesssim0.25K$ the solution due to phonons vanishes (Fig.~\ref{fig:sols}a), while it is positive in Fig.~\ref{fig:sols}b, implying an overall repulsive contribution to the plasmon-induced gap equation. This analysis also suggests an interesting possible route towards enhancing $T_C$ due to plasmons. If $\omega_{\tn{pl}}$ can be brought closer to the chemical potential, whilst maintaining the same strength of Coulomb interactions in a system, then it is possible to raise $T_C$.

We conclude the analysis of the plasmon-pole approximation and its effect on pairing by pointing out that suppressing the momentum dependence of the form-factors leads to an enhancement of $T_C$; see curves labeled (ii) and (iv) in Fig.~\ref{fig:plasmon}c. This can be understood by realizing that the Bloch wavefunction overlap depends on the underlying ``fidget-spinner'' structure of the energy contours, c.f. Fig.\ref{fig:fig_1}c, which can suppress certain scattering processes and thereby lower $T_C$.

The plasmon-pole approximation captures most of the features we obtain within the full Eliashberg calculation (Fig.~\ref{fig:fig_3}). However, it does not immediately lead to a simple explanation for the sharp peak associated with the (plasmon-)dome near $\nu \approx 2-3$. It is natural to ask if this feature arises solely due to a change in the dielectric function as a function of $\nu$. To explore this possibility, we compute $T_C$ by fixing the dielectric function corresponding to $\nu\approx 2.45$ (associated with the peak of the plasmon-dome in Fig.~\ref{fig:fig_3}a) and not varying it as a function of $\nu$; this corresponds to the approximation denoted as (i) above. We find that the overall shape of the resulting dome is completely identical to the full computation (result not shown). As explained previously, this behavior stems from the dielectric function of MATBG being dominated by inter-band transitions, which are largely insensitive to the filling. 

Taking all of these observations into account, the peak of the dome at a filling of $\nu \approx 2-3$ comes from an interplay of the form-factors along with the dielectric function. This conclusion stems from the analysis leading to Fig.~\ref{fig:plasmon}d, which demonstrates that the plasmon-pole approximation, even with the appropriate form-factors, can not reproduce the sharp peak at $\nu\approx 2-3$. 

To assess this further, we now focus on the $q$ dependence of the form-factors. To that end we plot in Fig.~\ref{fig:plasmon}b, the following quantity
\be
F(q,\mu) = \int_{\tn{MBZ}} d^2 \vec{k} \int d \theta_{\vec{k}'}~ k' \lvert \Lambda(\vec{k}',\vec{k}) \rvert^2 \delta(E_{\xi,\vec{k}}) \delta(E_{\xi,\vec{k'}})
\,,\label{eq:coherence_factor_overlap}
\ee
where $q=|\vec{k}-\vec{k'}|$ and $\delta(E_{\xi,\vec{k}})$, $\delta(E_{\xi,\vec{k'}})$ constrain the two states to lie on the Fermi-surface (within mesh resolution); $\theta_{\vec{k}'}$ denotes angle of $\vec{k}'$. 
The decrease in $F(q,\mu)$ with $q$ reflects the density dependence of the orbital hybridization of the Bloch wave-functions in the relevant bands. For comparison, $F(q,\mu)$ would be momentum independent and equal to unity, when there is no orbital hybridization (i.e. when the band and orbital basis are identical). In contrast, it is known to diminish rapidly with the momentum exchange, $q$, in a Dirac system~\cite{ruhman2017pairing}. Therefore, $F(q,\mu)$ extends to higher momentum close to $\nu \approx 2-3$, where orbital hybridization is minimized, taking a maximum at $\nu\approx 2.45$. In turn, this maximizes the product with the screened interaction from, Fig.\ref{fig:plasmon}a, leading to a higher $T_C$. {According to our analysis the strong non-monotonic dependence of $T_C$ on filling, when resulting from a plasmon mechanism,  is a  consequence of the form factors. This should be compared with the earlier work relying on a plasmon mechanism \cite{SA19}, where these form-factors are not included.}

Finally, we analyze the cooperative effect of both phonons and plasmons on pairing; the resulting $T_C$ vs. $\nu$ is shown for the combined (as well as individual) effect of phonons and plasmons in Figs.~\ref{fig:fig_3}(a)-(c). As before, we include the effects of $m$G umklapp processes with $m=0-2$, which has a strong effect on the phonon-mediated contribution but barely affects the plasmonic mechanism. We find that with the inclusion of up to $2$G processes,  the umklapp-driven phonon-mediated mechanism clearly dominates over the plasmon mechanism. {\footnote{The bare coupling constants for the Coulomb and electron-phonon interactions are chosen to be close to the experimentally reported values relevant for MATBG; we choose $N=20$.}} However, both mechanisms work in a cooperative fashion to give rise to pairing.

\subsection{Role of external screening}
\label{subsec:screen}

Our analysis thus far has shed light on the distinct features associated with a purely electronic vs. a purely phonon-based mechanism, as well as their combined effect, on the emergence of SC in MATBG. In light of the recent experiments \cite{saito2019decoupling,stepanov2019interplay,liu2020tuning} that have studied the role of an external screening layer on the filling dependence of $T_c$, let us explore the effect of similar screening within our setup.

We begin by studying the effects of a metallic gate, coupled to MATBG via Coulomb interactions, on pairing. The only role played by the gate is in a further renormalization of the Coulomb interaction, Eq.~\eqref{eq:V_C}, via the dielectric constant, $\ve_{\tn{RPA}}$. In the limit where the density of states associated with the metallic gate is higher than that of the screened substrate, the effect of the gate can be incorporated by modifying the bare Coulomb interaction as:
\be
\frac{2\pi e^2}{q}\to \frac{2\pi e^2}{q}\left(1-e^{-2q d}\right)\,,\label{eq:gate_screening}
\ee
where $d$ is the distance between MATBG and the gate. As a result, the dynamically screened interaction (Eq.~\eqref{eq:V_C}) now changes to
\begin{align}
\V^{\tn{C}}_{\xi,\xi'}(\vec{q},i\w) &= {1\over N}{2\pi e^2\over \ve_{\mrm{RPA}}(\bs q ,i\w) q} \left(1-e^{-2q d}\right)\delta_{\xi,-\xi'},\\ \nn
\epsilon_{\mrm{RPA}}(\bs q ,i\w) &= \kappa-\frac{2\pi e^2}{q} \Pi_{ee}(\vec{q},i\omega)\left(1-e^{-2q d}\right)\,.
\end{align}
The presence of the metallic gate results in an overall suppression of the Coulomb interaction, that scales exponentially with $d$. Thus, processes involving $q \lesssim 1/2d$ are not responsible for mediating SC. 

\begin{figure}[!tb]
    \centering
    \includegraphics[width=\linewidth]{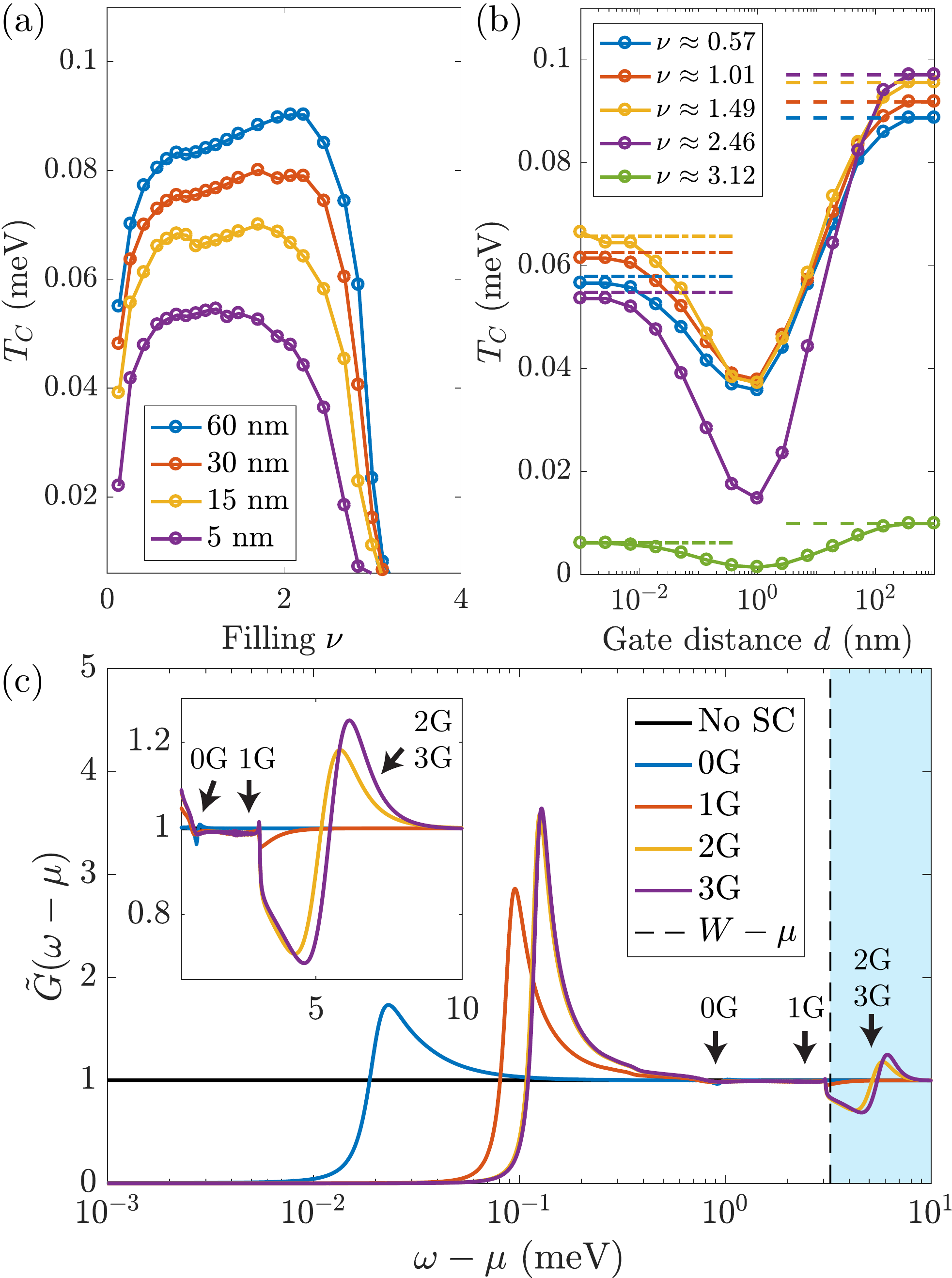}
    \caption{(a) Effect of screening by a metallic gate for four different values of $d$ (nm) on $T_C$ with the inclusion of $2$G umklapp processes ($N=20$). Due to the cooperative interplay of phonons and plasmons (see main text) we find that the gate suppresses $T_C$ for a range of fillings. (b) Effect of screening for few fillings over a wide range of gate distance. The dashed-dotted lines correspond to a purely phonon-driven mechanism, while the dashed lines include the cooperative effects due to both plasmons and phonons (see also Fig.~\ref{fig:fig_3}c). Note the non-monotonic dependence of $T_C$ on $d$ as well as the wide range of $d$ values over which $T_C$ varies; see discussion in the main text for implications on plasmon-mediated pairing. (c) Normalized differential conductance from Eq.\eqref{eq:diff_cond} computed at $T = T_C/10$ for each curve for phonon-mediated superconductivity. All umklapp processes result in conventional BCS gap behavior and exhibit resonances corresponding to relevant $m$G umklapp processes (indicated with arrows), c.f. Fig.\ref{fig:phonon_figure}b. This modulation of $\tilde{G}(\omega-\mu)$ allows to reconstruct the form of the superconducting pairing. Dashed line indicates the electronic bandwidth, which by virtue of the model, gives rise to a signal due to $2$G, $3$G phonons inside the gap between the flat and non-flat band, respectively (shaded area). As a result of mesh induced broadening, $N_0(\omega-\mu)$ has a small finite value, and so does $\tilde{G}$, inside the gap region.}
    \label{fig:fig_4}
\end{figure}

In Fig.~\ref{fig:fig_4}a, we plot $T_C$ as a function of filling ($\nu>0$) for a few different values of $d$ (with the $2$G umklapp processes included). As discussed previously in the context of Fig.~\ref{fig:fig_3}, the dynamically screened Coulomb interaction assists the phonon-driven mechanism in a cooperative fashion, especially for momenta smaller than the characteristic scale of the MBZ (see Fig.~\ref{fig:plasmon}a). A screening of the form in Eq.~\eqref{eq:gate_screening} is expected to result in a strong suppression of $T_C$ for a wide range of fillings, but especially so in the range of fillings where the plasmon contributes the most to pairing. This behavior is indeed observed in Fig.~\ref{fig:fig_4}a, where the gate leads to a suppression of superconductivity over the entire range of fillings, with perhaps a slightly more pronounced effect near fillings $\nu\approx 2-3$. 

In addition to the direct experimental relevance  \cite{saito2019decoupling,stepanov2019interplay} for understanding the effect of screening from a metallic gate on the strength of Coulomb interaction in Eq.\eqref{eq:gate_screening}, the above setup also helps elucidate the length-scales that participate in plasmon mediated pairing. Specifically, a metallic gate located a distance $d$ away suppresses the interaction over length-scales of $O(2d)$. Depending on how this length-scale compares with the other characteristic length-scales in the problem, which include e.g. the scale associated with twist-angle inhomogenities etc., will determine whether electron-electron interactions contribute towards pairing. This requires a careful analysis of the necessary length-scales associated with the plasmons that are required for plasmon-mediated pairing to be observable experimentally.

To address the question raised above, we study the effect of varying $d$ over many orders of magnitude in Fig.~\ref{fig:fig_4}b. For $d\to \infty$ (i.e. significantly larger than any other relevant length-scale in the problem), we expect to reproduce the previously obtained values for $T_C$ in Fig.~\ref{fig:fig_3}c for both the phonon and plasmon mediated interaction. On the other hand as $d\to 0$, we expect the contribution due to the plasmon to be suppressed significantly, such that $T_C$ will almost entirely be determined by only the phonon-mediated interaction. We observe that while low-momentum processes corresponding to distances $2d \approx 200~ {\rm nm} \approx 14.5 L_M$ are necessary for the plasmons to maximally assist in pairing, the plasmon substantially assists the phonon-driven mechanism already at lengthscales starting $2d \approx 50~ {\rm nm} \approx 3.5 L_M$.

Interestingly, we also observe that $T_C$ depends non-monotonically on the gate-distance, $d$, as shown in Fig.~\ref{fig:fig_4}b. As noted above, bringing the screening layer closer to MATBG starting from a very large separation, leads to a drop in the value of $T_C$ as a result of the suppression of plasmon-induced pairing.  
However, once $d\approx 10$ nm, the value for $T_C$ starts to drop beyond what is expected even from a purely phonon-mediated mechanism (i.e. $d\to 0$), eventually reaching a minimum value at $d\approx 1$ nm. The additional reduction of $T_C$ arises as a result of the metallic gate suppressing the attractive contribution to the pairing coming from momenta $q \ll K\sim{G}/2$, c.f. Fig.\ref{fig:plasmon}, leaving only the purely repulsive part. The minimum value for $T_C$ is attained approximately at the same value of $d$ at all fillings; this can be understood from the filling independence of the dielectric function for MATBG, which is  dominated by the interband transitions (see Sec.~\ref{subsec:plasmon_sc}). Beyond this value of $d$, decreasing it further leads to an increase in $T_C$, in agreement with the expectations in a purely phonon-mediated attraction \cite{PhysRev.167.331}.

As mentioned in the introduction, experimental setups that include a gate to control the strength of the Coulomb repulsion have already been realized in several groups. Namely, Refs.~\cite{stepanov2019interplay,saito2019decoupling,liu2020tuning} find that the gate tends to suppress the correlated state significantly, while superconductivity remains largely unaffected. As such these results are broadly consistent with our conclusions regarding superconductivity within our model. In these experiments, and in fact in majority of MATBG experiments to date, the gate (or, the screening substrate) is closer to MATBG than $30$ nm and thus our model would predict superconductivity to be primarily phonon-driven. As such therefore if the thickness of the substrate is increased, our analysis predicts an increase in critical temperature if plasmons participate strongly in mediating the pairing. We note however that the precise onset of the non-monotonic gate dependence seen in Fig. \ref{fig:fig_4}b is dependent on the finer details of the screening mechanism. For example, the extent to which the density of states for the metallic gate compares with that of MATBG, or, whether further modifications to the RPA dielectric function (such as local field effects) can become pronounced for $q$ greater than moir{\'e} momentum lengthscale can affect these details. Finally, we also note that this filling independent onset of the non-monotonic mechanism is indicative of a geometric scale dictating properties of a dielectric function --- this corresponds to the interband transitions that are set by the flat-band bandwidth, $W$. To that end, if flat-bands were to become broader, e.g. due to the presence of Hartree corrections, we would then expect a dependence on filling to appear. 

\subsection{Spectroscopic signature of umklapp phonons}
\label{subsec:spec}

Let us finally discuss an experimental fingerprint associated with the phonon umklapp processes and their role in pairing. In Sec.~\ref{subsec:phonon}, we have demonstrated how the scattering of umklapp phonons contributes significantly to pairing. Consequently, we anticipate these modes to appear as resonant features in the single-particle density of states at the energies where the spectral weight of the phonons is pronounced~\cite{Schrieffer1963} (see Fig.~\ref{fig:phonon_figure}b). Consider the normalized differential conductance, defined as 
\beq
\tilde{G}(\omega-\mu) = \frac{N_{\tn{SC}}(\omega-\mu)}{N_0(\omega-\mu)},\label{eq:diff_cond}
\eeq
where $N_{\tn{SC}}(\omega-\mu)$ is the density of states in the superconducting state and $N_0(\omega-\mu)$ is the corresponding density of states in the normal state; the result is shown in Fig.~\ref{fig:fig_4}c. To obtain the superconducting density of states, we extended the gap equation, Eq.~\eqref{eq:sc_gap_equation}, to its complete non-linear form. This allows us to calculate not only the critical temperature $T_C$, but also the actual superconducting gap. For more details on the specific computational aspects associated with solving the non-linear gap equation as well as the details of the analytic continuation to real frequencies necessary for calculation of the spectral function, see Appendices~\ref{app:numsol} and \ref{app:dos}. For all $m$G processes, we see a well-defined $s$-wave superconducting behavior with the usual coherence-peak at the onset of the pairing gap; the higher the $T_C$, the larger is the gap and hence the onset of the peak. Therefore, we observe the coherence peak shifting in agreement with umklapp processes enhancing the $T_C$. Away from the peak, $\tilde{G}(\omega-\mu)$ exhibits the classic square root singularity until it reaches the $0$G and $1$G phonon-resonance features (indicated with the first two, left-most arrows). We discuss now the final signature seen in the differential conductance.

We observe sharp ``dip-hump'' resonance-like features at frequencies of the order of 5 meV, which corresponds precisely to the energy where the density of states deduced from $\alpha^2F(\omega)$ for the $2$G and $3$G umklapp phonon processes is the largest, as seen earlier in  Fig.~\ref{fig:phonon_figure}b. These features are spectral fingerprints of the phonon modes ~\cite{Schrieffer1963,Marsiglio2008,PhysRevB.94.224515,Swartz1475}, which are most visible for modes that are strongly coupled. Therefore, our conclusion that the exchange of umklapp phonons is the strongest contributor to pairing in MATBG leads to the experimental prediction of such "dip-hump" features at these specific frequencies. This frequency of the "dip-hump" features does not rely on the specific choice of electronic band structure model, as it is only related to the sound velocity in graphene and the MBZ size. However, the visibility of the features depends strongly on the electronic density of states background, which is here much enhanced due to the band gap. We caution however that similar-looking resonances, refereed in the literature as replica bands, can also be present in the spectroscopic signal and bear little relation to superconductivity ---  for a detailed analysis we refer to Refs.~\cite{PhysRevB.97.060501,Rademaker_2016}.

\section{Discussion and Outlook}
\label{sec:dis}
In this work, we have focused on the pairing instabilities in MATBG within the framework of Eliashberg approximation. We have focused on two sources of interaction, mediated by acoustic phonons of the original graphene layers, and the dynamically screened Coulomb repulsion, respectively. Interestingly, we find that both sources contribute to pairing in a cooperative fashion with a relatively similar strength, Fig.~\ref{fig:fig_3}. However, the screened Coulomb repulsion plays a role in a narrow range of density, while the phonons contribute to pairing over a wide range of fillings that nearly extends over the entire narrow band. One of the key findings of our work is the prominent role played by the umklapp processes in the phonon mediated interaction in enhancing the transition temperature, $T_c$. Umklapp processes that scatter states up to the third MBZ (``$m$G'' with $m=3$) have a marked effect on the enhancement of $T_c$ and eventually saturate for any higher order processes ($m>3$), c.f. Fig.\ref{fig:phonon_figure}a. On the other hand, the dynamics associated with the screened Coulomb repulsion, while being relatively insensitive to the umklapp processes, plays a cooperative role in pairing for a wide range of fillings (i.e. by aiding the phonon-mediated mechanism), but in particular in the vicinity of $\nu \approx \pm 2-3$. As a result, the superconducting $T_C$ exhibits a non-monotonic dependence as a function of the distance to a nearby metallic (screening) gate. The last observation is dependent on both the properties of the screening material and the extent to which long-wavelength plasmons can participate in SC pairing.

{It is natural to ask if there are sharp experimental signatures associated with any of the scattering processes considered above in MATBG. We have demonstrated that the electron-phonon interaction leaves behind an imprint in a spectroscopic tunneling experiment. As was predicted by Ref.~\cite{Schrieffer1963} processes where phonons are emitted in the tunneling process lead to these fingerprints, which thus appear at a frequency $\D(0)+\w_0$, where $\w_0$ is the frequency of the bosonic mode contributing to pairing.  Thus, specifically for the umklapp phonon processes (which correspond to optical-like modes arising from a folding of the original acoustic branch in the MBZ), the resonant features are present at frequencies that are independent of the details associated with the electronic band structure. The ``dip-hump'' features in the tunneling density of states are most visible when they appear inside the bandgap between the nearly flat and dispersive electronic bands associated with MATBG. }

Whenever possible, we have explicitly pointed out the universal, model-independent aspects of our predictions, in contrast to the features that rely on the non-universal aspects of the model. We stress that the importance of phonon-umklapp processes in enhancing the pairing temperature is a model-independent feature. It relies solely on the geometric properties of the MBZ as well as on the presence of a slowly varying moir{\'e} interlayer potential that gives rise to slowly vanishing form-factors (Bloch wavefunction overlaps). As such, we therefore expect to see features in the differential conductance at the phonon resonance frequencies, which may or may not lie inside the bandgap --- the latter property being model specific. Likewise, for the plasmon-mediated mechanism for superconductivity, we can identify various universal features. Firstly, plasmons that originate from the flat bands necessary have a characteristic frequency that is of scale similar to the flat-band bandwidth (or, chemical potential). Therefore, the dynamical Coulomb screening is bound to play an important role in describing superconductivity, i.e. the effect of Coulomb interaction cannot be simply reduced to a static repulsion. Secondly, as a result of the microscopic origin of the behavior of the polarization function (i.e. interband transitions between flat-bands at small momenta and between flat/dispersive bands at large momenta), it is evident that the plasmon-mediated mechanism is weakly affected by umklapp processes. The extent to which the plasmon dome however exhibits a narrow peak at high fillings $\nu\approx 2-3$ is dependent on the extent to which the continuum model \cite{MK18} accurately captures behavior of the form-factors in MATBG.

There are a number of interesting directions that remain to be explored, based on the formalism developed here. A natural extension would be to include the filling-dependent bandstructure renormalization arising from the interaction itself \cite{Guinea13174,2020arXiv200414784G} and then study the pairing instabilities as a result of the same interactions. While we have considered the effects of a single acoustic phonon mode on pairing in this study, MATBG hosts multiple in-plane as well as out-of-plane phonon modes \cite{MK19,HO19};  some of these modes also develop small gaps as a result of folding due to the moir\'e potential. Studying the combined effects of these modes in the presence of umklapp scattering on pairing and on the resonant tunneling spectra is clearly an interesting problem. 

Recent theoretical works have indicated the possibility of low-energy goldstone modes associated with spontaneously broken continuous symmetries near some of the correlated insulating phases at commensurate fillings \cite{TS1,AV20,VK20}. It would be interesting to explore and clarify to what extent the umklapp processes considered in this work in the context of phonons are also important for these goldstone modes of electronic origin in general --- the recent analysis in Ref.~\cite{VK20} includes some such (``$2$G'') processes for one specific example. It would also be interesting to study the effect of screening gates on the associated phenomenology. With regard to the relevance of the umklapp processes considered here on other aspects of the phenomenology, it is likely that they also play an important role in electrical charge transport. To what extent these might be relevant for some of the recent unconventional results reported in Refs.~\cite{PhysRevLett.124.076801,Polshyn2019} is an interesting open question.

Another interesting question concerns the unavoidable role of disorder, on the superconducting phenomenology in MATBG \cite{Uri2020,SI20,Yoo2019}. Disorder can significantly affect the ability of the flat band to screen the Coulomb interaction, which will modify the dynamics of the Coulomb interaction and the results we have obtained here. Disorder in MATBG comes in many avatars and includes long-ranged impurities which are poorly screened at low energy, as well as various forms of correlated disorder and perhaps, the most dominant source being the unconventional ``twist-angle'' disorder \cite{Uri2020,SI20}.

Within our weak-coupling approach, we have focused on s-wave superconductivity with a uniform phase over the entire Fermi surface. While disorder might prompt such a state, recent experiments  suggest that the superconducting state near $\nu = -2$ breaks the rotational symmetry \cite{PJH20}. This implies that there is a strong indication that the SC order parameter is not purely s-wave and, moreover, belongs to a multi-component representation. Thus, it is important to understand under what conditions such a state is preferred and whether the mixture of Coulomb repulsion and umklapp phonons may favor such a state. Nonetheless, it should also be mentioned that the same umklapp enhancement mechanism for pairing considered here can also play a role in unconventional pairing states and we only considered the effect of these sources of attraction on the s-wave channel for simplicity.

We end by noting that in two spatial dimensions, the true SC transition is described by the Berezinskii-Kosterlitz-Thouless (BKT) transition, with $T_c=\pi D_s^-/2$, where $D_s^-$ is the superfluid stiffness at $T\rightarrow T_c^-$. In this paper, we identify $T_c$ with the scale associated with the formation of Cooper pairs, instead of the phase-ordering scale, which serves as an upper bound on the transition temperature. The geometric properties associated with the non-exponentially localizable Wannier functions for nearly flat bands \cite{Torma15,Bernevig19,Torma19,Rossi19,DC19} leads to an additional contribution to $D_s$. We leave a detailed study of such effects within our present framework for future work. 

\acknowledgements
C.L. acknowledges support from the STC Center for Integrated Quantum Materials, NSF Grant No. DMR-1231319, and from the Gordon and Betty Moore Foundation through Grant GBMF8682. D.C. is supported by a faculty startup grant at Cornell University. J.R. acknowledges funding by the Israeli Science Foundation under grant number 994/19.  

\begin{widetext}
\appendix
\section{Continuum model}
\label{app:bandstructure}

In this appendix, we provide additional details of the bandstructure model used in the calculations presented in the main text. We use the continuum model introduced in Ref.\cite{MK18} and restated here for completeness. The Hamiltonian for a valley $\xi=-1,1$ and spin $\sigma=\downarrow,\uparrow=-1,1$ takes the form
\be
H^{(\xi,\sigma)}=\begin{pmatrix} 
H_1 & U^\dagger \\
U & H_2 
\end{pmatrix}\label{eq:ham_cont}
\ee
in the basis of $(A_1, B_1, A_2, B_2)$ sites of the original two layers ($l=1,2$). The matrices $H_{l}$  correspond to the intralayer Hamiltonian of the layer $l$ and are explicitly given as
\be
H_{1}= -\frac{\hbar v}{a}\begin{pmatrix} 
0 &e^{-i\xi \theta/2} k_{-} a + \frac{4 \pi}{3} \\
e^{i\xi \theta/2} k_{+} a + \frac{4 \pi}{3} & 0
\end{pmatrix}\,,\quad H_{2}= -\frac{\hbar v}{a}\begin{pmatrix} 
0 &e^{i\xi \theta/2} k_{-} a + \frac{4 \pi}{3} \\
e^{-i\xi \theta/2} k_{+} a + \frac{4 \pi}{3} & 0
\end{pmatrix}
\ee
where $k_{\pm}=\xi k_x \pm i k_y$ and $k_x$,$k_y$ are crystal lattice momenta measured with respect to the original $\Gamma$ points of the graphene layers. The $4\pi/3$ terms in the matrix are remnants of the low-energy expansion of the graphene monolayer Hamiltonians around the $\bs K$ and $\bs K'$ points of the original layers. The MBZ of MATBG is defined as in the inset of Fig.\ref{fig:fig_1}a with the two reciprocal lattice vectors being
\be
    \vec{G}_1^{M} = -\frac{2\pi}{\sqrt{3} L_M} \begin{pmatrix} 
1 \\
\sqrt{3} 
\end{pmatrix}\,,\quad \vec{G}_2^{M} = \frac{4\pi}{\sqrt{3} L_M} \begin{pmatrix} 
1 \\
0 
\end{pmatrix}\,.
\ee
Here, we use the moir\'e real space lattice constant, $L_M = a/2\sin(\theta/2)$, and $\theta=1.05^\circ$. The matrix, $U$, is the effective moir{\'e} interlayer coupling given by:
\begin{align}
U = \begin{pmatrix}
u & u'\\
u' & u\\\end{pmatrix}+\begin{pmatrix}
u & u'\nu^{-\xi}\\
u'\nu^{\xi} & u\\\end{pmatrix}e^{i\xi \vec{G}_{1}^M\cdot\vec{r}}+\nn
&+\begin{pmatrix}
u & u'\nu^{\xi}\\
u'\nu^{-\xi} & u\\\end{pmatrix}e^{i\xi \left(\vec{G}_{1}^M+\vec{G}_{2}^M\right)\cdot\vec{r}}\,,
\end{align}
where $\nu=e^{i2\pi/3}$. We take the energy scale as $\hbar v/a = 2.1354$ eV and the lattice constant $a = 0.246$ nm. The interlayer coupling terms $u$ and $u'$ are taken as $u=0.0797$ eV and $u'=0.0975$ eV. For a detailed analysis of the origins of the Hamiltonian and discussion of the significance and numerical value of the coefficients, we refer the reader to Ref.~\cite{MK18} and references therein. In practice, the integers $m_1$ and $m_2$ in total cover at most only around $\sim 60$ combinations, which stems from using the cutoff procedure explained in Ref.~\cite{MK18}. We stress that no qualitative change to the bandstructure would be observed if the cutoff were to be increased. This is because majority of the spectral weight is in fact present only in the $m_1,m_2\in [-3,3]$ range covering $49$ possible combinations - that fact lies also behind the saturation of the phonon $T_C$ domes with the inclusion of $3$G umklapp processes, Fig.~\ref{fig:phonon_figure}a. The bandstructure for the two flat bands for valley $\xi=1$ is shown in Fig.\ref{fig:fig_1}b. We refer to the bands Fig.~\ref{fig:fig_1}b as the flat bands each with a bandwidth $W \approx 4$ meV. Accordingly, all other bands are called non-flat bands and are separated from the flat-bands by a bandgap, $\Delta_{\rm band} \approx 12$ meV.

\section{Dielectric function of TBG}
\label{app:RPA}

{In this appendix we detail the procedure used to obtain the dielectric function $\epsilon(\vec{q},i \omega)$. We also discuss key properties of the dynamical dielectric function and its connection to plasmon properties in TBG. }

{To find the dielectric function, we start from its conventional definition as explained in the main text:}
\begin{equation}
    \epsilon_{RPA}(\vec{q},\omega) = \kappa - 2 \pi e^2 \Pi_{ee}(\vec{q}, \omega)/q
\end{equation}
{and approximate the polarization function within the random phase approximation\cite{mahan2000many-particle}:}
\begin{equation}
\Pi_{ee}(\vec{q},\omega) = \sum_{\vec{k},\gamma,\gamma'} \frac{ (f_{\gamma,\vec k+\vec q}-f_{\gamma',\vec k})\Lambda_{\gamma\gamma'}(\vec{k}+\vec{q},\vec{k})}{E_{\vec{k}+\vec{q}\{\gamma\}}-E_{\vec{k}\{\gamma'\}}-\omega-i0}
,
\label{eq:pol_tight_binding}
\end{equation}
{where $\sum_{\vec{k}}$ denotes integration over the Brillouin zone and the composite indices $\gamma,\gamma'$ run over electron bands, valley and spins. Here $f_{\gamma,\vec k}$ is the Fermi-Dirac distribution for a state with energy $E_{\vec{k}\{\gamma\}}$, and $\Lambda_{\g\g'}(\bs p,\bs k)=\delta_{\xi\xi'}\delta_{\s\s'}\left\langle \bs p,\{\g\}\left|e^{i (\vec{p-k})\cdot \vec{r}} \right|\bs k,\{\g'\}\right \rangle$ describes the overlap between the Bloch eigenstates as introduced in the main text.}

{A plot of a dielectric function for few fillings is shown in Fig.\ref{fig:app_fig_die_fun}. As in Ref.\cite{Lewandowski20869} we find a plasmon mode that pierces through the p-h continuum and rises above it. At low momenta, $q \ll k_F$ (here $k_F$ is Fermi momentum), plasmon disperses as $\omega_{pl}(q) \propto \sqrt{q}$ as expected of a 2D plasmon. At large momenta comparable to a moir\'e reciprocal lattice vector, $q \approx G/2$, as mentioned in the text and explained in Ref.\cite{Lewandowski20869},  dispersion of the plasmon is set by the bandwidth $W$ of the nearly-flat band, and the gap $\Delta_{band}$ between flat and dispersive bands as $\omega_{pl}\approx \sqrt{W \Delta_{band}}$. }

{To compute the dielectric function $\epsilon_{RPA}(\vec{q},i\omega)$ we calculate first corresponding dielectric function at real-frequencies and then employ Kramers-Kronig relations. This approach yields precisely the same dielectric function as that of Eq.\eqref{eq:pol_tight_binding} upon replacement $\omega +i0 \to i\omega$. We resorted to using this procedure as in a typical pairing calculation $\epsilon_{RPA}(\vec{q},i\omega)$ has to be evaluated multiple times for different Matsubara frequencies. Given that evaluation of the dielectric function is the most demanding step of the $T_C$ calculation this approach proved to be most efficient as it involved computing dielectric function once for a given filling.}

\begin{figure}
    \centering
    \includegraphics[width=\linewidth]{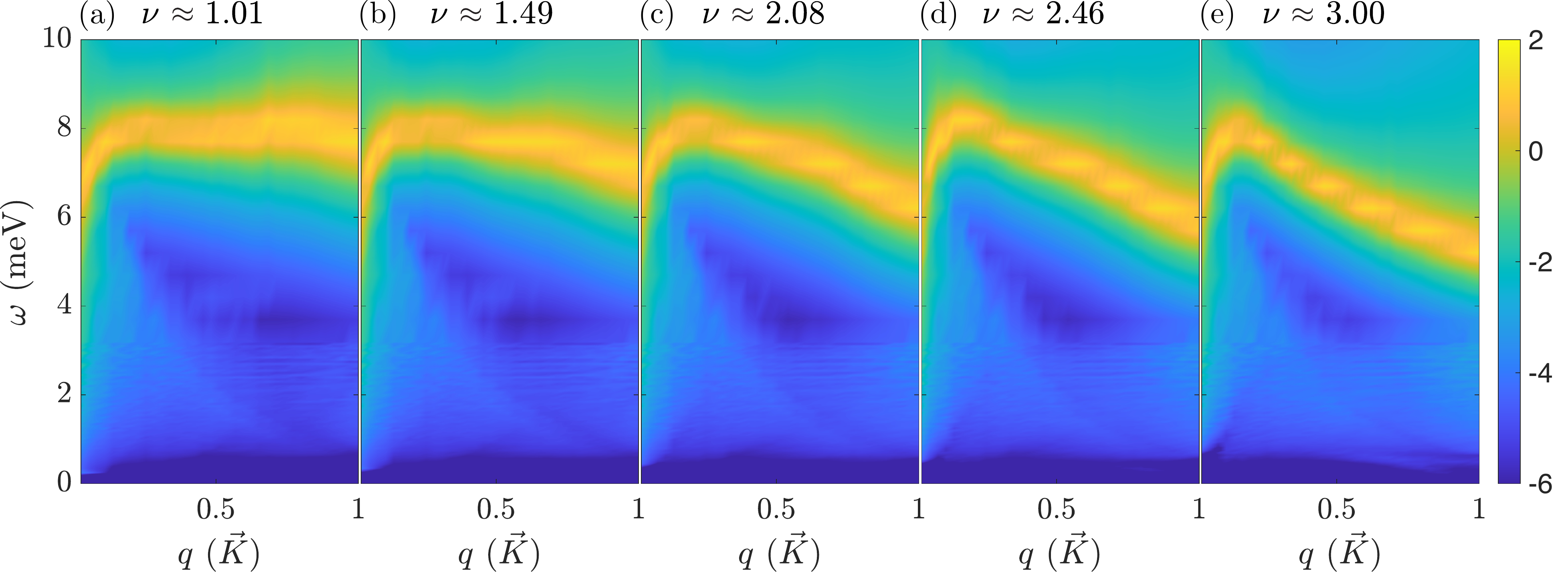}
    \caption{{(a-e) Logarithm of electron loss function $\log[\Im [-1/\epsilon_{RPA}(\vec{q},\omega)]]$ for a range of fillings. Blue regions correspond to the particle-hole continuum, whilst the bright yellow region is the plasmon mode.}}
    \label{fig:app_fig_die_fun}
\end{figure}

\section{Numerical solution of the (non-)linear gap equations}
\label{app:numsol}

Here we detail the procedure used to solve the linearized gap equation introduced in the main text. In the second half of this section we extend this calculation to the full non-linear gap problem, which is then employed in the process of analytic continuation, necessary for computation of the tunneling density of states.

As explained in the main text, the linearized gap equation is posed as an eigenvalue problem, Eq. \eqref{eq:sc_gap_equation}:
\begin{align}
 \Delta(i\omega,{k}) = -T \sum_{\nu} \sum_{{p}} K(i\w,{k};i\nu,{p})\Delta(i\nu,{p}) \label{eq:app_sc_gap_equation},
\end{align}
where the kernel $K(...)$ is given by 
\begin{align}\label{eq:app_Kernel}
K(i\w,{k};i\nu,{p}) \equiv \frac{1}{(2\pi)^2} \int d\W_{\bs p} \V^{\tn{int}}_{-\xi,\xi}(\bs k-\bs p,i\omega-i\nu) \frac{\Lambda( \bs p ,\bs k)\Lambda(-\bs p,\bs k) }{\nu^2+{E}^2_{\xi,\vec{p}}}
\end{align}
and where $\int d\W_{\bs p}$ denotes integration over the angle between vectors $\vec{k}$ and $\vec{p}$ for a fixed direction of $\vec{k}$. To proceed with the analysis, we first make a simplifying assumption that the superconducting order parameter is a function of Matsubara frequency, $i\omega$, and magnitude of the momentum, $k$, but not on the angle, respectively. This allows us to solve the resulting integral equation on a two-dimensional grid of $k\times\omega$ points. With this assumption, the eigenvalue problem simplifies then to
\begin{align}
 \Delta(i\omega,k) = -T \sum_{\nu} \sum_{k} K(i\w,k;i\nu,p)\Delta(i\nu,p) 
 \label{eq:app_sc_gap_equation_mom_mag}
\end{align}
with the momentum direction-averaged kernel, described in detail below, is given by
\begin{align}
K(i\w,k;i\nu,p) \equiv {1\over (2\pi)^2} \int d\W_{\bs p} \V^{\tn{int}}_{-\xi,\xi}(|\bs k-\bs p|,i\omega-i\nu) \frac{\Lambda( \bs p ,\bs k)\Lambda(-\bs p,\bs k) }{\nu^2+{E}^2_{\xi,\vec{p}}}
\label{eq:app_Kernel_mom_mag}
\end{align}
In the above kernel we have made the assumption that the interaction $V^{\tn{int}}(...)$ has no explicit dependence on the angle associated with the momentum, and only depends on its magnitude. This simplification allows us to compute the dynamical dielectric function along just one momentum direction and then estimate it along the other directions by simply comparing magnitudes of $|\vec{k}-\vec{p}|$. 

We start the calculation by pre-computing a 2D MBZ mesh of points, with their associated Bloch wavefunctions and energies. In the calculations we use a mesh of $\sim{8000}$ MBZ points. To carry out an angle-average of the kernel from Eq.~\eqref{eq:app_Kernel_mom_mag}, we first fix a particular direction of vector $\vec{k}$ (upon verifying that the conclusions are not dependent on the specific direction) and then identify all $\vec{p}$ points that are of magnitude $p$ (within a resolution admitted by the mesh). We then estimate the angular average Eq.~\eqref{eq:app_Kernel_mom_mag} by averaging over these $\vec{p}$ points --- in practice, $\sim 50-100$ points are used for each pair of $k$ and $p$ momentum values.

To determine the critical temperature, we seek the temperature $T$ for which Eq.~\eqref{eq:app_sc_gap_equation_mom_mag} has an eigenvalue of unity. In practice, we carry out a bisection method search for a $T$ giving an eigenvalue within $\pm 0.001$ of the unity. In the calculations we use a linearly spaced grid of $30$ $k$ points ranging from the center of the MBZ $\bar{\Gamma}$ to the  $\bar{K}$ point. For the Matsubara grid we employed both an exact Matsubara frequency summation as well as an approximate scheme, upon verifying that the approximate Matsubara grid agrees (in determining the value of the critical temperature) with the exact summation to 3 significant figures. The approximate Matsubara grid was chosen to consist of 10 first Matsubara frequencies followed by 20 linearly spaced frequencies starting from the 11th Matsubara frequency to the UV cutoff. Several UV cutoffs were tested: multiples of plasma frequency, multiples of the Debye frequency, or, multiples of the system bandwidth. All cutoffs were found consistent with each other provided that the UV cutoff exceeds, roughly, $30$ meV for the bandstructure used.

To determine a self-consistent gap, we extend the calculation by modifying the kernel in the integral equation Eq. \eqref{eq:app_sc_gap_equation_mom_mag} as
\begin{align}
K(i\w,k;i\nu,p)\to K_{\tn{SC}}(i\w,k;i\nu,p) \equiv {1\over (2\pi)^2} \int d\W_{\bs p} \V^{\tn{int}}_{-\xi,\xi}(|\bs k-\bs p|,i\omega-i\nu) \frac{\Lambda( \bs p ,\bs k)\Lambda(-\bs p,\bs k) }{\nu^2+{E}^2_{\xi,\vec{p}}+\Delta^2(i\nu,\vec{p})}
\label{eq:app_Kernel_mom_mag_SC}
\end{align}
where $\Delta(i\nu,\vec{p})$ is now the self-consistent gap. Starting with the eigenvector of the linearized gap equation as an input, we then self-consistently solve the gap equation. For temperatures, $T \ll T_C$, as used in Fig.~\ref{fig:fig_4}c, the gap converges within $10-20$ iterations to below $0.1\%$ total relative error (difference between successive self-consistent steps). The resulting gap function follows the BCS result; see Fig.\ref{fig:app_fig_gap_equation}(a). We note that in the above procedure for obtaining the self-consistent gap we neglect other Eliashberg equations, specifically the interaction correction to quasiparticle weight, $Z$. This procedure is justified in the large-$N$ limit as explained in the main text as these corrections would be subleading in $N$.

\begin{figure}[!htb]
    \centering
    \includegraphics[width=\linewidth]{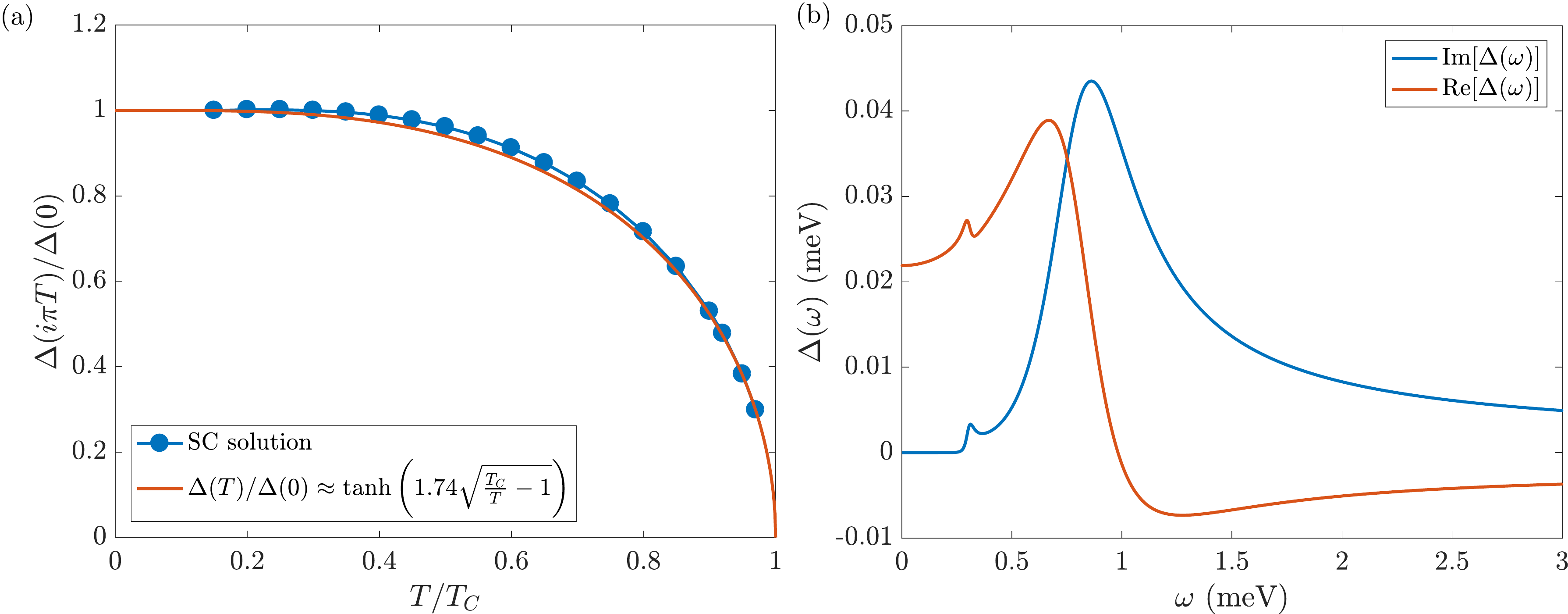}
    \caption{(a) Superconducting gap as a function of temperature obtained from the self-consistent calculation. Plotted quantity (blue) corresponds to an average of the superconducting gap $\Delta(i\nu,\vec{p})$ over the Fermi surface. Orange line corresponds to the BCS interpolating solution (accurate withing few percent). (b) Analytically continued superconducting gap $\Delta(\omega)$ as a function of real frequency obtained from the self-consistent calculation. As in (a) the gap was averaged over the Fermi surface. In (a),(b) we consider phonon mediated superconducting with only 0G processes at a characteristic filling of $\nu = 1$. Note how ${\rm Im}[\Delta(\omega)]$ is zero until gap opens, followed by a peak at the characteristic frequency for 0G processes, as expected from Fig.~\ref{fig:phonon_figure}b. }
    \label{fig:app_fig_gap_equation}
\end{figure}

\section{Analytic continuation and single particle density of states}
\label{app:dos}
To determine the tunneling density of states it is necessary to obtain the self-consistent gap as a function of real frequency \cite{Marsiglio2008,PhysRevB.87.024505}. To avoid this complex procedure, few schemes of carrying out an analytic continuation from the Matsubara frequency to real frequencies have been introduced in literature with the two most common relying on either Pad{\'e} approximants \cite{10.1007/BF00655090,LEAVENS1985137}, or, an iterative solution to the Eliashberg equations \cite{PhysRevB.37.4965}. Due to the ease of its implementation we employ a Pad{\'e} approximation scheme, in particular, the approach detailed in Ref.~\cite{PhysRevB.61.5147}. We stress however that the qualitative features, the ``dip-humps'', of Fig.\ref{fig:fig_4}c will remain unaffected as they stem from the additional peaks of the el-ph spectral function.

As an input, we start with the self-consistent gap  obtained following the procedure detailed in the previous section. This gap is then approximated with the help of Pad{\'e} approximant  and analytically continued to real frequencies. An example of real frequency dependence of a gap is shown in Fig. \ref{fig:app_fig_gap_equation}(b). This self-consistent gap defined at real-frequencies is then used to compute the spectral function \cite{coleman_2015},
\be
\mathcal{A}_{\{\gamma\}}(\omega,\vec{k}) = -\frac{1}{\pi}{\rm Im}\left[\frac{(\omega+i0)+E_{\vec{k},\{\gamma\}}}{(\omega+i0)^2-E^2_{\vec{k},\{\gamma\}}-\Delta^2(\omega)}\right]\,,
\ee
and the tunneling density of states $N_{\tn{SC}}(\omega-\mu)$,
\be
N_{SC}(\omega-\mu) = \sum_{\{\gamma\}} \int_{MBZ} d^2 \vec{k}~ \mathcal{A}_{\{\gamma\}}(\omega,\vec{k}),
\ee
shown in Fig.~\ref{fig:fig_4}c.

\end{widetext}

\bibliography{references}

\begin{thebibliography}{90}%
\makeatletter
\providecommand \@ifxundefined [1]{%
 \@ifx{#1\undefined}
}%
\providecommand \@ifnum [1]{%
 \ifnum #1\expandafter \@firstoftwo
 \else \expandafter \@secondoftwo
 \fi
}%
\providecommand \@ifx [1]{%
 \ifx #1\expandafter \@firstoftwo
 \else \expandafter \@secondoftwo
 \fi
}%
\providecommand \natexlab [1]{#1}%
\providecommand \enquote  [1]{``#1''}%
\providecommand \bibnamefont  [1]{#1}%
\providecommand \bibfnamefont [1]{#1}%
\providecommand \citenamefont [1]{#1}%
\providecommand \href@noop [0]{\@secondoftwo}%
\providecommand \href [0]{\begingroup \@sanitize@url \@href}%
\providecommand \@href[1]{\@@startlink{#1}\@@href}%
\providecommand \@@href[1]{\endgroup#1\@@endlink}%
\providecommand \@sanitize@url [0]{\catcode `\\12\catcode `\$12\catcode
  `\&12\catcode `\#12\catcode `\^12\catcode `\_12\catcode `\%12\relax}%
\providecommand \@@startlink[1]{}%
\providecommand \@@endlink[0]{}%
\providecommand \url  [0]{\begingroup\@sanitize@url \@url }%
\providecommand \@url [1]{\endgroup\@href {#1}{\urlprefix }}%
\providecommand \urlprefix  [0]{URL }%
\providecommand \Eprint [0]{\href }%
\providecommand \doibase [0]{http://dx.doi.org/}%
\providecommand \selectlanguage [0]{\@gobble}%
\providecommand \bibinfo  [0]{\@secondoftwo}%
\providecommand \bibfield  [0]{\@secondoftwo}%
\providecommand \translation [1]{[#1]}%
\providecommand \BibitemOpen [0]{}%
\providecommand \bibitemStop [0]{}%
\providecommand \bibitemNoStop [0]{.\EOS\space}%
\providecommand \EOS [0]{\spacefactor3000\relax}%
\providecommand \BibitemShut  [1]{\csname bibitem#1\endcsname}%
\let\auto@bib@innerbib\@empty
\bibitem [{\citenamefont {Cao}\ \emph {et~al.}(2018{\natexlab{a}})\citenamefont
  {Cao}, \citenamefont {Fatemi}, \citenamefont {Fang}, \citenamefont
  {Watanabe}, \citenamefont {Taniguchi}, \citenamefont {Kaxiras},\ and\
  \citenamefont {Jarillo-Herrero}}]{cao2}%
  \BibitemOpen
  \bibfield  {author} {\bibinfo {author} {\bibfnamefont {Yuan}\ \bibnamefont
  {Cao}}, \bibinfo {author} {\bibfnamefont {Valla}\ \bibnamefont {Fatemi}},
  \bibinfo {author} {\bibfnamefont {Shiang}\ \bibnamefont {Fang}}, \bibinfo
  {author} {\bibfnamefont {Kenji}\ \bibnamefont {Watanabe}}, \bibinfo {author}
  {\bibfnamefont {Takashi}\ \bibnamefont {Taniguchi}}, \bibinfo {author}
  {\bibfnamefont {Efthimios}\ \bibnamefont {Kaxiras}}, \ and\ \bibinfo {author}
  {\bibfnamefont {Pablo}\ \bibnamefont {Jarillo-Herrero}},\ }\bibfield  {title}
  {\enquote {\bibinfo {title} {Unconventional superconductivity in magic-angle
  graphene superlattices},}\ }\href {https://doi.org/10.1038/nature26160}
  {\bibfield  {journal} {\bibinfo  {journal} {Nature}\ }\textbf {\bibinfo
  {volume} {556}},\ \bibinfo {pages} {43 EP --} (\bibinfo {year}
  {2018}{\natexlab{a}})}\BibitemShut {NoStop}%
\bibitem [{\citenamefont {Cao}\ \emph {et~al.}(2018{\natexlab{b}})\citenamefont
  {Cao}, \citenamefont {Fatemi}, \citenamefont {Demir}, \citenamefont {Fang},
  \citenamefont {Tomarken}, \citenamefont {Luo}, \citenamefont
  {Sanchez-Yamagishi}, \citenamefont {Watanabe}, \citenamefont {Taniguchi},
  \citenamefont {Kaxiras}, \citenamefont {Ashoori},\ and\ \citenamefont
  {Jarillo-Herrero}}]{cao1}%
  \BibitemOpen
  \bibfield  {author} {\bibinfo {author} {\bibfnamefont {Yuan}\ \bibnamefont
  {Cao}}, \bibinfo {author} {\bibfnamefont {Valla}\ \bibnamefont {Fatemi}},
  \bibinfo {author} {\bibfnamefont {Ahmet}\ \bibnamefont {Demir}}, \bibinfo
  {author} {\bibfnamefont {Shiang}\ \bibnamefont {Fang}}, \bibinfo {author}
  {\bibfnamefont {Spencer~L.}\ \bibnamefont {Tomarken}}, \bibinfo {author}
  {\bibfnamefont {Jason~Y.}\ \bibnamefont {Luo}}, \bibinfo {author}
  {\bibfnamefont {Javier~D.}\ \bibnamefont {Sanchez-Yamagishi}}, \bibinfo
  {author} {\bibfnamefont {Kenji}\ \bibnamefont {Watanabe}}, \bibinfo {author}
  {\bibfnamefont {Takashi}\ \bibnamefont {Taniguchi}}, \bibinfo {author}
  {\bibfnamefont {Efthimios}\ \bibnamefont {Kaxiras}}, \bibinfo {author}
  {\bibfnamefont {Ray~C.}\ \bibnamefont {Ashoori}}, \ and\ \bibinfo {author}
  {\bibfnamefont {Pablo}\ \bibnamefont {Jarillo-Herrero}},\ }\bibfield  {title}
  {\enquote {\bibinfo {title} {Correlated insulator behaviour at half-filling
  in magic-angle graphene superlattices},}\ }\href
  {https://doi.org/10.1038/nature26154} {\bibfield  {journal} {\bibinfo
  {journal} {Nature}\ }\textbf {\bibinfo {volume} {556}},\ \bibinfo {pages} {80
  EP --} (\bibinfo {year} {2018}{\natexlab{b}})}\BibitemShut {NoStop}%
\bibitem [{\citenamefont {Yankowitz}\ \emph {et~al.}(2019)\citenamefont
  {Yankowitz}, \citenamefont {Chen}, \citenamefont {Polshyn}, \citenamefont
  {Zhang}, \citenamefont {Watanabe}, \citenamefont {Taniguchi}, \citenamefont
  {Graf}, \citenamefont {Young},\ and\ \citenamefont {Dean}}]{Yankowitz1059}%
  \BibitemOpen
  \bibfield  {author} {\bibinfo {author} {\bibfnamefont {Matthew}\ \bibnamefont
  {Yankowitz}}, \bibinfo {author} {\bibfnamefont {Shaowen}\ \bibnamefont
  {Chen}}, \bibinfo {author} {\bibfnamefont {Hryhoriy}\ \bibnamefont
  {Polshyn}}, \bibinfo {author} {\bibfnamefont {Yuxuan}\ \bibnamefont {Zhang}},
  \bibinfo {author} {\bibfnamefont {K.}~\bibnamefont {Watanabe}}, \bibinfo
  {author} {\bibfnamefont {T.}~\bibnamefont {Taniguchi}}, \bibinfo {author}
  {\bibfnamefont {David}\ \bibnamefont {Graf}}, \bibinfo {author}
  {\bibfnamefont {Andrea~F.}\ \bibnamefont {Young}}, \ and\ \bibinfo {author}
  {\bibfnamefont {Cory~R.}\ \bibnamefont {Dean}},\ }\bibfield  {title}
  {\enquote {\bibinfo {title} {Tuning superconductivity in twisted bilayer
  graphene},}\ }\href {\doibase 10.1126/science.aav1910} {\bibfield  {journal}
  {\bibinfo  {journal} {Science}\ }\textbf {\bibinfo {volume} {363}},\ \bibinfo
  {pages} {1059--1064} (\bibinfo {year} {2019})}\BibitemShut {NoStop}%
\bibitem [{\citenamefont {Lu}\ \emph {et~al.}(2019)\citenamefont {Lu},
  \citenamefont {Stepanov}, \citenamefont {Yang}, \citenamefont {Xie},
  \citenamefont {Aamir}, \citenamefont {Das}, \citenamefont {Urgell},
  \citenamefont {Watanabe}, \citenamefont {Taniguchi}, \citenamefont {Zhang},
  \citenamefont {Bachtold}, \citenamefont {MacDonald},\ and\ \citenamefont
  {Efetov}}]{efetov2019}%
  \BibitemOpen
  \bibfield  {author} {\bibinfo {author} {\bibfnamefont {Xiaobo}\ \bibnamefont
  {Lu}}, \bibinfo {author} {\bibfnamefont {Petr}\ \bibnamefont {Stepanov}},
  \bibinfo {author} {\bibfnamefont {Wei}\ \bibnamefont {Yang}}, \bibinfo
  {author} {\bibfnamefont {Ming}\ \bibnamefont {Xie}}, \bibinfo {author}
  {\bibfnamefont {Mohammed~Ali}\ \bibnamefont {Aamir}}, \bibinfo {author}
  {\bibfnamefont {Ipsita}\ \bibnamefont {Das}}, \bibinfo {author}
  {\bibfnamefont {Carles}\ \bibnamefont {Urgell}}, \bibinfo {author}
  {\bibfnamefont {Kenji}\ \bibnamefont {Watanabe}}, \bibinfo {author}
  {\bibfnamefont {Takashi}\ \bibnamefont {Taniguchi}}, \bibinfo {author}
  {\bibfnamefont {Guangyu}\ \bibnamefont {Zhang}}, \bibinfo {author}
  {\bibfnamefont {Adrian}\ \bibnamefont {Bachtold}}, \bibinfo {author}
  {\bibfnamefont {Allan~H.}\ \bibnamefont {MacDonald}}, \ and\ \bibinfo
  {author} {\bibfnamefont {Dmitri~K.}\ \bibnamefont {Efetov}},\ }\bibfield
  {title} {\enquote {\bibinfo {title} {Superconductors, orbital magnets and
  correlated states in magic-angle bilayer graphene},}\ }\href {\doibase
  10.1038/s41586-019-1695-0} {\bibfield  {journal} {\bibinfo  {journal}
  {Nature}\ }\textbf {\bibinfo {volume} {574}},\ \bibinfo {pages} {653--657}
  (\bibinfo {year} {2019})}\BibitemShut {NoStop}%
\bibitem [{\citenamefont {Balents}\ \emph {et~al.}(2020)\citenamefont
  {Balents}, \citenamefont {Dean}, \citenamefont {Efetov},\ and\ \citenamefont
  {Young}}]{Balents2020}%
  \BibitemOpen
  \bibfield  {author} {\bibinfo {author} {\bibfnamefont {Leon}\ \bibnamefont
  {Balents}}, \bibinfo {author} {\bibfnamefont {Cory~R.}\ \bibnamefont {Dean}},
  \bibinfo {author} {\bibfnamefont {Dmitri~K.}\ \bibnamefont {Efetov}}, \ and\
  \bibinfo {author} {\bibfnamefont {Andrea~F.}\ \bibnamefont {Young}},\
  }\bibfield  {title} {\enquote {\bibinfo {title} {Superconductivity and strong
  correlations in moir{\'e}flat bands},}\ }\href {\doibase
  10.1038/s41567-020-0906-9} {\bibfield  {journal} {\bibinfo  {journal} {Nature
  Physics}\ }\textbf {\bibinfo {volume} {16}},\ \bibinfo {pages} {725--733}
  (\bibinfo {year} {2020})}\BibitemShut {NoStop}%
\bibitem [{\citenamefont {Lopes~dos Santos}\ \emph {et~al.}(2007)\citenamefont
  {Lopes~dos Santos}, \citenamefont {Peres},\ and\ \citenamefont
  {Castro~Neto}}]{Castro}%
  \BibitemOpen
  \bibfield  {author} {\bibinfo {author} {\bibfnamefont {J.~M.~B.}\
  \bibnamefont {Lopes~dos Santos}}, \bibinfo {author} {\bibfnamefont
  {N.~M.~R.}\ \bibnamefont {Peres}}, \ and\ \bibinfo {author} {\bibfnamefont
  {A.~H.}\ \bibnamefont {Castro~Neto}},\ }\bibfield  {title} {\enquote
  {\bibinfo {title} {Graphene bilayer with a twist: Electronic structure},}\
  }\href {\doibase 10.1103/PhysRevLett.99.256802} {\bibfield  {journal}
  {\bibinfo  {journal} {Phys. Rev. Lett.}\ }\textbf {\bibinfo {volume} {99}},\
  \bibinfo {pages} {256802} (\bibinfo {year} {2007})}\BibitemShut {NoStop}%
\bibitem [{\citenamefont {Su\'arez~Morell}\ \emph {et~al.}(2010)\citenamefont
  {Su\'arez~Morell}, \citenamefont {Correa}, \citenamefont {Vargas},
  \citenamefont {Pacheco},\ and\ \citenamefont {Barticevic}}]{suarez}%
  \BibitemOpen
  \bibfield  {author} {\bibinfo {author} {\bibfnamefont {E.}~\bibnamefont
  {Su\'arez~Morell}}, \bibinfo {author} {\bibfnamefont {J.~D.}\ \bibnamefont
  {Correa}}, \bibinfo {author} {\bibfnamefont {P.}~\bibnamefont {Vargas}},
  \bibinfo {author} {\bibfnamefont {M.}~\bibnamefont {Pacheco}}, \ and\
  \bibinfo {author} {\bibfnamefont {Z.}~\bibnamefont {Barticevic}},\ }\bibfield
   {title} {\enquote {\bibinfo {title} {Flat bands in slightly twisted bilayer
  graphene: Tight-binding calculations},}\ }\href {\doibase
  10.1103/PhysRevB.82.121407} {\bibfield  {journal} {\bibinfo  {journal} {Phys.
  Rev. B}\ }\textbf {\bibinfo {volume} {82}},\ \bibinfo {pages} {121407}
  (\bibinfo {year} {2010})}\BibitemShut {NoStop}%
\bibitem [{\citenamefont {Bistritzer}\ and\ \citenamefont
  {MacDonald}(2011)}]{macdonald11}%
  \BibitemOpen
  \bibfield  {author} {\bibinfo {author} {\bibfnamefont {Rafi}\ \bibnamefont
  {Bistritzer}}\ and\ \bibinfo {author} {\bibfnamefont {Allan~H.}\ \bibnamefont
  {MacDonald}},\ }\bibfield  {title} {\enquote {\bibinfo {title} {Moir{\'e}
  bands in twisted double-layer graphene},}\ }\href {\doibase
  10.1073/pnas.1108174108} {\bibfield  {journal} {\bibinfo  {journal}
  {Proceedings of the National Academy of Sciences}\ }\textbf {\bibinfo
  {volume} {108}},\ \bibinfo {pages} {12233--12237} (\bibinfo {year}
  {2011})}\BibitemShut {NoStop}%
\bibitem [{\citenamefont {Kerelsky}\ \emph {et~al.}(2019)\citenamefont
  {Kerelsky}, \citenamefont {McGilly}, \citenamefont {Kennes}, \citenamefont
  {Xian}, \citenamefont {Yankowitz}, \citenamefont {Chen}, \citenamefont
  {Watanabe}, \citenamefont {Taniguchi}, \citenamefont {Hone}, \citenamefont
  {Dean}, \citenamefont {Rubio},\ and\ \citenamefont {Pasupathy}}]{AP19}%
  \BibitemOpen
  \bibfield  {author} {\bibinfo {author} {\bibfnamefont {Alexander}\
  \bibnamefont {Kerelsky}}, \bibinfo {author} {\bibfnamefont {Leo~J.}\
  \bibnamefont {McGilly}}, \bibinfo {author} {\bibfnamefont {Dante~M.}\
  \bibnamefont {Kennes}}, \bibinfo {author} {\bibfnamefont {Lede}\ \bibnamefont
  {Xian}}, \bibinfo {author} {\bibfnamefont {Matthew}\ \bibnamefont
  {Yankowitz}}, \bibinfo {author} {\bibfnamefont {Shaowen}\ \bibnamefont
  {Chen}}, \bibinfo {author} {\bibfnamefont {K.}~\bibnamefont {Watanabe}},
  \bibinfo {author} {\bibfnamefont {T.}~\bibnamefont {Taniguchi}}, \bibinfo
  {author} {\bibfnamefont {James}\ \bibnamefont {Hone}}, \bibinfo {author}
  {\bibfnamefont {Cory}\ \bibnamefont {Dean}}, \bibinfo {author} {\bibfnamefont
  {Angel}\ \bibnamefont {Rubio}}, \ and\ \bibinfo {author} {\bibfnamefont
  {Abhay~N.}\ \bibnamefont {Pasupathy}},\ }\bibfield  {title} {\enquote
  {\bibinfo {title} {Maximized electron interactions at the magic angle in
  twisted bilayer graphene},}\ }\href {\doibase 10.1038/s41586-019-1431-9}
  {\bibfield  {journal} {\bibinfo  {journal} {Nature}\ }\textbf {\bibinfo
  {volume} {572}},\ \bibinfo {pages} {95--100} (\bibinfo {year}
  {2019})}\BibitemShut {NoStop}%
\bibitem [{\citenamefont {Choi}\ \emph {et~al.}(2019)\citenamefont {Choi},
  \citenamefont {Kemmer}, \citenamefont {Peng}, \citenamefont {Thomson},
  \citenamefont {Arora}, \citenamefont {Polski}, \citenamefont {Zhang},
  \citenamefont {Ren}, \citenamefont {Alicea}, \citenamefont {Refael},
  \citenamefont {von Oppen}, \citenamefont {Watanabe}, \citenamefont
  {Taniguchi},\ and\ \citenamefont {Nadj-Perge}}]{SNP19}%
  \BibitemOpen
  \bibfield  {author} {\bibinfo {author} {\bibfnamefont {Youngjoon}\
  \bibnamefont {Choi}}, \bibinfo {author} {\bibfnamefont {Jeannette}\
  \bibnamefont {Kemmer}}, \bibinfo {author} {\bibfnamefont {Yang}\ \bibnamefont
  {Peng}}, \bibinfo {author} {\bibfnamefont {Alex}\ \bibnamefont {Thomson}},
  \bibinfo {author} {\bibfnamefont {Harpreet}\ \bibnamefont {Arora}}, \bibinfo
  {author} {\bibfnamefont {Robert}\ \bibnamefont {Polski}}, \bibinfo {author}
  {\bibfnamefont {Yiran}\ \bibnamefont {Zhang}}, \bibinfo {author}
  {\bibfnamefont {Hechen}\ \bibnamefont {Ren}}, \bibinfo {author}
  {\bibfnamefont {Jason}\ \bibnamefont {Alicea}}, \bibinfo {author}
  {\bibfnamefont {Gil}\ \bibnamefont {Refael}}, \bibinfo {author}
  {\bibfnamefont {Felix}\ \bibnamefont {von Oppen}}, \bibinfo {author}
  {\bibfnamefont {Kenji}\ \bibnamefont {Watanabe}}, \bibinfo {author}
  {\bibfnamefont {Takashi}\ \bibnamefont {Taniguchi}}, \ and\ \bibinfo {author}
  {\bibfnamefont {Stevan}\ \bibnamefont {Nadj-Perge}},\ }\bibfield  {title}
  {\enquote {\bibinfo {title} {Electronic correlations in twisted bilayer
  graphene near the magic angle},}\ }\href {\doibase 10.1038/s41567-019-0606-5}
  {\bibfield  {journal} {\bibinfo  {journal} {Nature Physics}\ }\textbf
  {\bibinfo {volume} {15}},\ \bibinfo {pages} {1174--1180} (\bibinfo {year}
  {2019})}\BibitemShut {NoStop}%
\bibitem [{\citenamefont {Jiang}\ \emph {et~al.}(2019)\citenamefont {Jiang},
  \citenamefont {Lai}, \citenamefont {Watanabe}, \citenamefont {Taniguchi},
  \citenamefont {Haule}, \citenamefont {Mao},\ and\ \citenamefont
  {Andrei}}]{EA19}%
  \BibitemOpen
  \bibfield  {author} {\bibinfo {author} {\bibfnamefont {Yuhang}\ \bibnamefont
  {Jiang}}, \bibinfo {author} {\bibfnamefont {Xinyuan}\ \bibnamefont {Lai}},
  \bibinfo {author} {\bibfnamefont {Kenji}\ \bibnamefont {Watanabe}}, \bibinfo
  {author} {\bibfnamefont {Takashi}\ \bibnamefont {Taniguchi}}, \bibinfo
  {author} {\bibfnamefont {Kristjan}\ \bibnamefont {Haule}}, \bibinfo {author}
  {\bibfnamefont {Jinhai}\ \bibnamefont {Mao}}, \ and\ \bibinfo {author}
  {\bibfnamefont {Eva~Y.}\ \bibnamefont {Andrei}},\ }\bibfield  {title}
  {\enquote {\bibinfo {title} {Charge order and broken rotational symmetry in
  magic-angle twisted bilayer graphene},}\ }\href {\doibase
  10.1038/s41586-019-1460-4} {\bibfield  {journal} {\bibinfo  {journal}
  {Nature}\ }\textbf {\bibinfo {volume} {573}},\ \bibinfo {pages} {91--95}
  (\bibinfo {year} {2019})}\BibitemShut {NoStop}%
\bibitem [{\citenamefont {Zondiner}\ \emph {et~al.}(2020)\citenamefont
  {Zondiner}, \citenamefont {Rozen}, \citenamefont {Rodan-Legrain},
  \citenamefont {Cao}, \citenamefont {Queiroz}, \citenamefont {Taniguchi},
  \citenamefont {Watanabe}, \citenamefont {Oreg}, \citenamefont {von Oppen},
  \citenamefont {Stern}, \citenamefont {Berg}, \citenamefont
  {Jarillo-Herrero},\ and\ \citenamefont {Ilani}}]{SI20}%
  \BibitemOpen
  \bibfield  {author} {\bibinfo {author} {\bibfnamefont {U.}~\bibnamefont
  {Zondiner}}, \bibinfo {author} {\bibfnamefont {A.}~\bibnamefont {Rozen}},
  \bibinfo {author} {\bibfnamefont {D.}~\bibnamefont {Rodan-Legrain}}, \bibinfo
  {author} {\bibfnamefont {Y.}~\bibnamefont {Cao}}, \bibinfo {author}
  {\bibfnamefont {R.}~\bibnamefont {Queiroz}}, \bibinfo {author} {\bibfnamefont
  {T.}~\bibnamefont {Taniguchi}}, \bibinfo {author} {\bibfnamefont
  {K.}~\bibnamefont {Watanabe}}, \bibinfo {author} {\bibfnamefont
  {Y.}~\bibnamefont {Oreg}}, \bibinfo {author} {\bibfnamefont {F.}~\bibnamefont
  {von Oppen}}, \bibinfo {author} {\bibfnamefont {Ady}\ \bibnamefont {Stern}},
  \bibinfo {author} {\bibfnamefont {E.}~\bibnamefont {Berg}}, \bibinfo {author}
  {\bibfnamefont {P.}~\bibnamefont {Jarillo-Herrero}}, \ and\ \bibinfo {author}
  {\bibfnamefont {S.}~\bibnamefont {Ilani}},\ }\bibfield  {title} {\enquote
  {\bibinfo {title} {Cascade of phase transitions and dirac revivals in
  magic-angle graphene},}\ }\href {\doibase 10.1038/s41586-020-2373-y}
  {\bibfield  {journal} {\bibinfo  {journal} {Nature}\ }\textbf {\bibinfo
  {volume} {582}},\ \bibinfo {pages} {203--208} (\bibinfo {year}
  {2020})}\BibitemShut {NoStop}%
\bibitem [{\citenamefont {Wong}\ \emph {et~al.}(2020)\citenamefont {Wong},
  \citenamefont {Nuckolls}, \citenamefont {Oh}, \citenamefont {Lian},
  \citenamefont {Xie}, \citenamefont {Jeon}, \citenamefont {Watanabe},
  \citenamefont {Taniguchi}, \citenamefont {Bernevig},\ and\ \citenamefont
  {Yazdani}}]{AY20}%
  \BibitemOpen
  \bibfield  {author} {\bibinfo {author} {\bibfnamefont {Dillon}\ \bibnamefont
  {Wong}}, \bibinfo {author} {\bibfnamefont {Kevin~P.}\ \bibnamefont
  {Nuckolls}}, \bibinfo {author} {\bibfnamefont {Myungchul}\ \bibnamefont
  {Oh}}, \bibinfo {author} {\bibfnamefont {Biao}\ \bibnamefont {Lian}},
  \bibinfo {author} {\bibfnamefont {Yonglong}\ \bibnamefont {Xie}}, \bibinfo
  {author} {\bibfnamefont {Sangjun}\ \bibnamefont {Jeon}}, \bibinfo {author}
  {\bibfnamefont {Kenji}\ \bibnamefont {Watanabe}}, \bibinfo {author}
  {\bibfnamefont {Takashi}\ \bibnamefont {Taniguchi}}, \bibinfo {author}
  {\bibfnamefont {B.~Andrei}\ \bibnamefont {Bernevig}}, \ and\ \bibinfo
  {author} {\bibfnamefont {Ali}\ \bibnamefont {Yazdani}},\ }\bibfield  {title}
  {\enquote {\bibinfo {title} {Cascade of electronic transitions in magic-angle
  twisted bilayer graphene},}\ }\href {\doibase 10.1038/s41586-020-2339-0}
  {\bibfield  {journal} {\bibinfo  {journal} {Nature}\ }\textbf {\bibinfo
  {volume} {582}},\ \bibinfo {pages} {198--202} (\bibinfo {year}
  {2020})}\BibitemShut {NoStop}%
\bibitem [{\citenamefont {{Cao}}\ \emph {et~al.}(2020)\citenamefont {{Cao}},
  \citenamefont {{Rodan-Legrain}}, \citenamefont {{Park}}, \citenamefont {{Noah
  Yuan}}, \citenamefont {{Watanabe}}, \citenamefont {{Taniguchi}},
  \citenamefont {{Fernandes}}, \citenamefont {{Fu}},\ and\ \citenamefont
  {{Jarillo-Herrero}}}]{PJH20}%
  \BibitemOpen
  \bibfield  {author} {\bibinfo {author} {\bibfnamefont {Yuan}\ \bibnamefont
  {{Cao}}}, \bibinfo {author} {\bibfnamefont {Daniel}\ \bibnamefont
  {{Rodan-Legrain}}}, \bibinfo {author} {\bibfnamefont {Jeong~Min}\
  \bibnamefont {{Park}}}, \bibinfo {author} {\bibfnamefont {Fanqi}\
  \bibnamefont {{Noah Yuan}}}, \bibinfo {author} {\bibfnamefont {Kenji}\
  \bibnamefont {{Watanabe}}}, \bibinfo {author} {\bibfnamefont {Takashi}\
  \bibnamefont {{Taniguchi}}}, \bibinfo {author} {\bibfnamefont {Rafael~M.}\
  \bibnamefont {{Fernandes}}}, \bibinfo {author} {\bibfnamefont {Liang}\
  \bibnamefont {{Fu}}}, \ and\ \bibinfo {author} {\bibfnamefont {Pablo}\
  \bibnamefont {{Jarillo-Herrero}}},\ }\bibfield  {title} {\enquote {\bibinfo
  {title} {{Nematicity and Competing Orders in Superconducting Magic-Angle
  Graphene}},}\ }\href@noop {} {\bibfield  {journal} {\bibinfo  {journal}
  {arXiv e-prints}\ ,\ \bibinfo {eid} {arXiv:2004.04148}} (\bibinfo {year}
  {2020})},\ \Eprint {http://arxiv.org/abs/2004.04148} {arXiv:2004.04148
  [cond-mat.mes-hall]} \BibitemShut {NoStop}%
\bibitem [{\citenamefont {Cao}\ \emph {et~al.}(2016)\citenamefont {Cao},
  \citenamefont {Luo}, \citenamefont {Fatemi}, \citenamefont {Fang},
  \citenamefont {Sanchez-Yamagishi}, \citenamefont {Watanabe}, \citenamefont
  {Taniguchi}, \citenamefont {Kaxiras},\ and\ \citenamefont
  {Jarillo-Herrero}}]{Cao2016}%
  \BibitemOpen
  \bibfield  {author} {\bibinfo {author} {\bibfnamefont {Y.}~\bibnamefont
  {Cao}}, \bibinfo {author} {\bibfnamefont {J.~Y.}\ \bibnamefont {Luo}},
  \bibinfo {author} {\bibfnamefont {V.}~\bibnamefont {Fatemi}}, \bibinfo
  {author} {\bibfnamefont {S.}~\bibnamefont {Fang}}, \bibinfo {author}
  {\bibfnamefont {J.~D.}\ \bibnamefont {Sanchez-Yamagishi}}, \bibinfo {author}
  {\bibfnamefont {K.}~\bibnamefont {Watanabe}}, \bibinfo {author}
  {\bibfnamefont {T.}~\bibnamefont {Taniguchi}}, \bibinfo {author}
  {\bibfnamefont {E.}~\bibnamefont {Kaxiras}}, \ and\ \bibinfo {author}
  {\bibfnamefont {P.}~\bibnamefont {Jarillo-Herrero}},\ }\bibfield  {title}
  {\enquote {\bibinfo {title} {Superlattice-induced insulating states and
  valley-protected orbits in twisted bilayer graphene},}\ }\href {\doibase
  10.1103/PhysRevLett.117.116804} {\bibfield  {journal} {\bibinfo  {journal}
  {Phys. Rev. Lett.}\ }\textbf {\bibinfo {volume} {117}},\ \bibinfo {pages}
  {116804} (\bibinfo {year} {2016})}\BibitemShut {NoStop}%
\bibitem [{\citenamefont {Keimer}\ \emph {et~al.}(2015)\citenamefont {Keimer},
  \citenamefont {Kivelson}, \citenamefont {Norman}, \citenamefont {Uchida},\
  and\ \citenamefont {Zaanen}}]{Keimer15}%
  \BibitemOpen
  \bibfield  {author} {\bibinfo {author} {\bibfnamefont {B.}~\bibnamefont
  {Keimer}}, \bibinfo {author} {\bibfnamefont {S.~A.}\ \bibnamefont
  {Kivelson}}, \bibinfo {author} {\bibfnamefont {M.~R.}\ \bibnamefont
  {Norman}}, \bibinfo {author} {\bibfnamefont {S.}~\bibnamefont {Uchida}}, \
  and\ \bibinfo {author} {\bibfnamefont {J.}~\bibnamefont {Zaanen}},\
  }\bibfield  {title} {\enquote {\bibinfo {title} {From quantum matter to
  high-temperature superconductivity in copper oxides},}\ }\href
  {http://dx.doi.org/10.1038/nature14165} {\bibfield  {journal} {\bibinfo
  {journal} {Nature}\ }\textbf {\bibinfo {volume} {518}},\ \bibinfo {pages}
  {179--186} (\bibinfo {year} {2015})}\BibitemShut {NoStop}%
\bibitem [{\citenamefont {Po}\ \emph {et~al.}(2018)\citenamefont {Po},
  \citenamefont {Zou}, \citenamefont {Vishwanath},\ and\ \citenamefont
  {Senthil}}]{TS1}%
  \BibitemOpen
  \bibfield  {author} {\bibinfo {author} {\bibfnamefont {Hoi~Chun}\
  \bibnamefont {Po}}, \bibinfo {author} {\bibfnamefont {Liujun}\ \bibnamefont
  {Zou}}, \bibinfo {author} {\bibfnamefont {Ashvin}\ \bibnamefont
  {Vishwanath}}, \ and\ \bibinfo {author} {\bibfnamefont {T.}~\bibnamefont
  {Senthil}},\ }\bibfield  {title} {\enquote {\bibinfo {title} {Origin of mott
  insulating behavior and superconductivity in twisted bilayer graphene},}\
  }\href {\doibase 10.1103/PhysRevX.8.031089} {\bibfield  {journal} {\bibinfo
  {journal} {Phys. Rev. X}\ }\textbf {\bibinfo {volume} {8}},\ \bibinfo {pages}
  {031089} (\bibinfo {year} {2018})}\BibitemShut {NoStop}%
\bibitem [{\citenamefont {Zou}\ \emph {et~al.}(2018)\citenamefont {Zou},
  \citenamefont {Po}, \citenamefont {Vishwanath},\ and\ \citenamefont
  {Senthil}}]{TS2}%
  \BibitemOpen
  \bibfield  {author} {\bibinfo {author} {\bibfnamefont {Liujun}\ \bibnamefont
  {Zou}}, \bibinfo {author} {\bibfnamefont {Hoi~Chun}\ \bibnamefont {Po}},
  \bibinfo {author} {\bibfnamefont {Ashvin}\ \bibnamefont {Vishwanath}}, \ and\
  \bibinfo {author} {\bibfnamefont {T.}~\bibnamefont {Senthil}},\ }\bibfield
  {title} {\enquote {\bibinfo {title} {Band structure of twisted bilayer
  graphene: Emergent symmetries, commensurate approximants, and wannier
  obstructions},}\ }\href {\doibase 10.1103/PhysRevB.98.085435} {\bibfield
  {journal} {\bibinfo  {journal} {Phys. Rev. B}\ }\textbf {\bibinfo {volume}
  {98}},\ \bibinfo {pages} {085435} (\bibinfo {year} {2018})}\BibitemShut
  {NoStop}%
\bibitem [{\citenamefont {Po}\ \emph {et~al.}(2019)\citenamefont {Po},
  \citenamefont {Zou}, \citenamefont {Senthil},\ and\ \citenamefont
  {Vishwanath}}]{TS3}%
  \BibitemOpen
  \bibfield  {author} {\bibinfo {author} {\bibfnamefont {Hoi~Chun}\
  \bibnamefont {Po}}, \bibinfo {author} {\bibfnamefont {Liujun}\ \bibnamefont
  {Zou}}, \bibinfo {author} {\bibfnamefont {T.}~\bibnamefont {Senthil}}, \ and\
  \bibinfo {author} {\bibfnamefont {Ashvin}\ \bibnamefont {Vishwanath}},\
  }\bibfield  {title} {\enquote {\bibinfo {title} {Faithful tight-binding
  models and fragile topology of magic-angle bilayer graphene},}\ }\href
  {\doibase 10.1103/PhysRevB.99.195455} {\bibfield  {journal} {\bibinfo
  {journal} {Phys. Rev. B}\ }\textbf {\bibinfo {volume} {99}},\ \bibinfo
  {pages} {195455} (\bibinfo {year} {2019})}\BibitemShut {NoStop}%
\bibitem [{\citenamefont {Song}\ \emph {et~al.}(2019)\citenamefont {Song},
  \citenamefont {Wang}, \citenamefont {Shi}, \citenamefont {Li}, \citenamefont
  {Fang},\ and\ \citenamefont {Bernevig}}]{AB1}%
  \BibitemOpen
  \bibfield  {author} {\bibinfo {author} {\bibfnamefont {Zhida}\ \bibnamefont
  {Song}}, \bibinfo {author} {\bibfnamefont {Zhijun}\ \bibnamefont {Wang}},
  \bibinfo {author} {\bibfnamefont {Wujun}\ \bibnamefont {Shi}}, \bibinfo
  {author} {\bibfnamefont {Gang}\ \bibnamefont {Li}}, \bibinfo {author}
  {\bibfnamefont {Chen}\ \bibnamefont {Fang}}, \ and\ \bibinfo {author}
  {\bibfnamefont {B.~Andrei}\ \bibnamefont {Bernevig}},\ }\bibfield  {title}
  {\enquote {\bibinfo {title} {All magic angles in twisted bilayer graphene are
  topological},}\ }\href {\doibase 10.1103/PhysRevLett.123.036401} {\bibfield
  {journal} {\bibinfo  {journal} {Phys. Rev. Lett.}\ }\textbf {\bibinfo
  {volume} {123}},\ \bibinfo {pages} {036401} (\bibinfo {year}
  {2019})}\BibitemShut {NoStop}%
\bibitem [{\citenamefont {Ahn}\ \emph {et~al.}(2019)\citenamefont {Ahn},
  \citenamefont {Park},\ and\ \citenamefont {Yang}}]{BJY1}%
  \BibitemOpen
  \bibfield  {author} {\bibinfo {author} {\bibfnamefont {Junyeong}\
  \bibnamefont {Ahn}}, \bibinfo {author} {\bibfnamefont {Sungjoon}\
  \bibnamefont {Park}}, \ and\ \bibinfo {author} {\bibfnamefont {Bohm-Jung}\
  \bibnamefont {Yang}},\ }\bibfield  {title} {\enquote {\bibinfo {title}
  {Failure of nielsen-ninomiya theorem and fragile topology in two-dimensional
  systems with space-time inversion symmetry: Application to twisted bilayer
  graphene at magic angle},}\ }\href {\doibase 10.1103/PhysRevX.9.021013}
  {\bibfield  {journal} {\bibinfo  {journal} {Phys. Rev. X}\ }\textbf {\bibinfo
  {volume} {9}},\ \bibinfo {pages} {021013} (\bibinfo {year}
  {2019})}\BibitemShut {NoStop}%
\bibitem [{\citenamefont {Koshino}\ and\ \citenamefont {Son}(2019)}]{MK19}%
  \BibitemOpen
  \bibfield  {author} {\bibinfo {author} {\bibfnamefont {Mikito}\ \bibnamefont
  {Koshino}}\ and\ \bibinfo {author} {\bibfnamefont {Young-Woo}\ \bibnamefont
  {Son}},\ }\bibfield  {title} {\enquote {\bibinfo {title} {Moir\'e phonons in
  twisted bilayer graphene},}\ }\href {\doibase 10.1103/PhysRevB.100.075416}
  {\bibfield  {journal} {\bibinfo  {journal} {Phys. Rev. B}\ }\textbf {\bibinfo
  {volume} {100}},\ \bibinfo {pages} {075416} (\bibinfo {year}
  {2019})}\BibitemShut {NoStop}%
\bibitem [{\citenamefont {Ochoa}(2019)}]{HO19}%
  \BibitemOpen
  \bibfield  {author} {\bibinfo {author} {\bibfnamefont {H\'ector}\
  \bibnamefont {Ochoa}},\ }\bibfield  {title} {\enquote {\bibinfo {title}
  {Moir\'e-pattern fluctuations and electron-phason coupling in twisted bilayer
  graphene},}\ }\href {\doibase 10.1103/PhysRevB.100.155426} {\bibfield
  {journal} {\bibinfo  {journal} {Phys. Rev. B}\ }\textbf {\bibinfo {volume}
  {100}},\ \bibinfo {pages} {155426} (\bibinfo {year} {2019})}\BibitemShut
  {NoStop}%
\bibitem [{\citenamefont {Lewandowski}\ and\ \citenamefont
  {Levitov}(2019)}]{Lewandowski20869}%
  \BibitemOpen
  \bibfield  {author} {\bibinfo {author} {\bibfnamefont {Cyprian}\ \bibnamefont
  {Lewandowski}}\ and\ \bibinfo {author} {\bibfnamefont {Leonid}\ \bibnamefont
  {Levitov}},\ }\bibfield  {title} {\enquote {\bibinfo {title} {Intrinsically
  undamped plasmon modes in narrow electron bands},}\ }\href {\doibase
  10.1073/pnas.1909069116} {\bibfield  {journal} {\bibinfo  {journal}
  {Proceedings of the National Academy of Sciences}\ }\textbf {\bibinfo
  {volume} {116}},\ \bibinfo {pages} {20869--20874} (\bibinfo {year}
  {2019})}\BibitemShut {NoStop}%
\bibitem [{\citenamefont {Stepanov}\ \emph {et~al.}(2019)\citenamefont
  {Stepanov}, \citenamefont {Das}, \citenamefont {Lu}, \citenamefont
  {Fahimniya}, \citenamefont {Watanabe}, \citenamefont {Taniguchi},
  \citenamefont {Koppens}, \citenamefont {Lischner}, \citenamefont {Levitov},\
  and\ \citenamefont {Efetov}}]{stepanov2019interplay}%
  \BibitemOpen
  \bibfield  {author} {\bibinfo {author} {\bibfnamefont {Petr}\ \bibnamefont
  {Stepanov}}, \bibinfo {author} {\bibfnamefont {Ipsita}\ \bibnamefont {Das}},
  \bibinfo {author} {\bibfnamefont {Xiaobo}\ \bibnamefont {Lu}}, \bibinfo
  {author} {\bibfnamefont {Ali}\ \bibnamefont {Fahimniya}}, \bibinfo {author}
  {\bibfnamefont {Kenji}\ \bibnamefont {Watanabe}}, \bibinfo {author}
  {\bibfnamefont {Takashi}\ \bibnamefont {Taniguchi}}, \bibinfo {author}
  {\bibfnamefont {Frank H.~L.}\ \bibnamefont {Koppens}}, \bibinfo {author}
  {\bibfnamefont {Johannes}\ \bibnamefont {Lischner}}, \bibinfo {author}
  {\bibfnamefont {Leonid}\ \bibnamefont {Levitov}}, \ and\ \bibinfo {author}
  {\bibfnamefont {Dmitri~K.}\ \bibnamefont {Efetov}},\ }\href@noop {} {\enquote
  {\bibinfo {title} {The interplay of insulating and superconducting orders in
  magic-angle graphene bilayers},}\ } (\bibinfo {year} {2019}),\ \Eprint
  {http://arxiv.org/abs/1911.09198} {arXiv:1911.09198 [cond-mat.supr-con]}
  \BibitemShut {NoStop}%
\bibitem [{\citenamefont {Saito}\ \emph {et~al.}(2020)\citenamefont {Saito},
  \citenamefont {Ge}, \citenamefont {Watanabe}, \citenamefont {Taniguchi},\
  and\ \citenamefont {Young}}]{saito2019decoupling}%
  \BibitemOpen
  \bibfield  {author} {\bibinfo {author} {\bibfnamefont {Yu}~\bibnamefont
  {Saito}}, \bibinfo {author} {\bibfnamefont {Jingyuan}\ \bibnamefont {Ge}},
  \bibinfo {author} {\bibfnamefont {Kenji}\ \bibnamefont {Watanabe}}, \bibinfo
  {author} {\bibfnamefont {Takashi}\ \bibnamefont {Taniguchi}}, \ and\ \bibinfo
  {author} {\bibfnamefont {Andrea~F.}\ \bibnamefont {Young}},\ }\bibfield
  {title} {\enquote {\bibinfo {title} {Independent superconductors and
  correlated insulators in twisted bilayer graphene},}\ }\href {\doibase
  10.1038/s41567-020-0928-3} {\bibfield  {journal} {\bibinfo  {journal} {Nature
  Physics}\ }\textbf {\bibinfo {volume} {16}},\ \bibinfo {pages} {926--930}
  (\bibinfo {year} {2020})}\BibitemShut {NoStop}%
\bibitem [{\citenamefont {Senthil}(2020)}]{senthil2020jccm}%
  \BibitemOpen
  \bibfield  {author} {\bibinfo {author} {\bibfnamefont {T.}~\bibnamefont
  {Senthil}},\ }\bibfield  {title} {\enquote {\bibinfo {title} {What drives
  superconductivity in twisted bilayer graphene?}}\ }\href {\doibase
  10.36471/jccm{\_}may{\_}2020{\_}03} {\bibfield  {journal} {\bibinfo
  {journal} {Journal Club for Condensed Matter Physics}\ } (\bibinfo {year}
  {2020}),\ 10.36471/jccm{\_}may{\_}2020{\_}03}\BibitemShut {NoStop}%
\bibitem [{\citenamefont {Liu}\ \emph {et~al.}(2021)\citenamefont {Liu},
  \citenamefont {Wang}, \citenamefont {Watanabe}, \citenamefont {Taniguchi},
  \citenamefont {Vafek},\ and\ \citenamefont {Li}}]{liu2020tuning}%
  \BibitemOpen
  \bibfield  {author} {\bibinfo {author} {\bibfnamefont {Xiaoxue}\ \bibnamefont
  {Liu}}, \bibinfo {author} {\bibfnamefont {Zhi}\ \bibnamefont {Wang}},
  \bibinfo {author} {\bibfnamefont {K.}~\bibnamefont {Watanabe}}, \bibinfo
  {author} {\bibfnamefont {T.}~\bibnamefont {Taniguchi}}, \bibinfo {author}
  {\bibfnamefont {Oskar}\ \bibnamefont {Vafek}}, \ and\ \bibinfo {author}
  {\bibfnamefont {J.~I.~A.}\ \bibnamefont {Li}},\ }\bibfield  {title} {\enquote
  {\bibinfo {title} {Tuning electron correlation in magic-angle twisted bilayer
  graphene using coulomb screening},}\ }\href {\doibase
  10.1126/science.abb8754} {\bibfield  {journal} {\bibinfo  {journal}
  {Science}\ }\textbf {\bibinfo {volume} {371}},\ \bibinfo {pages} {1261--1265}
  (\bibinfo {year} {2021})},\ \Eprint
  {http://arxiv.org/abs/https://science.sciencemag.org/content/371/6535/1261.full.pdf}
  {https://science.sciencemag.org/content/371/6535/1261.full.pdf} \BibitemShut
  {NoStop}%
\bibitem [{\citenamefont {Arora}\ \emph {et~al.}(2020)\citenamefont {Arora},
  \citenamefont {Polski}, \citenamefont {Zhang}, \citenamefont {Thomson},
  \citenamefont {Choi}, \citenamefont {Kim}, \citenamefont {Lin}, \citenamefont
  {Wilson}, \citenamefont {Xu}, \citenamefont {Chu}, \citenamefont {Watanabe},
  \citenamefont {Taniguchi}, \citenamefont {Alicea},\ and\ \citenamefont
  {Nadj-Perge}}]{10.1038/s41586-020-2473-8}%
  \BibitemOpen
  \bibfield  {author} {\bibinfo {author} {\bibfnamefont {Harpreet~Singh}\
  \bibnamefont {Arora}}, \bibinfo {author} {\bibfnamefont {Robert}\
  \bibnamefont {Polski}}, \bibinfo {author} {\bibfnamefont {Yiran}\
  \bibnamefont {Zhang}}, \bibinfo {author} {\bibfnamefont {Alex}\ \bibnamefont
  {Thomson}}, \bibinfo {author} {\bibfnamefont {Youngjoon}\ \bibnamefont
  {Choi}}, \bibinfo {author} {\bibfnamefont {Hyunjin}\ \bibnamefont {Kim}},
  \bibinfo {author} {\bibfnamefont {Zhong}\ \bibnamefont {Lin}}, \bibinfo
  {author} {\bibfnamefont {Ilham~Zaky}\ \bibnamefont {Wilson}}, \bibinfo
  {author} {\bibfnamefont {Xiaodong}\ \bibnamefont {Xu}}, \bibinfo {author}
  {\bibfnamefont {Jiun-Haw}\ \bibnamefont {Chu}}, \bibinfo {author}
  {\bibfnamefont {Kenji}\ \bibnamefont {Watanabe}}, \bibinfo {author}
  {\bibfnamefont {Takashi}\ \bibnamefont {Taniguchi}}, \bibinfo {author}
  {\bibfnamefont {Jason}\ \bibnamefont {Alicea}}, \ and\ \bibinfo {author}
  {\bibfnamefont {Stevan}\ \bibnamefont {Nadj-Perge}},\ }\bibfield  {title}
  {\enquote {\bibinfo {title} {Superconductivity in metallic twisted bilayer
  graphene stabilized by wse2},}\ }\href {\doibase 10.1038/s41586-020-2473-8}
  {\bibfield  {journal} {\bibinfo  {journal} {Nature}\ }\textbf {\bibinfo
  {volume} {583}},\ \bibinfo {pages} {379--384} (\bibinfo {year}
  {2020})}\BibitemShut {NoStop}%
\bibitem [{\citenamefont {Isobe}\ \emph {et~al.}(2018)\citenamefont {Isobe},
  \citenamefont {Yuan},\ and\ \citenamefont {Fu}}]{Fu18}%
  \BibitemOpen
  \bibfield  {author} {\bibinfo {author} {\bibfnamefont {Hiroki}\ \bibnamefont
  {Isobe}}, \bibinfo {author} {\bibfnamefont {Noah F.~Q.}\ \bibnamefont
  {Yuan}}, \ and\ \bibinfo {author} {\bibfnamefont {Liang}\ \bibnamefont
  {Fu}},\ }\bibfield  {title} {\enquote {\bibinfo {title} {Unconventional
  superconductivity and density waves in twisted bilayer graphene},}\ }\href
  {\doibase 10.1103/PhysRevX.8.041041} {\bibfield  {journal} {\bibinfo
  {journal} {Phys. Rev. X}\ }\textbf {\bibinfo {volume} {8}},\ \bibinfo {pages}
  {041041} (\bibinfo {year} {2018})}\BibitemShut {NoStop}%
\bibitem [{\citenamefont {Peltonen}\ \emph {et~al.}(2018)\citenamefont
  {Peltonen}, \citenamefont {Ojaj\"arvi},\ and\ \citenamefont
  {Heikkil\"a}}]{TH18a}%
  \BibitemOpen
  \bibfield  {author} {\bibinfo {author} {\bibfnamefont {Teemu~J.}\
  \bibnamefont {Peltonen}}, \bibinfo {author} {\bibfnamefont {Risto}\
  \bibnamefont {Ojaj\"arvi}}, \ and\ \bibinfo {author} {\bibfnamefont
  {Tero~T.}\ \bibnamefont {Heikkil\"a}},\ }\bibfield  {title} {\enquote
  {\bibinfo {title} {Mean-field theory for superconductivity in twisted bilayer
  graphene},}\ }\href {\doibase 10.1103/PhysRevB.98.220504} {\bibfield
  {journal} {\bibinfo  {journal} {Phys. Rev. B}\ }\textbf {\bibinfo {volume}
  {98}},\ \bibinfo {pages} {220504} (\bibinfo {year} {2018})}\BibitemShut
  {NoStop}%
\bibitem [{\citenamefont {Wu}\ \emph {et~al.}(2018)\citenamefont {Wu},
  \citenamefont {MacDonald},\ and\ \citenamefont {Martin}}]{Wu18}%
  \BibitemOpen
  \bibfield  {author} {\bibinfo {author} {\bibfnamefont {Fengcheng}\
  \bibnamefont {Wu}}, \bibinfo {author} {\bibfnamefont {A.~H.}\ \bibnamefont
  {MacDonald}}, \ and\ \bibinfo {author} {\bibfnamefont {Ivar}\ \bibnamefont
  {Martin}},\ }\bibfield  {title} {\enquote {\bibinfo {title} {Theory of
  phonon-mediated superconductivity in twisted bilayer graphene},}\ }\href
  {\doibase 10.1103/PhysRevLett.121.257001} {\bibfield  {journal} {\bibinfo
  {journal} {Phys. Rev. Lett.}\ }\textbf {\bibinfo {volume} {121}},\ \bibinfo
  {pages} {257001} (\bibinfo {year} {2018})}\BibitemShut {NoStop}%
\bibitem [{\citenamefont {Kennes}\ \emph {et~al.}(2018)\citenamefont {Kennes},
  \citenamefont {Lischner},\ and\ \citenamefont {Karrasch}}]{CK18}%
  \BibitemOpen
  \bibfield  {author} {\bibinfo {author} {\bibfnamefont {Dante~M.}\
  \bibnamefont {Kennes}}, \bibinfo {author} {\bibfnamefont {Johannes}\
  \bibnamefont {Lischner}}, \ and\ \bibinfo {author} {\bibfnamefont
  {Christoph}\ \bibnamefont {Karrasch}},\ }\bibfield  {title} {\enquote
  {\bibinfo {title} {Strong correlations and $d+\mathit{id}$ superconductivity
  in twisted bilayer graphene},}\ }\href {\doibase 10.1103/PhysRevB.98.241407}
  {\bibfield  {journal} {\bibinfo  {journal} {Phys. Rev. B}\ }\textbf {\bibinfo
  {volume} {98}},\ \bibinfo {pages} {241407} (\bibinfo {year}
  {2018})}\BibitemShut {NoStop}%
\bibitem [{\citenamefont {Choi}\ and\ \citenamefont {Choi}(2018)}]{HJC18}%
  \BibitemOpen
  \bibfield  {author} {\bibinfo {author} {\bibfnamefont {Young~Woo}\
  \bibnamefont {Choi}}\ and\ \bibinfo {author} {\bibfnamefont {Hyoung~Joon}\
  \bibnamefont {Choi}},\ }\bibfield  {title} {\enquote {\bibinfo {title}
  {Strong electron-phonon coupling, electron-hole asymmetry, and
  nonadiabaticity in magic-angle twisted bilayer graphene},}\ }\href {\doibase
  10.1103/PhysRevB.98.241412} {\bibfield  {journal} {\bibinfo  {journal} {Phys.
  Rev. B}\ }\textbf {\bibinfo {volume} {98}},\ \bibinfo {pages} {241412}
  (\bibinfo {year} {2018})}\BibitemShut {NoStop}%
\bibitem [{\citenamefont {You}\ and\ \citenamefont
  {Vishwanath}(2019)}]{You2019}%
  \BibitemOpen
  \bibfield  {author} {\bibinfo {author} {\bibfnamefont {Yi-Zhuang}\
  \bibnamefont {You}}\ and\ \bibinfo {author} {\bibfnamefont {Ashvin}\
  \bibnamefont {Vishwanath}},\ }\bibfield  {title} {\enquote {\bibinfo {title}
  {Superconductivity from valley fluctuations and approximate so(4) symmetry in
  a weak coupling theory of twisted bilayer graphene},}\ }\href {\doibase
  10.1038/s41535-019-0153-4} {\bibfield  {journal} {\bibinfo  {journal} {npj
  Quantum Materials}\ }\textbf {\bibinfo {volume} {4}},\ \bibinfo {pages} {16}
  (\bibinfo {year} {2019})}\BibitemShut {NoStop}%
\bibitem [{\citenamefont {Lian}\ \emph {et~al.}(2019)\citenamefont {Lian},
  \citenamefont {Wang},\ and\ \citenamefont {Bernevig}}]{BAB19}%
  \BibitemOpen
  \bibfield  {author} {\bibinfo {author} {\bibfnamefont {Biao}\ \bibnamefont
  {Lian}}, \bibinfo {author} {\bibfnamefont {Zhijun}\ \bibnamefont {Wang}}, \
  and\ \bibinfo {author} {\bibfnamefont {B.~Andrei}\ \bibnamefont {Bernevig}},\
  }\bibfield  {title} {\enquote {\bibinfo {title} {Twisted bilayer graphene: A
  phonon-driven superconductor},}\ }\href {\doibase
  10.1103/PhysRevLett.122.257002} {\bibfield  {journal} {\bibinfo  {journal}
  {Phys. Rev. Lett.}\ }\textbf {\bibinfo {volume} {122}},\ \bibinfo {pages}
  {257002} (\bibinfo {year} {2019})}\BibitemShut {NoStop}%
\bibitem [{\citenamefont {Lin}\ and\ \citenamefont
  {Nandkishore}(2019)}]{RN19a}%
  \BibitemOpen
  \bibfield  {author} {\bibinfo {author} {\bibfnamefont {Yu-Ping}\ \bibnamefont
  {Lin}}\ and\ \bibinfo {author} {\bibfnamefont {Rahul~M.}\ \bibnamefont
  {Nandkishore}},\ }\bibfield  {title} {\enquote {\bibinfo {title} {Chiral
  twist on the high-${T}_{c}$ phase diagram in moir\'e heterostructures},}\
  }\href {\doibase 10.1103/PhysRevB.100.085136} {\bibfield  {journal} {\bibinfo
   {journal} {Phys. Rev. B}\ }\textbf {\bibinfo {volume} {100}},\ \bibinfo
  {pages} {085136} (\bibinfo {year} {2019})}\BibitemShut {NoStop}%
\bibitem [{\citenamefont {Samajdar}\ and\ \citenamefont
  {Scheurer}(2020)}]{PhysRevB.102.064501}%
  \BibitemOpen
  \bibfield  {author} {\bibinfo {author} {\bibfnamefont {Rhine}\ \bibnamefont
  {Samajdar}}\ and\ \bibinfo {author} {\bibfnamefont {Mathias~S.}\ \bibnamefont
  {Scheurer}},\ }\bibfield  {title} {\enquote {\bibinfo {title} {Microscopic
  pairing mechanism, order parameter, and disorder sensitivity in moir\'e
  superlattices: Applications to twisted double-bilayer graphene},}\ }\href
  {\doibase 10.1103/PhysRevB.102.064501} {\bibfield  {journal} {\bibinfo
  {journal} {Phys. Rev. B}\ }\textbf {\bibinfo {volume} {102}},\ \bibinfo
  {pages} {064501} (\bibinfo {year} {2020})}\BibitemShut {NoStop}%
\bibitem [{\citenamefont {Ojaj\"arvi}\ \emph {et~al.}(2018)\citenamefont
  {Ojaj\"arvi}, \citenamefont {Hyart}, \citenamefont {Silaev},\ and\
  \citenamefont {Heikkil\"a}}]{TH18b}%
  \BibitemOpen
  \bibfield  {author} {\bibinfo {author} {\bibfnamefont {Risto}\ \bibnamefont
  {Ojaj\"arvi}}, \bibinfo {author} {\bibfnamefont {Timo}\ \bibnamefont
  {Hyart}}, \bibinfo {author} {\bibfnamefont {Mihail~A.}\ \bibnamefont
  {Silaev}}, \ and\ \bibinfo {author} {\bibfnamefont {Tero~T.}\ \bibnamefont
  {Heikkil\"a}},\ }\bibfield  {title} {\enquote {\bibinfo {title} {Competition
  of electron-phonon mediated superconductivity and stoner magnetism on a flat
  band},}\ }\href {\doibase 10.1103/PhysRevB.98.054515} {\bibfield  {journal}
  {\bibinfo  {journal} {Phys. Rev. B}\ }\textbf {\bibinfo {volume} {98}},\
  \bibinfo {pages} {054515} (\bibinfo {year} {2018})}\BibitemShut {NoStop}%
\bibitem [{\citenamefont {Sharma}\ \emph {et~al.}(2020)\citenamefont {Sharma},
  \citenamefont {Trushin}, \citenamefont {Sushkov}, \citenamefont {Vignale},\
  and\ \citenamefont {Adam}}]{SA19}%
  \BibitemOpen
  \bibfield  {author} {\bibinfo {author} {\bibfnamefont {Girish}\ \bibnamefont
  {Sharma}}, \bibinfo {author} {\bibfnamefont {Maxim}\ \bibnamefont {Trushin}},
  \bibinfo {author} {\bibfnamefont {Oleg~P.}\ \bibnamefont {Sushkov}}, \bibinfo
  {author} {\bibfnamefont {Giovanni}\ \bibnamefont {Vignale}}, \ and\ \bibinfo
  {author} {\bibfnamefont {Shaffique}\ \bibnamefont {Adam}},\ }\bibfield
  {title} {\enquote {\bibinfo {title} {Superconductivity from collective
  excitations in magic-angle twisted bilayer graphene},}\ }\href {\doibase
  10.1103/PhysRevResearch.2.022040} {\bibfield  {journal} {\bibinfo  {journal}
  {Phys. Rev. Research}\ }\textbf {\bibinfo {volume} {2}},\ \bibinfo {pages}
  {022040} (\bibinfo {year} {2020})}\BibitemShut {NoStop}%
\bibitem [{\citenamefont {Schrodi}\ \emph {et~al.}(2020)\citenamefont
  {Schrodi}, \citenamefont {Aperis},\ and\ \citenamefont {Oppeneer}}]{PO20}%
  \BibitemOpen
  \bibfield  {author} {\bibinfo {author} {\bibfnamefont {Fabian}\ \bibnamefont
  {Schrodi}}, \bibinfo {author} {\bibfnamefont {Alex}\ \bibnamefont {Aperis}},
  \ and\ \bibinfo {author} {\bibfnamefont {Peter~M.}\ \bibnamefont
  {Oppeneer}},\ }\bibfield  {title} {\enquote {\bibinfo {title} {Prominent
  cooper pairing away from the fermi level and its spectroscopic signature in
  twisted bilayer graphene},}\ }\href {\doibase
  10.1103/PhysRevResearch.2.012066} {\bibfield  {journal} {\bibinfo  {journal}
  {Phys. Rev. Research}\ }\textbf {\bibinfo {volume} {2}},\ \bibinfo {pages}
  {012066} (\bibinfo {year} {2020})}\BibitemShut {NoStop}%
\bibitem [{\citenamefont {Khalaf}\ \emph {et~al.}(2020)\citenamefont {Khalaf},
  \citenamefont {Chatterjee}, \citenamefont {Bultinck}, \citenamefont
  {Zaletel},\ and\ \citenamefont {Vishwanath}}]{AV20}%
  \BibitemOpen
  \bibfield  {author} {\bibinfo {author} {\bibfnamefont {Eslam}\ \bibnamefont
  {Khalaf}}, \bibinfo {author} {\bibfnamefont {Shubhayu}\ \bibnamefont
  {Chatterjee}}, \bibinfo {author} {\bibfnamefont {Nick}\ \bibnamefont
  {Bultinck}}, \bibinfo {author} {\bibfnamefont {Michael~P.}\ \bibnamefont
  {Zaletel}}, \ and\ \bibinfo {author} {\bibfnamefont {Ashvin}\ \bibnamefont
  {Vishwanath}},\ }\href@noop {} {\enquote {\bibinfo {title} {Charged skyrmions
  and topological origin of superconductivity in magic angle graphene},}\ }
  (\bibinfo {year} {2020}),\ \Eprint {http://arxiv.org/abs/2004.00638}
  {arXiv:2004.00638 [cond-mat.str-el]} \BibitemShut {NoStop}%
\bibitem [{\citenamefont {Christos}\ \emph {et~al.}(2020)\citenamefont
  {Christos}, \citenamefont {Sachdev},\ and\ \citenamefont {Scheurer}}]{SS20}%
  \BibitemOpen
  \bibfield  {author} {\bibinfo {author} {\bibfnamefont {Maine}\ \bibnamefont
  {Christos}}, \bibinfo {author} {\bibfnamefont {Subir}\ \bibnamefont
  {Sachdev}}, \ and\ \bibinfo {author} {\bibfnamefont {Mathias}\ \bibnamefont
  {Scheurer}},\ }\href@noop {} {\enquote {\bibinfo {title} {Superconductivity,
  correlated insulators, and wess-zumino-witten terms in twisted bilayer
  graphene},}\ } (\bibinfo {year} {2020}),\ \Eprint
  {http://arxiv.org/abs/2007.00007} {arXiv:2007.00007 [cond-mat.str-el]}
  \BibitemShut {NoStop}%
\bibitem [{\citenamefont {Scalapino}(1969)}]{DJSeliash}%
  \BibitemOpen
  \bibfield  {author} {\bibinfo {author} {\bibfnamefont {Douglas~J}\
  \bibnamefont {Scalapino}},\ }\bibfield  {title} {\enquote {\bibinfo {title}
  {The electron-phonon interaction and strong-coupling superconductors},}\ }in\
  \href@noop {} {\emph {\bibinfo {booktitle} {Superconductivity, edited by R.D.
  Parks}}}\ (\bibinfo  {publisher} {Marcel Dekker, Inc., New York},\ \bibinfo
  {year} {1969})\ pp.\ \bibinfo {pages} {449--560}\BibitemShut {NoStop}%
\bibitem [{\citenamefont {Chou}\ \emph {et~al.}(2019)\citenamefont {Chou},
  \citenamefont {Lin}, \citenamefont {Das~Sarma},\ and\ \citenamefont
  {Nandkishore}}]{SDSRN19}%
  \BibitemOpen
  \bibfield  {author} {\bibinfo {author} {\bibfnamefont {Yang-Zhi}\
  \bibnamefont {Chou}}, \bibinfo {author} {\bibfnamefont {Yu-Ping}\
  \bibnamefont {Lin}}, \bibinfo {author} {\bibfnamefont {Sankar}\ \bibnamefont
  {Das~Sarma}}, \ and\ \bibinfo {author} {\bibfnamefont {Rahul~M.}\
  \bibnamefont {Nandkishore}},\ }\bibfield  {title} {\enquote {\bibinfo {title}
  {Superconductor versus insulator in twisted bilayer graphene},}\ }\href
  {\doibase 10.1103/PhysRevB.100.115128} {\bibfield  {journal} {\bibinfo
  {journal} {Phys. Rev. B}\ }\textbf {\bibinfo {volume} {100}},\ \bibinfo
  {pages} {115128} (\bibinfo {year} {2019})}\BibitemShut {NoStop}%
\bibitem [{\citenamefont {Wu}\ \emph {et~al.}(2019)\citenamefont {Wu},
  \citenamefont {Hwang},\ and\ \citenamefont {Das~Sarma}}]{PhysRevB.99.165112}%
  \BibitemOpen
  \bibfield  {author} {\bibinfo {author} {\bibfnamefont {Fengcheng}\
  \bibnamefont {Wu}}, \bibinfo {author} {\bibfnamefont {Euyheon}\ \bibnamefont
  {Hwang}}, \ and\ \bibinfo {author} {\bibfnamefont {Sankar}\ \bibnamefont
  {Das~Sarma}},\ }\bibfield  {title} {\enquote {\bibinfo {title}
  {Phonon-induced giant linear-in-$t$ resistivity in magic angle twisted
  bilayer graphene: Ordinary strangeness and exotic superconductivity},}\
  }\href {\doibase 10.1103/PhysRevB.99.165112} {\bibfield  {journal} {\bibinfo
  {journal} {Phys. Rev. B}\ }\textbf {\bibinfo {volume} {99}},\ \bibinfo
  {pages} {165112} (\bibinfo {year} {2019})}\BibitemShut {NoStop}%
\bibitem [{\citenamefont {Schrieffer}\ \emph {et~al.}(1963)\citenamefont
  {Schrieffer}, \citenamefont {Scalapino},\ and\ \citenamefont
  {Wilkins}}]{Schrieffer1963}%
  \BibitemOpen
  \bibfield  {author} {\bibinfo {author} {\bibfnamefont {J.~R.}\ \bibnamefont
  {Schrieffer}}, \bibinfo {author} {\bibfnamefont {D.~J.}\ \bibnamefont
  {Scalapino}}, \ and\ \bibinfo {author} {\bibfnamefont {J.~W.}\ \bibnamefont
  {Wilkins}},\ }\bibfield  {title} {\enquote {\bibinfo {title} {Effective
  tunneling density of states in superconductors},}\ }\href {\doibase
  10.1103/PhysRevLett.10.336} {\bibfield  {journal} {\bibinfo  {journal} {Phys.
  Rev. Lett.}\ }\textbf {\bibinfo {volume} {10}},\ \bibinfo {pages} {336--339}
  (\bibinfo {year} {1963})}\BibitemShut {NoStop}%
\bibitem [{\citenamefont {Koshino}\ \emph {et~al.}(2018)\citenamefont
  {Koshino}, \citenamefont {Yuan}, \citenamefont {Koretsune}, \citenamefont
  {Ochi}, \citenamefont {Kuroki},\ and\ \citenamefont {Fu}}]{MK18}%
  \BibitemOpen
  \bibfield  {author} {\bibinfo {author} {\bibfnamefont {Mikito}\ \bibnamefont
  {Koshino}}, \bibinfo {author} {\bibfnamefont {Noah F.~Q.}\ \bibnamefont
  {Yuan}}, \bibinfo {author} {\bibfnamefont {Takashi}\ \bibnamefont
  {Koretsune}}, \bibinfo {author} {\bibfnamefont {Masayuki}\ \bibnamefont
  {Ochi}}, \bibinfo {author} {\bibfnamefont {Kazuhiko}\ \bibnamefont {Kuroki}},
  \ and\ \bibinfo {author} {\bibfnamefont {Liang}\ \bibnamefont {Fu}},\
  }\bibfield  {title} {\enquote {\bibinfo {title} {Maximally localized wannier
  orbitals and the extended hubbard model for twisted bilayer graphene},}\
  }\href {\doibase 10.1103/PhysRevX.8.031087} {\bibfield  {journal} {\bibinfo
  {journal} {Phys. Rev. X}\ }\textbf {\bibinfo {volume} {8}},\ \bibinfo {pages}
  {031087} (\bibinfo {year} {2018})}\BibitemShut {NoStop}%
\bibitem [{\citenamefont {Efetov}\ and\ \citenamefont
  {Kim}(2010)}]{PhysRevLett.105.256805}%
  \BibitemOpen
  \bibfield  {author} {\bibinfo {author} {\bibfnamefont {Dmitri~K.}\
  \bibnamefont {Efetov}}\ and\ \bibinfo {author} {\bibfnamefont {Philip}\
  \bibnamefont {Kim}},\ }\bibfield  {title} {\enquote {\bibinfo {title}
  {Controlling electron-phonon interactions in graphene at ultrahigh carrier
  densities},}\ }\href {\doibase 10.1103/PhysRevLett.105.256805} {\bibfield
  {journal} {\bibinfo  {journal} {Phys. Rev. Lett.}\ }\textbf {\bibinfo
  {volume} {105}},\ \bibinfo {pages} {256805} (\bibinfo {year}
  {2010})}\BibitemShut {NoStop}%
\bibitem [{\citenamefont {Chen}\ \emph {et~al.}(2008)\citenamefont {Chen},
  \citenamefont {Jang}, \citenamefont {Xiao}, \citenamefont {Ishigami},\ and\
  \citenamefont {Fuhrer}}]{Chen2008}%
  \BibitemOpen
  \bibfield  {author} {\bibinfo {author} {\bibfnamefont {Jian-Hao}\
  \bibnamefont {Chen}}, \bibinfo {author} {\bibfnamefont {Chaun}\ \bibnamefont
  {Jang}}, \bibinfo {author} {\bibfnamefont {Shudong}\ \bibnamefont {Xiao}},
  \bibinfo {author} {\bibfnamefont {Masa}\ \bibnamefont {Ishigami}}, \ and\
  \bibinfo {author} {\bibfnamefont {Michael~S.}\ \bibnamefont {Fuhrer}},\
  }\bibfield  {title} {\enquote {\bibinfo {title} {Intrinsic and extrinsic
  performance limits of graphene devices on sio2},}\ }\href {\doibase
  10.1038/nnano.2008.58} {\bibfield  {journal} {\bibinfo  {journal} {Nature
  Nanotechnology}\ }\textbf {\bibinfo {volume} {3}},\ \bibinfo {pages}
  {206--209} (\bibinfo {year} {2008})}\BibitemShut {NoStop}%
\bibitem [{\citenamefont {Takada}(1992)}]{takada1992insignificance}%
  \BibitemOpen
  \bibfield  {author} {\bibinfo {author} {\bibfnamefont {Yasutami}\
  \bibnamefont {Takada}},\ }\bibfield  {title} {\enquote {\bibinfo {title}
  {Insignificance of vertex corrections in the plasmon mechanism of
  superconductivity at low electron concentration: Migdal’s theorem in the
  antiadiabatic region},}\ }\href@noop {} {\bibfield  {journal} {\bibinfo
  {journal} {Journal of the Physical Society of Japan}\ }\textbf {\bibinfo
  {volume} {61}},\ \bibinfo {pages} {3849--3852} (\bibinfo {year}
  {1992})}\BibitemShut {NoStop}%
\bibitem [{\citenamefont {Guinea}\ and\ \citenamefont
  {Walet}(2018)}]{Guinea13174}%
  \BibitemOpen
  \bibfield  {author} {\bibinfo {author} {\bibfnamefont {Francisco}\
  \bibnamefont {Guinea}}\ and\ \bibinfo {author} {\bibfnamefont {Niels~R.}\
  \bibnamefont {Walet}},\ }\bibfield  {title} {\enquote {\bibinfo {title}
  {Electrostatic effects, band distortions, and superconductivity in twisted
  graphene bilayers},}\ }\href {\doibase 10.1073/pnas.1810947115} {\bibfield
  {journal} {\bibinfo  {journal} {Proceedings of the National Academy of
  Sciences}\ }\textbf {\bibinfo {volume} {115}},\ \bibinfo {pages}
  {13174--13179} (\bibinfo {year} {2018})}\BibitemShut {NoStop}%
\bibitem [{\citenamefont {Goodwin}\ \emph {et~al.}(2020)\citenamefont
  {Goodwin}, \citenamefont {Vitale}, \citenamefont {Liang}, \citenamefont
  {Mostofi},\ and\ \citenamefont {Lischner}}]{2020arXiv200414784G}%
  \BibitemOpen
  \bibfield  {author} {\bibinfo {author} {\bibfnamefont {Zachary A~H}\
  \bibnamefont {Goodwin}}, \bibinfo {author} {\bibfnamefont {Valerio}\
  \bibnamefont {Vitale}}, \bibinfo {author} {\bibfnamefont {Xia}\ \bibnamefont
  {Liang}}, \bibinfo {author} {\bibfnamefont {Arash~A}\ \bibnamefont
  {Mostofi}}, \ and\ \bibinfo {author} {\bibfnamefont {Johannes}\ \bibnamefont
  {Lischner}},\ }\bibfield  {title} {\enquote {\bibinfo {title} {Hartree theory
  calculations of quasiparticle properties in twisted bilayer graphene},}\
  }\href {\doibase 10.1088/2516-1075/ab9f94} {\bibfield  {journal} {\bibinfo
  {journal} {Electronic Structure}\ }\textbf {\bibinfo {volume} {2}},\ \bibinfo
  {pages} {034001} (\bibinfo {year} {2020})}\BibitemShut {NoStop}%
\bibitem [{\citenamefont {Phan}\ and\ \citenamefont
  {Chubukov}(2020)}]{PhysRevB.101.024503}%
  \BibitemOpen
  \bibfield  {author} {\bibinfo {author} {\bibfnamefont {Dan}\ \bibnamefont
  {Phan}}\ and\ \bibinfo {author} {\bibfnamefont {Andrey~V.}\ \bibnamefont
  {Chubukov}},\ }\bibfield  {title} {\enquote {\bibinfo {title} {Kohn-luttinger
  correction to ${T}_{c}$ in a phonon superconductor},}\ }\href {\doibase
  10.1103/PhysRevB.101.024503} {\bibfield  {journal} {\bibinfo  {journal}
  {Phys. Rev. B}\ }\textbf {\bibinfo {volume} {101}},\ \bibinfo {pages}
  {024503} (\bibinfo {year} {2020})}\BibitemShut {NoStop}%
\bibitem [{\citenamefont {Gastiasoro}\ \emph {et~al.}(2019)\citenamefont
  {Gastiasoro}, \citenamefont {Chubukov},\ and\ \citenamefont
  {Fernandes}}]{PhysRevB.99.094524}%
  \BibitemOpen
  \bibfield  {author} {\bibinfo {author} {\bibfnamefont {Maria~N.}\
  \bibnamefont {Gastiasoro}}, \bibinfo {author} {\bibfnamefont {Andrey~V.}\
  \bibnamefont {Chubukov}}, \ and\ \bibinfo {author} {\bibfnamefont
  {Rafael~M.}\ \bibnamefont {Fernandes}},\ }\bibfield  {title} {\enquote
  {\bibinfo {title} {Phonon-mediated superconductivity in low carrier-density
  systems},}\ }\href {\doibase 10.1103/PhysRevB.99.094524} {\bibfield
  {journal} {\bibinfo  {journal} {Phys. Rev. B}\ }\textbf {\bibinfo {volume}
  {99}},\ \bibinfo {pages} {094524} (\bibinfo {year} {2019})}\BibitemShut
  {NoStop}%
\bibitem [{\citenamefont {Takada}(1978)}]{doi:10.1143/JPSJ.45.786}%
  \BibitemOpen
  \bibfield  {author} {\bibinfo {author} {\bibfnamefont {Yasutami}\
  \bibnamefont {Takada}},\ }\bibfield  {title} {\enquote {\bibinfo {title}
  {Plasmon mechanism of superconductivity in two- and three-dimensional
  electron systems},}\ }\href {\doibase 10.1143/JPSJ.45.786} {\bibfield
  {journal} {\bibinfo  {journal} {Journal of the Physical Society of Japan}\
  }\textbf {\bibinfo {volume} {45}},\ \bibinfo {pages} {786--794} (\bibinfo
  {year} {1978})},\ \Eprint
  {http://arxiv.org/abs/https://doi.org/10.1143/JPSJ.45.786}
  {https://doi.org/10.1143/JPSJ.45.786} \BibitemShut {NoStop}%
\bibitem [{\citenamefont {Kang}\ and\ \citenamefont {Vafek}(2018)}]{OV18}%
  \BibitemOpen
  \bibfield  {author} {\bibinfo {author} {\bibfnamefont {Jian}\ \bibnamefont
  {Kang}}\ and\ \bibinfo {author} {\bibfnamefont {Oskar}\ \bibnamefont
  {Vafek}},\ }\bibfield  {title} {\enquote {\bibinfo {title} {Symmetry,
  maximally localized wannier states, and a low-energy model for twisted
  bilayer graphene narrow bands},}\ }\href {\doibase 10.1103/PhysRevX.8.031088}
  {\bibfield  {journal} {\bibinfo  {journal} {Phys. Rev. X}\ }\textbf {\bibinfo
  {volume} {8}},\ \bibinfo {pages} {031088} (\bibinfo {year}
  {2018})}\BibitemShut {NoStop}%
\bibitem [{\citenamefont {Bardeen}\ \emph {et~al.}(1957)\citenamefont
  {Bardeen}, \citenamefont {Cooper},\ and\ \citenamefont
  {Schrieffer}}]{PhysRev.108.1175}%
  \BibitemOpen
  \bibfield  {author} {\bibinfo {author} {\bibfnamefont {J.}~\bibnamefont
  {Bardeen}}, \bibinfo {author} {\bibfnamefont {L.~N.}\ \bibnamefont {Cooper}},
  \ and\ \bibinfo {author} {\bibfnamefont {J.~R.}\ \bibnamefont {Schrieffer}},\
  }\bibfield  {title} {\enquote {\bibinfo {title} {Theory of
  superconductivity},}\ }\href {\doibase 10.1103/PhysRev.108.1175} {\bibfield
  {journal} {\bibinfo  {journal} {Phys. Rev.}\ }\textbf {\bibinfo {volume}
  {108}},\ \bibinfo {pages} {1175--1204} (\bibinfo {year} {1957})}\BibitemShut
  {NoStop}%
\bibitem [{\citenamefont {McMillan}(1968)}]{PhysRev.167.331}%
  \BibitemOpen
  \bibfield  {author} {\bibinfo {author} {\bibfnamefont {W.~L.}\ \bibnamefont
  {McMillan}},\ }\bibfield  {title} {\enquote {\bibinfo {title} {Transition
  temperature of strong-coupled superconductors},}\ }\href {\doibase
  10.1103/PhysRev.167.331} {\bibfield  {journal} {\bibinfo  {journal} {Phys.
  Rev.}\ }\textbf {\bibinfo {volume} {167}},\ \bibinfo {pages} {331--344}
  (\bibinfo {year} {1968})}\BibitemShut {NoStop}%
\bibitem [{\citenamefont {Gor’kov}\ and\ \citenamefont
  {Melik-Barkhudarov}(1961)}]{gor1961contribution}%
  \BibitemOpen
  \bibfield  {author} {\bibinfo {author} {\bibfnamefont {LP}~\bibnamefont
  {Gor’kov}}\ and\ \bibinfo {author} {\bibfnamefont {TK}~\bibnamefont
  {Melik-Barkhudarov}},\ }\bibfield  {title} {\enquote {\bibinfo {title}
  {Contribution to the theory of superfluidity in an imperfect fermi gas},}\
  }\href@noop {} {\bibfield  {journal} {\bibinfo  {journal} {Sov. Phys. JETP}\
  }\textbf {\bibinfo {volume} {13}},\ \bibinfo {pages} {1018} (\bibinfo {year}
  {1961})}\BibitemShut {NoStop}%
\bibitem [{\citenamefont {Ikeda}\ \emph {et~al.}(1992)\citenamefont {Ikeda},
  \citenamefont {Ogasawara},\ and\ \citenamefont {Sugihara}}]{ikeda1992migdal}%
  \BibitemOpen
  \bibfield  {author} {\bibinfo {author} {\bibfnamefont {MA}~\bibnamefont
  {Ikeda}}, \bibinfo {author} {\bibfnamefont {A}~\bibnamefont {Ogasawara}}, \
  and\ \bibinfo {author} {\bibfnamefont {M}~\bibnamefont {Sugihara}},\
  }\bibfield  {title} {\enquote {\bibinfo {title} {On migdal's theorem},}\
  }\href@noop {} {\bibfield  {journal} {\bibinfo  {journal} {Physics Letters
  A}\ }\textbf {\bibinfo {volume} {170}},\ \bibinfo {pages} {319--324}
  (\bibinfo {year} {1992})}\BibitemShut {NoStop}%
\bibitem [{\citenamefont {Kozii}\ \emph {et~al.}(2019)\citenamefont {Kozii},
  \citenamefont {Bi},\ and\ \citenamefont {Ruhman}}]{Kozii2019}%
  \BibitemOpen
  \bibfield  {author} {\bibinfo {author} {\bibfnamefont {Vladyslav}\
  \bibnamefont {Kozii}}, \bibinfo {author} {\bibfnamefont {Zhen}\ \bibnamefont
  {Bi}}, \ and\ \bibinfo {author} {\bibfnamefont {Jonathan}\ \bibnamefont
  {Ruhman}},\ }\bibfield  {title} {\enquote {\bibinfo {title}
  {Superconductivity near a ferroelectric quantum critical point in
  ultralow-density dirac materials},}\ }\href {\doibase
  10.1103/PhysRevX.9.031046} {\bibfield  {journal} {\bibinfo  {journal} {Phys.
  Rev. X}\ }\textbf {\bibinfo {volume} {9}},\ \bibinfo {pages} {031046}
  (\bibinfo {year} {2019})}\BibitemShut {NoStop}%
\bibitem [{\citenamefont {Ruhman}\ and\ \citenamefont
  {Lee}(2017)}]{ruhman2017pairing}%
  \BibitemOpen
  \bibfield  {author} {\bibinfo {author} {\bibfnamefont {Jonathan}\
  \bibnamefont {Ruhman}}\ and\ \bibinfo {author} {\bibfnamefont {Patrick~A.}\
  \bibnamefont {Lee}},\ }\bibfield  {title} {\enquote {\bibinfo {title}
  {Pairing from dynamically screened coulomb repulsion in bismuth},}\ }\href
  {\doibase 10.1103/PhysRevB.96.235107} {\bibfield  {journal} {\bibinfo
  {journal} {Phys. Rev. B}\ }\textbf {\bibinfo {volume} {96}},\ \bibinfo
  {pages} {235107} (\bibinfo {year} {2017})}\BibitemShut {NoStop}%
\bibitem [{\citenamefont {Coleman}(2015)}]{coleman_2015}%
  \BibitemOpen
  \bibfield  {author} {\bibinfo {author} {\bibfnamefont {Piers}\ \bibnamefont
  {Coleman}},\ }\href {\doibase 10.1017/CBO9781139020916} {\emph {\bibinfo
  {title} {Introduction to Many-Body Physics}}}\ (\bibinfo  {publisher}
  {Cambridge University Press},\ \bibinfo {year} {2015})\BibitemShut {NoStop}%
\bibitem [{\citenamefont {Stern}(1967)}]{PhysRevLett.18.546}%
  \BibitemOpen
  \bibfield  {author} {\bibinfo {author} {\bibfnamefont {Frank}\ \bibnamefont
  {Stern}},\ }\bibfield  {title} {\enquote {\bibinfo {title} {Polarizability of
  a two-dimensional electron gas},}\ }\href {\doibase
  10.1103/PhysRevLett.18.546} {\bibfield  {journal} {\bibinfo  {journal} {Phys.
  Rev. Lett.}\ }\textbf {\bibinfo {volume} {18}},\ \bibinfo {pages} {546--548}
  (\bibinfo {year} {1967})}\BibitemShut {NoStop}%
\bibitem [{\citenamefont {Pizarro}\ \emph {et~al.}(2019)\citenamefont
  {Pizarro}, \citenamefont {R\"osner}, \citenamefont {Thomale}, \citenamefont
  {Valent\'{\i}},\ and\ \citenamefont {Wehling}}]{PhysRevB.100.161102}%
  \BibitemOpen
  \bibfield  {author} {\bibinfo {author} {\bibfnamefont {J.~M.}\ \bibnamefont
  {Pizarro}}, \bibinfo {author} {\bibfnamefont {M.}~\bibnamefont {R\"osner}},
  \bibinfo {author} {\bibfnamefont {R.}~\bibnamefont {Thomale}}, \bibinfo
  {author} {\bibfnamefont {R.}~\bibnamefont {Valent\'{\i}}}, \ and\ \bibinfo
  {author} {\bibfnamefont {T.~O.}\ \bibnamefont {Wehling}},\ }\bibfield
  {title} {\enquote {\bibinfo {title} {Internal screening and dielectric
  engineering in magic-angle twisted bilayer graphene},}\ }\href {\doibase
  10.1103/PhysRevB.100.161102} {\bibfield  {journal} {\bibinfo  {journal}
  {Phys. Rev. B}\ }\textbf {\bibinfo {volume} {100}},\ \bibinfo {pages}
  {161102} (\bibinfo {year} {2019})}\BibitemShut {NoStop}%
\bibitem [{\citenamefont {Goodwin}\ \emph {et~al.}(2019)\citenamefont
  {Goodwin}, \citenamefont {Corsetti}, \citenamefont {Mostofi},\ and\
  \citenamefont {Lischner}}]{PhysRevB.100.235424}%
  \BibitemOpen
  \bibfield  {author} {\bibinfo {author} {\bibfnamefont {Zachary A.~H.}\
  \bibnamefont {Goodwin}}, \bibinfo {author} {\bibfnamefont {Fabiano}\
  \bibnamefont {Corsetti}}, \bibinfo {author} {\bibfnamefont {Arash~A.}\
  \bibnamefont {Mostofi}}, \ and\ \bibinfo {author} {\bibfnamefont {Johannes}\
  \bibnamefont {Lischner}},\ }\bibfield  {title} {\enquote {\bibinfo {title}
  {Attractive electron-electron interactions from internal screening in
  magic-angle twisted bilayer graphene},}\ }\href {\doibase
  10.1103/PhysRevB.100.235424} {\bibfield  {journal} {\bibinfo  {journal}
  {Phys. Rev. B}\ }\textbf {\bibinfo {volume} {100}},\ \bibinfo {pages}
  {235424} (\bibinfo {year} {2019})}\BibitemShut {NoStop}%
\bibitem [{\citenamefont {Hwang}\ and\ \citenamefont
  {Das~Sarma}(2007)}]{PhysRevB.75.205418}%
  \BibitemOpen
  \bibfield  {author} {\bibinfo {author} {\bibfnamefont {E.~H.}\ \bibnamefont
  {Hwang}}\ and\ \bibinfo {author} {\bibfnamefont {S.}~\bibnamefont
  {Das~Sarma}},\ }\bibfield  {title} {\enquote {\bibinfo {title} {Dielectric
  function, screening, and plasmons in two-dimensional graphene},}\ }\href
  {\doibase 10.1103/PhysRevB.75.205418} {\bibfield  {journal} {\bibinfo
  {journal} {Phys. Rev. B}\ }\textbf {\bibinfo {volume} {75}},\ \bibinfo
  {pages} {205418} (\bibinfo {year} {2007})}\BibitemShut {NoStop}%
\bibitem [{\citenamefont {Wunsch}\ \emph {et~al.}(2006)\citenamefont {Wunsch},
  \citenamefont {Stauber}, \citenamefont {Sols},\ and\ \citenamefont
  {Guinea}}]{Wunsch_2006}%
  \BibitemOpen
  \bibfield  {author} {\bibinfo {author} {\bibfnamefont {B}~\bibnamefont
  {Wunsch}}, \bibinfo {author} {\bibfnamefont {T}~\bibnamefont {Stauber}},
  \bibinfo {author} {\bibfnamefont {F}~\bibnamefont {Sols}}, \ and\ \bibinfo
  {author} {\bibfnamefont {F}~\bibnamefont {Guinea}},\ }\bibfield  {title}
  {\enquote {\bibinfo {title} {Dynamical polarization of graphene at finite
  doping},}\ }\href {\doibase 10.1088/1367-2630/8/12/318} {\bibfield  {journal}
  {\bibinfo  {journal} {New Journal of Physics}\ }\textbf {\bibinfo {volume}
  {8}},\ \bibinfo {pages} {318--318} (\bibinfo {year} {2006})}\BibitemShut
  {NoStop}%
\bibitem [{\citenamefont {Marsiglio}\ and\ \citenamefont
  {Carbotte}(2008)}]{Marsiglio2008}%
  \BibitemOpen
  \bibfield  {author} {\bibinfo {author} {\bibfnamefont {F.}~\bibnamefont
  {Marsiglio}}\ and\ \bibinfo {author} {\bibfnamefont {J.~P.}\ \bibnamefont
  {Carbotte}},\ }\enquote {\bibinfo {title} {Electron-phonon
  superconductivity},}\ in\ \href {\doibase 10.1007/978-3-540-73253-2_3} {\emph
  {\bibinfo {booktitle} {Superconductivity: Conventional and Unconventional
  Superconductors}}},\ \bibinfo {editor} {edited by\ \bibinfo {editor}
  {\bibfnamefont {K.~H.}\ \bibnamefont {Bennemann}}\ and\ \bibinfo {editor}
  {\bibfnamefont {John~B.}\ \bibnamefont {Ketterson}}}\ (\bibinfo  {publisher}
  {Springer Berlin Heidelberg},\ \bibinfo {address} {Berlin, Heidelberg},\
  \bibinfo {year} {2008})\ pp.\ \bibinfo {pages} {73--162}\BibitemShut
  {NoStop}%
\bibitem [{\citenamefont {Ruhman}\ and\ \citenamefont
  {Lee}(2016)}]{PhysRevB.94.224515}%
  \BibitemOpen
  \bibfield  {author} {\bibinfo {author} {\bibfnamefont {Jonathan}\
  \bibnamefont {Ruhman}}\ and\ \bibinfo {author} {\bibfnamefont {Patrick~A.}\
  \bibnamefont {Lee}},\ }\bibfield  {title} {\enquote {\bibinfo {title}
  {Superconductivity at very low density: The case of strontium titanate},}\
  }\href {\doibase 10.1103/PhysRevB.94.224515} {\bibfield  {journal} {\bibinfo
  {journal} {Phys. Rev. B}\ }\textbf {\bibinfo {volume} {94}},\ \bibinfo
  {pages} {224515} (\bibinfo {year} {2016})}\BibitemShut {NoStop}%
\bibitem [{\citenamefont {Swartz}\ \emph {et~al.}(2018)\citenamefont {Swartz},
  \citenamefont {Inoue}, \citenamefont {Merz}, \citenamefont {Hikita},
  \citenamefont {Raghu}, \citenamefont {Devereaux}, \citenamefont {Johnston},\
  and\ \citenamefont {Hwang}}]{Swartz1475}%
  \BibitemOpen
  \bibfield  {author} {\bibinfo {author} {\bibfnamefont {Adrian~G.}\
  \bibnamefont {Swartz}}, \bibinfo {author} {\bibfnamefont {Hisashi}\
  \bibnamefont {Inoue}}, \bibinfo {author} {\bibfnamefont {Tyler~A.}\
  \bibnamefont {Merz}}, \bibinfo {author} {\bibfnamefont {Yasuyuki}\
  \bibnamefont {Hikita}}, \bibinfo {author} {\bibfnamefont {Srinivas}\
  \bibnamefont {Raghu}}, \bibinfo {author} {\bibfnamefont {Thomas~P.}\
  \bibnamefont {Devereaux}}, \bibinfo {author} {\bibfnamefont {Steven}\
  \bibnamefont {Johnston}}, \ and\ \bibinfo {author} {\bibfnamefont
  {Harold~Y.}\ \bibnamefont {Hwang}},\ }\bibfield  {title} {\enquote {\bibinfo
  {title} {Polaronic behavior in a weak-coupling superconductor},}\ }\href
  {\doibase 10.1073/pnas.1713916115} {\bibfield  {journal} {\bibinfo  {journal}
  {Proceedings of the National Academy of Sciences}\ }\textbf {\bibinfo
  {volume} {115}},\ \bibinfo {pages} {1475--1480} (\bibinfo {year} {2018})},\
  \Eprint
  {http://arxiv.org/abs/https://www.pnas.org/content/115/7/1475.full.pdf}
  {https://www.pnas.org/content/115/7/1475.full.pdf} \BibitemShut {NoStop}%
\bibitem [{\citenamefont {Aperis}\ and\ \citenamefont
  {Oppeneer}(2018)}]{PhysRevB.97.060501}%
  \BibitemOpen
  \bibfield  {author} {\bibinfo {author} {\bibfnamefont {Alex}\ \bibnamefont
  {Aperis}}\ and\ \bibinfo {author} {\bibfnamefont {Peter~M.}\ \bibnamefont
  {Oppeneer}},\ }\bibfield  {title} {\enquote {\bibinfo {title} {Multiband
  full-bandwidth anisotropic eliashberg theory of interfacial electron-phonon
  coupling and high-$t_c$ superconductivity in
  $\mathbf{FeSe}/{\mathbf{srtio}}_{\mathbf{3}}$},}\ }\href {\doibase
  10.1103/PhysRevB.97.060501} {\bibfield  {journal} {\bibinfo  {journal} {Phys.
  Rev. B}\ }\textbf {\bibinfo {volume} {97}},\ \bibinfo {pages} {060501}
  (\bibinfo {year} {2018})}\BibitemShut {NoStop}%
\bibitem [{\citenamefont {Rademaker}\ \emph {et~al.}(2016)\citenamefont
  {Rademaker}, \citenamefont {Wang}, \citenamefont {Berlijn},\ and\
  \citenamefont {Johnston}}]{Rademaker_2016}%
  \BibitemOpen
  \bibfield  {author} {\bibinfo {author} {\bibfnamefont {Louk}\ \bibnamefont
  {Rademaker}}, \bibinfo {author} {\bibfnamefont {Yan}\ \bibnamefont {Wang}},
  \bibinfo {author} {\bibfnamefont {Tom}\ \bibnamefont {Berlijn}}, \ and\
  \bibinfo {author} {\bibfnamefont {Steve}\ \bibnamefont {Johnston}},\
  }\bibfield  {title} {\enquote {\bibinfo {title} {Enhanced superconductivity
  due to forward scattering in {FeSe} thin films on {SrTiO}3substrates},}\
  }\href {\doibase 10.1088/1367-2630/18/2/022001} {\bibfield  {journal}
  {\bibinfo  {journal} {New Journal of Physics}\ }\textbf {\bibinfo {volume}
  {18}},\ \bibinfo {pages} {022001} (\bibinfo {year} {2016})}\BibitemShut
  {NoStop}%
\bibitem [{\citenamefont {{Kozii}}\ \emph {et~al.}(2020)\citenamefont
  {{Kozii}}, \citenamefont {{Zaletel}},\ and\ \citenamefont
  {{Bultinck}}}]{VK20}%
  \BibitemOpen
  \bibfield  {author} {\bibinfo {author} {\bibfnamefont {Vladyslav}\
  \bibnamefont {{Kozii}}}, \bibinfo {author} {\bibfnamefont {Michael~P.}\
  \bibnamefont {{Zaletel}}}, \ and\ \bibinfo {author} {\bibfnamefont {Nick}\
  \bibnamefont {{Bultinck}}},\ }\bibfield  {title} {\enquote {\bibinfo {title}
  {{Superconductivity in a doped valley coherent insulator in magic angle
  graphene: Goldstone-mediated pairing and Kohn-Luttinger mechanism}},}\
  }\href@noop {} {\bibfield  {journal} {\bibinfo  {journal} {arXiv e-prints}\
  ,\ \bibinfo {eid} {arXiv:2005.12961}} (\bibinfo {year} {2020})},\ \Eprint
  {http://arxiv.org/abs/2005.12961} {arXiv:2005.12961 [cond-mat.str-el]}
  \BibitemShut {NoStop}%
\bibitem [{\citenamefont {Cao}\ \emph {et~al.}(2020)\citenamefont {Cao},
  \citenamefont {Chowdhury}, \citenamefont {Rodan-Legrain}, \citenamefont
  {Rubies-Bigorda}, \citenamefont {Watanabe}, \citenamefont {Taniguchi},
  \citenamefont {Senthil},\ and\ \citenamefont
  {Jarillo-Herrero}}]{PhysRevLett.124.076801}%
  \BibitemOpen
  \bibfield  {author} {\bibinfo {author} {\bibfnamefont {Yuan}\ \bibnamefont
  {Cao}}, \bibinfo {author} {\bibfnamefont {Debanjan}\ \bibnamefont
  {Chowdhury}}, \bibinfo {author} {\bibfnamefont {Daniel}\ \bibnamefont
  {Rodan-Legrain}}, \bibinfo {author} {\bibfnamefont {Oriol}\ \bibnamefont
  {Rubies-Bigorda}}, \bibinfo {author} {\bibfnamefont {Kenji}\ \bibnamefont
  {Watanabe}}, \bibinfo {author} {\bibfnamefont {Takashi}\ \bibnamefont
  {Taniguchi}}, \bibinfo {author} {\bibfnamefont {T.}~\bibnamefont {Senthil}},
  \ and\ \bibinfo {author} {\bibfnamefont {Pablo}\ \bibnamefont
  {Jarillo-Herrero}},\ }\bibfield  {title} {\enquote {\bibinfo {title} {Strange
  metal in magic-angle graphene with near planckian dissipation},}\ }\href
  {\doibase 10.1103/PhysRevLett.124.076801} {\bibfield  {journal} {\bibinfo
  {journal} {Phys. Rev. Lett.}\ }\textbf {\bibinfo {volume} {124}},\ \bibinfo
  {pages} {076801} (\bibinfo {year} {2020})}\BibitemShut {NoStop}%
\bibitem [{\citenamefont {Polshyn}\ \emph {et~al.}(2019)\citenamefont
  {Polshyn}, \citenamefont {Yankowitz}, \citenamefont {Chen}, \citenamefont
  {Zhang}, \citenamefont {Watanabe}, \citenamefont {Taniguchi}, \citenamefont
  {Dean},\ and\ \citenamefont {Young}}]{Polshyn2019}%
  \BibitemOpen
  \bibfield  {author} {\bibinfo {author} {\bibfnamefont {Hryhoriy}\
  \bibnamefont {Polshyn}}, \bibinfo {author} {\bibfnamefont {Matthew}\
  \bibnamefont {Yankowitz}}, \bibinfo {author} {\bibfnamefont {Shaowen}\
  \bibnamefont {Chen}}, \bibinfo {author} {\bibfnamefont {Yuxuan}\ \bibnamefont
  {Zhang}}, \bibinfo {author} {\bibfnamefont {K.}~\bibnamefont {Watanabe}},
  \bibinfo {author} {\bibfnamefont {T.}~\bibnamefont {Taniguchi}}, \bibinfo
  {author} {\bibfnamefont {Cory~R.}\ \bibnamefont {Dean}}, \ and\ \bibinfo
  {author} {\bibfnamefont {Andrea~F.}\ \bibnamefont {Young}},\ }\bibfield
  {title} {\enquote {\bibinfo {title} {Large linear-in-temperature resistivity
  in twisted bilayer graphene},}\ }\href {\doibase 10.1038/s41567-019-0596-3}
  {\bibfield  {journal} {\bibinfo  {journal} {Nature Physics}\ }\textbf
  {\bibinfo {volume} {15}},\ \bibinfo {pages} {1011--1016} (\bibinfo {year}
  {2019})}\BibitemShut {NoStop}%
\bibitem [{\citenamefont {Uri}\ \emph {et~al.}(2020)\citenamefont {Uri},
  \citenamefont {Grover}, \citenamefont {Cao}, \citenamefont {Crosse},
  \citenamefont {Bagani}, \citenamefont {Rodan-Legrain}, \citenamefont
  {Myasoedov}, \citenamefont {Watanabe}, \citenamefont {Taniguchi},
  \citenamefont {Moon}, \citenamefont {Koshino}, \citenamefont
  {Jarillo-Herrero},\ and\ \citenamefont {Zeldov}}]{Uri2020}%
  \BibitemOpen
  \bibfield  {author} {\bibinfo {author} {\bibfnamefont {A.}~\bibnamefont
  {Uri}}, \bibinfo {author} {\bibfnamefont {S.}~\bibnamefont {Grover}},
  \bibinfo {author} {\bibfnamefont {Y.}~\bibnamefont {Cao}}, \bibinfo {author}
  {\bibfnamefont {J.~A.}\ \bibnamefont {Crosse}}, \bibinfo {author}
  {\bibfnamefont {K.}~\bibnamefont {Bagani}}, \bibinfo {author} {\bibfnamefont
  {D.}~\bibnamefont {Rodan-Legrain}}, \bibinfo {author} {\bibfnamefont
  {Y.}~\bibnamefont {Myasoedov}}, \bibinfo {author} {\bibfnamefont
  {K.}~\bibnamefont {Watanabe}}, \bibinfo {author} {\bibfnamefont
  {T.}~\bibnamefont {Taniguchi}}, \bibinfo {author} {\bibfnamefont
  {P.}~\bibnamefont {Moon}}, \bibinfo {author} {\bibfnamefont {M.}~\bibnamefont
  {Koshino}}, \bibinfo {author} {\bibfnamefont {P.}~\bibnamefont
  {Jarillo-Herrero}}, \ and\ \bibinfo {author} {\bibfnamefont {E.}~\bibnamefont
  {Zeldov}},\ }\bibfield  {title} {\enquote {\bibinfo {title} {Mapping the
  twist-angle disorder and landau levels in magic-angle graphene},}\ }\href
  {\doibase 10.1038/s41586-020-2255-3} {\bibfield  {journal} {\bibinfo
  {journal} {Nature}\ }\textbf {\bibinfo {volume} {581}},\ \bibinfo {pages}
  {47--52} (\bibinfo {year} {2020})}\BibitemShut {NoStop}%
\bibitem [{\citenamefont {Yoo}\ \emph {et~al.}(2019)\citenamefont {Yoo},
  \citenamefont {Engelke}, \citenamefont {Carr}, \citenamefont {Fang},
  \citenamefont {Zhang}, \citenamefont {Cazeaux}, \citenamefont {Sung},
  \citenamefont {Hovden}, \citenamefont {Tsen}, \citenamefont {Taniguchi},
  \citenamefont {Watanabe}, \citenamefont {Yi}, \citenamefont {Kim},
  \citenamefont {Luskin}, \citenamefont {Tadmor}, \citenamefont {Kaxiras},\
  and\ \citenamefont {Kim}}]{Yoo2019}%
  \BibitemOpen
  \bibfield  {author} {\bibinfo {author} {\bibfnamefont {Hyobin}\ \bibnamefont
  {Yoo}}, \bibinfo {author} {\bibfnamefont {Rebecca}\ \bibnamefont {Engelke}},
  \bibinfo {author} {\bibfnamefont {Stephen}\ \bibnamefont {Carr}}, \bibinfo
  {author} {\bibfnamefont {Shiang}\ \bibnamefont {Fang}}, \bibinfo {author}
  {\bibfnamefont {Kuan}\ \bibnamefont {Zhang}}, \bibinfo {author}
  {\bibfnamefont {Paul}\ \bibnamefont {Cazeaux}}, \bibinfo {author}
  {\bibfnamefont {Suk~Hyun}\ \bibnamefont {Sung}}, \bibinfo {author}
  {\bibfnamefont {Robert}\ \bibnamefont {Hovden}}, \bibinfo {author}
  {\bibfnamefont {Adam~W.}\ \bibnamefont {Tsen}}, \bibinfo {author}
  {\bibfnamefont {Takashi}\ \bibnamefont {Taniguchi}}, \bibinfo {author}
  {\bibfnamefont {Kenji}\ \bibnamefont {Watanabe}}, \bibinfo {author}
  {\bibfnamefont {Gyu-Chul}\ \bibnamefont {Yi}}, \bibinfo {author}
  {\bibfnamefont {Miyoung}\ \bibnamefont {Kim}}, \bibinfo {author}
  {\bibfnamefont {Mitchell}\ \bibnamefont {Luskin}}, \bibinfo {author}
  {\bibfnamefont {Ellad~B.}\ \bibnamefont {Tadmor}}, \bibinfo {author}
  {\bibfnamefont {Efthimios}\ \bibnamefont {Kaxiras}}, \ and\ \bibinfo {author}
  {\bibfnamefont {Philip}\ \bibnamefont {Kim}},\ }\bibfield  {title} {\enquote
  {\bibinfo {title} {Atomic and electronic reconstruction at the van der waals
  interface in twisted bilayer graphene},}\ }\href {\doibase
  10.1038/s41563-019-0346-z} {\bibfield  {journal} {\bibinfo  {journal} {Nature
  Materials}\ }\textbf {\bibinfo {volume} {18}},\ \bibinfo {pages} {448--453}
  (\bibinfo {year} {2019})}\BibitemShut {NoStop}%
\bibitem [{\citenamefont {Peotta}\ and\ \citenamefont
  {T{\"o}rm{\"a}}(2015)}]{Torma15}%
  \BibitemOpen
  \bibfield  {author} {\bibinfo {author} {\bibfnamefont {Sebastiano}\
  \bibnamefont {Peotta}}\ and\ \bibinfo {author} {\bibfnamefont {P{\"a}ivi}\
  \bibnamefont {T{\"o}rm{\"a}}},\ }\bibfield  {title} {\enquote {\bibinfo
  {title} {Superfluidity in topologically nontrivial flat bands},}\ }\href
  {https://doi.org/10.1038/ncomms9944} {\bibfield  {journal} {\bibinfo
  {journal} {Nature Communications}\ }\textbf {\bibinfo {volume} {6}},\
  \bibinfo {pages} {8944} (\bibinfo {year} {2015})}\BibitemShut {NoStop}%
\bibitem [{\citenamefont {Xie}\ \emph {et~al.}(2020)\citenamefont {Xie},
  \citenamefont {Song}, \citenamefont {Lian},\ and\ \citenamefont
  {Bernevig}}]{Bernevig19}%
  \BibitemOpen
  \bibfield  {author} {\bibinfo {author} {\bibfnamefont {Fang}\ \bibnamefont
  {Xie}}, \bibinfo {author} {\bibfnamefont {Zhida}\ \bibnamefont {Song}},
  \bibinfo {author} {\bibfnamefont {Biao}\ \bibnamefont {Lian}}, \ and\
  \bibinfo {author} {\bibfnamefont {B.~Andrei}\ \bibnamefont {Bernevig}},\
  }\bibfield  {title} {\enquote {\bibinfo {title} {Topology-bounded superfluid
  weight in twisted bilayer graphene},}\ }\href {\doibase
  10.1103/PhysRevLett.124.167002} {\bibfield  {journal} {\bibinfo  {journal}
  {Phys. Rev. Lett.}\ }\textbf {\bibinfo {volume} {124}},\ \bibinfo {pages}
  {167002} (\bibinfo {year} {2020})}\BibitemShut {NoStop}%
\bibitem [{\citenamefont {Julku}\ \emph {et~al.}(2020)\citenamefont {Julku},
  \citenamefont {Peltonen}, \citenamefont {Liang}, \citenamefont {Heikkil\"a},\
  and\ \citenamefont {T\"orm\"a}}]{Torma19}%
  \BibitemOpen
  \bibfield  {author} {\bibinfo {author} {\bibfnamefont {A.}~\bibnamefont
  {Julku}}, \bibinfo {author} {\bibfnamefont {T.~J.}\ \bibnamefont {Peltonen}},
  \bibinfo {author} {\bibfnamefont {L.}~\bibnamefont {Liang}}, \bibinfo
  {author} {\bibfnamefont {T.~T.}\ \bibnamefont {Heikkil\"a}}, \ and\ \bibinfo
  {author} {\bibfnamefont {P.}~\bibnamefont {T\"orm\"a}},\ }\bibfield  {title}
  {\enquote {\bibinfo {title} {Superfluid weight and
  berezinskii-kosterlitz-thouless transition temperature of twisted bilayer
  graphene},}\ }\href {\doibase 10.1103/PhysRevB.101.060505} {\bibfield
  {journal} {\bibinfo  {journal} {Phys. Rev. B}\ }\textbf {\bibinfo {volume}
  {101}},\ \bibinfo {pages} {060505} (\bibinfo {year} {2020})}\BibitemShut
  {NoStop}%
\bibitem [{\citenamefont {Hu}\ \emph {et~al.}(2019)\citenamefont {Hu},
  \citenamefont {Hyart}, \citenamefont {Pikulin},\ and\ \citenamefont
  {Rossi}}]{Rossi19}%
  \BibitemOpen
  \bibfield  {author} {\bibinfo {author} {\bibfnamefont {Xiang}\ \bibnamefont
  {Hu}}, \bibinfo {author} {\bibfnamefont {Timo}\ \bibnamefont {Hyart}},
  \bibinfo {author} {\bibfnamefont {Dmitry~I.}\ \bibnamefont {Pikulin}}, \ and\
  \bibinfo {author} {\bibfnamefont {Enrico}\ \bibnamefont {Rossi}},\ }\bibfield
   {title} {\enquote {\bibinfo {title} {Geometric and conventional contribution
  to the superfluid weight in twisted bilayer graphene},}\ }\href {\doibase
  10.1103/PhysRevLett.123.237002} {\bibfield  {journal} {\bibinfo  {journal}
  {Phys. Rev. Lett.}\ }\textbf {\bibinfo {volume} {123}},\ \bibinfo {pages}
  {237002} (\bibinfo {year} {2019})}\BibitemShut {NoStop}%
\bibitem [{\citenamefont {{Hofmann}}\ \emph {et~al.}(2019)\citenamefont
  {{Hofmann}}, \citenamefont {{Berg}},\ and\ \citenamefont
  {{Chowdhury}}}]{DC19}%
  \BibitemOpen
  \bibfield  {author} {\bibinfo {author} {\bibfnamefont {Johannes~S.}\
  \bibnamefont {{Hofmann}}}, \bibinfo {author} {\bibfnamefont {Erez}\
  \bibnamefont {{Berg}}}, \ and\ \bibinfo {author} {\bibfnamefont {Debanjan}\
  \bibnamefont {{Chowdhury}}},\ }\bibfield  {title} {\enquote {\bibinfo {title}
  {{Superconductivity, pseudogap, and phase separation in topological flat
  bands: a quantum Monte Carlo study}},}\ }\href@noop {} {\bibfield  {journal}
  {\bibinfo  {journal} {arXiv e-prints}\ ,\ \bibinfo {eid} {arXiv:1912.08848}}
  (\bibinfo {year} {2019})},\ \Eprint {http://arxiv.org/abs/1912.08848}
  {arXiv:1912.08848 [cond-mat.str-el]} \BibitemShut {NoStop}%
\bibitem [{\citenamefont {Mahan}(2000)}]{mahan2000many-particle}%
  \BibitemOpen
  \bibfield  {author} {\bibinfo {author} {\bibfnamefont {Gerald}\ \bibnamefont
  {Mahan}},\ }\href@noop {} {\emph {\bibinfo {title} {Many-particle physics}}}\
  (\bibinfo  {publisher} {Kluwer Academic/Plenum Publishers},\ \bibinfo
  {address} {New York},\ \bibinfo {year} {2000})\BibitemShut {NoStop}%
\bibitem [{\citenamefont {Margine}\ and\ \citenamefont
  {Giustino}(2013)}]{PhysRevB.87.024505}%
  \BibitemOpen
  \bibfield  {author} {\bibinfo {author} {\bibfnamefont {E.~R.}\ \bibnamefont
  {Margine}}\ and\ \bibinfo {author} {\bibfnamefont {F.}~\bibnamefont
  {Giustino}},\ }\bibfield  {title} {\enquote {\bibinfo {title} {Anisotropic
  migdal-eliashberg theory using wannier functions},}\ }\href {\doibase
  10.1103/PhysRevB.87.024505} {\bibfield  {journal} {\bibinfo  {journal} {Phys.
  Rev. B}\ }\textbf {\bibinfo {volume} {87}},\ \bibinfo {pages} {024505}
  (\bibinfo {year} {2013})}\BibitemShut {NoStop}%
\bibitem [{\citenamefont {Vidberg}\ and\ \citenamefont
  {Serene}(1977)}]{10.1007/BF00655090}%
  \BibitemOpen
  \bibfield  {author} {\bibinfo {author} {\bibfnamefont {H.~J.}\ \bibnamefont
  {Vidberg}}\ and\ \bibinfo {author} {\bibfnamefont {J.~W.}\ \bibnamefont
  {Serene}},\ }\bibfield  {title} {\enquote {\bibinfo {title} {Solving the
  eliashberg equations by means ofn-point pad{\'e}approximants},}\ }\href
  {\doibase 10.1007/BF00655090} {\bibfield  {journal} {\bibinfo  {journal}
  {Journal of Low Temperature Physics}\ }\textbf {\bibinfo {volume} {29}},\
  \bibinfo {pages} {179--192} (\bibinfo {year} {1977})}\BibitemShut {NoStop}%
\bibitem [{\citenamefont {Leavens}\ and\ \citenamefont
  {Ritchie}(1985)}]{LEAVENS1985137}%
  \BibitemOpen
  \bibfield  {author} {\bibinfo {author} {\bibfnamefont {C.R.}\ \bibnamefont
  {Leavens}}\ and\ \bibinfo {author} {\bibfnamefont {D.S.}\ \bibnamefont
  {Ritchie}},\ }\bibfield  {title} {\enquote {\bibinfo {title} {Extension of
  the n-point pad{\'e} approximants solution of the eliashberg equations to t
  $~$ tc},}\ }\href {\doibase https://doi.org/10.1016/0038-1098(85)90112-7}
  {\bibfield  {journal} {\bibinfo  {journal} {Solid State Communications}\
  }\textbf {\bibinfo {volume} {53}},\ \bibinfo {pages} {137 -- 142} (\bibinfo
  {year} {1985})}\BibitemShut {NoStop}%
\bibitem [{\citenamefont {Marsiglio}\ \emph {et~al.}(1988)\citenamefont
  {Marsiglio}, \citenamefont {Schossmann},\ and\ \citenamefont
  {Carbotte}}]{PhysRevB.37.4965}%
  \BibitemOpen
  \bibfield  {author} {\bibinfo {author} {\bibfnamefont {F.}~\bibnamefont
  {Marsiglio}}, \bibinfo {author} {\bibfnamefont {M.}~\bibnamefont
  {Schossmann}}, \ and\ \bibinfo {author} {\bibfnamefont {J.~P.}\ \bibnamefont
  {Carbotte}},\ }\bibfield  {title} {\enquote {\bibinfo {title} {Iterative
  analytic continuation of the electron self-energy to the real axis},}\ }\href
  {\doibase 10.1103/PhysRevB.37.4965} {\bibfield  {journal} {\bibinfo
  {journal} {Phys. Rev. B}\ }\textbf {\bibinfo {volume} {37}},\ \bibinfo
  {pages} {4965--4969} (\bibinfo {year} {1988})}\BibitemShut {NoStop}%
\bibitem [{\citenamefont {Beach}\ \emph {et~al.}(2000)\citenamefont {Beach},
  \citenamefont {Gooding},\ and\ \citenamefont {Marsiglio}}]{PhysRevB.61.5147}%
  \BibitemOpen
  \bibfield  {author} {\bibinfo {author} {\bibfnamefont {K.~S.~D.}\
  \bibnamefont {Beach}}, \bibinfo {author} {\bibfnamefont {R.~J.}\ \bibnamefont
  {Gooding}}, \ and\ \bibinfo {author} {\bibfnamefont {F.}~\bibnamefont
  {Marsiglio}},\ }\bibfield  {title} {\enquote {\bibinfo {title} {Reliable
  pad\'e analytical continuation method based on a high-accuracy symbolic
  computation algorithm},}\ }\href {\doibase 10.1103/PhysRevB.61.5147}
  {\bibfield  {journal} {\bibinfo  {journal} {Phys. Rev. B}\ }\textbf {\bibinfo
  {volume} {61}},\ \bibinfo {pages} {5147--5157} (\bibinfo {year}
  {2000})}\BibitemShut {NoStop}%
\end{thebibliography}%

\end{document}